\documentclass[twocolumn]{aa}

\usepackage{graphicx}
\usepackage{natbib}
\bibpunct{(}{)}{;}{a}{}{,}   
\usepackage{txfonts}
\usepackage{xspace}
\usepackage{tabularx}
\usepackage{etoolbox} 
\usepackage{color}
\makeatletter
\patchcmd{\AA@hyperref}
  {\AddToHook{begindocument}{showkeys}{before}{nameref}}
  {}{}{}
\makeatother

\newcommand{\ha} {\mbox{H$\alpha$}}
\newcommand{\hb} {\mbox{H$\beta$}}
\newcommand{\hg} {\mbox{H$\gamma$}}
\newcommand{\hd} {\mbox{H$\delta$}}
\newcommand{\Nai}{\ion{Na}{i}}
\newcommand{\Naid}{\ion{Na}{id}}
\newcommand{\Feii} {\ion{Fe}{ii}}

\newcommand{\Caii} {\ion{Ca}{ii}}

\newcommand{\Scii} {\ion{Sc}{ii}}
\newcommand{\Baii} {\ion{Ba}{ii}}

\begin{document}
\title{SN 2025aedz: A typical short-plateau type IIP supernova with rapid post-peak decline }

   \author{Bo Wang\inst{1},
         Luhan Li\inst{1},
         Chengyuan Wu\inst{1},
         Jujia Zhang\inst{1},
         Yongzhi Cai\inst{2},
         Yize Dong\inst{3},
         Xuefei Chen\inst{1},
          \and
          Zhanwen Han\inst{1}     
          }
   \institute{
   International Centre of Supernovae (ICESUN), Yunnan Key Laboratory of Supernova Research, Yunnan Observatories, Chinese Academy of Sciences, Kunming 650216, China;
   {\it wangbo@ynao.ac.cn; liluhan@ynao.ac.cn; zhanwenhan@ynao.ac.cn}
    \and
    INAF - Osservatorio Astronomico di Padova, Vicolo dell’Osservatorio 5, 35122 Padova, Italy
     \and
     Center for Astrophysics | Harvard \& Smithsonian, 60 Garden Street, Cambridge, MA 02138-1516, USA
    }

   \date{}

\abstract{
Type IIP supernovae (SNe IIP) are the most common subclass of core-collapse SNe in observations. 
However, SNe IIP with short plateaus of the order of tens of days are rarely observed. 
The progenitors for this kind of SN can help to address the red supergiant issue in stellar evolution. 
In this article, we report optical photometry and spectroscopy of SN\,2025aedz, a rapidly post-peak declining SN IIP with a typical short plateau. 
It exhibits a peak absolute magnitude of $M_r=-17.16\pm0.03$\,mag. 
The $r$-band light curve shows a steep early post-peak decline of $\sim5\,\mathrm{mag}\,(100\,\mathrm{d})^{-1}$ followed by a relatively short plateau, with a plateau duration of $\sim50\pm3$\,d. 
The overall spectral evolution is consistent with that of normal SNe~IIP, showing a blue continuum with prominent Balmer P-Cygni profiles during the photospheric phase, followed by the gradual strengthening of hydrogen and metal lines as the ejecta cools down, although the metal lines remain weak and the expansion velocities decline rapidly.
SN\,2025aedz is similar to the short-plateau SN\,2018gj in its overall evolution, whereas its pronounced early decline resembles that of SN\,2023ufx, which has the shortest plateau duration known so far.
The radioactive tail of the bolometric light curve implies a synthesized $^{56}$Ni mass of $\sim0.03\pm0.01\,M_\odot$. 
Motivated by the steep early decline, we performed radiation hydrodynamic modelling using \texttt{MESA+STELLA} by exploring different circumstellar material configurations to reproduce its early bolometric light curve. 
These simulations indicate that SN\,2025aedz originated from a progenitor with a relatively low-mass hydrogen envelope, possibly produced through enhanced mass loss or binary interaction, while the steep early decline is likely explained by additional luminosity from circumstellar interaction.
}


\titlerunning{SN~2025aedz}

\authorrunning{B. Wang et al.}

\maketitle
\nolinenumbers
\section{Introduction} 

Core-collapse supernovae (CC SNe) originate from the evolution of 
massive stars with initial masses higher than $\sim8\,M_\odot$,
leaving behind a neutron star or a black hole after a SN
explosion (see e.g. \citealt{Heger2003ApJ...591..288H,Smartt2009ARA&A, Burrows2024ApJ...964L..16B}).
They play a major role in many aspects of astrophysics, as follows:
(1) The observations of CC SNe can be used to constrain the star formation history of the Universe and to test the stellar evolution theory (e.g. \citealt{Madau1998MNRAS.297L..17M,Dahlen2004ApJ...613..189D,Smartt2015PASA...32...16S}). 
(2) During their explosions, they can be  used to detect some fundamental physics phenomena, such as neutrinos, shock breakout emissions and gravitational wave signals (e.g. \citealt{Li2024Natur,Zhang2024ApJ}). 
(3) They play a key role in the chemical evolution of galaxies owing to the main contribution of 
heavy elements to their host galaxies, especially for intermediate-mass elements (see e.g. \citealt{Matteucci1986,Woosley1995ApJS..101..181W,Arcones2023A&ARv..31....1A}).
(4) They are important kinetic-energy sources for the evolution of galaxies due to energetic explosions 
(see e.g. \citealt{Powell2011MNRAS,LiY2018ApJ}), and 
most of cosmic rays are thought to be accelerated by SN remnants  
(see e.g. \citealt{Helder2009Sci,Vink2020pesr.book.....V}).

On the basis of the difference of their explosion mechanisms, 
CC SNe can be divided into iron core-collapse (Fe CC) SNe and electron-capture (EC) SNe, 
in which Fe CC SNe occupy the vast majority (see e.g. \citealt{Hiramatsu2021NatAs...5..903H,Wang2026RAA....26c2001W}).
It has been thought that 
the observational diversity of CC SNe results from the inherent variety of their progenitors 
(for a review see \citealt{Smartt2009ARA&A}).
According to the difference of their observable characteristics (i.e., light curves and spectral lines),
CC SNe can be divided into hydrogen-deficient type Ib/c SNe (SNe Ib/c)  and hydrogen-rich type II SNe (SNe II), among which SNe II are the most commonly observed  (see e.g. \citealt{Filippenko1997ARA&A..35..309F,LiWeidong2011MNRAS.412.1441L,Parrent2014,Shivvers2017PASP..129e4201S}). 
SNe II show a wide diversity likely owing to the amount of hydrogen at the moment of core collapse, in which the most common subclasses are SNe IIP and IIL (see e.g. \citealt{Barbon1979A&A....72..287B,Filippenko1997ARA&A..35..309F,Arcavi2017hsn..book..239A}). 

SNe IIP have a durable plateau stage in luminosity after maximum, whereas SNe IIL present a linear decline after peak luminosity.
Although SNe IIP and IIL show distinct light-curve shapes, the properties of their photometry and spectra exhibit a continuous population of these events. 
Compared with SNe IIP, SNe IIL have a brighter peak luminosity,
faster declining light curves, shallow $\rm H\alpha$ absorption and higher expansion velocities (see e.g. \citealt{Anderson2014ApJ...786...67A,Sanders2015ApJ...799..208S,Galbany2016AJ....151...33G,LinHan2025MNRAS.540.2591L}).
By establishing a complete sample of 211 SNe within 40\,Mpc,
\citet{Ma2025A&A...698A.305M} estimated that SNe IIP account for about 70\% of SNe II.
Importantly,
it has been proposed that SNe IIP can be used to measure the cosmological distance, further constraining the Hubble constant (see e.g. \citealt{Kirshner1974ApJ...193...27K,Hamuy2002ApJ...566L..63H,LinHan2025MNRAS.540.2591L,LiLuhan2026ApJ..1002...68L}).

In observations, SNe IIP have diverse luminosities, with peak absolute magnitudes in the $V$ band ranging from $-14$ to $-18$\,mag (see e.g. \citealt{Anderson2014ApJ...786...67A,DasKaustav2025PASP..137d4203D}).
Especially, 
their light curves show a plateau stage for a few months,
followed by a rapidly declining tail powered by 
the radioactive decay of $^{56}\rm Co$$\rightarrow$$^{56}\rm Fe$
(see e.g. \citealt{Teja2023ApJ...954..155T,WangZeyi2026ApJ..1001..156W}).
The long-lasting plateau light curve is mainly powered by 
the recombination of ionized hydrogen in the expanding envelope 
(see e.g. \citealt{Popov1993ApJ...414..712P,Kasen2009ApJ...703.2205K}).
The plateau length in most SNe IIP ranges from $70$\,d to $120$\,d, in which the average length is around $100$\,d (see e.g. \citealt{Sahu2006MNRAS.372.1315S,Pastorello2009MNRAS.394.2266P,Hiramatsu2021ApJ...913...55H,Yang2021A&A...655A..90Y,Teja2022ApJ...930...34T,Ravi2025ApJ...982...12R,LiLuhan2026A&A...714A.430L}).

Supernovae with longer plateaus are mostly discovered as low-luminosity IIP events with progenitor masses 
of $\sim10-15\,M_\odot$ (see e.g. \citealt{Spiro2014MNRAS.439.2873S,LinHan2025MNRAS.540.2591L,LiLuhan2026ApJ..1002...68L,Das2026PASP..138b4204D,Teja2026ApJ..1005....4T}). 
It has been suggested that IIP events with short plateaus are rare in observations and remain poorly reproduced by theoretical models (see e.g. \citealt{Eldridge2018PASA...35...49E,Hiramatsu2021ApJ...913...55H}).
The short plateaus in SNe IIP are usually explained by stripping the hydrogen envelope of their progenitors via wind mass loss or presence of a secondary star. 

Red supergiants (RSGs) have been directly identified 
in pre-explosion images as the progenitors of $\sim20$ SNe IIP, 
among which the initial masses are in the range of $\sim8-17\,M_\odot$ (see e.g. \citealt{Maund2005MNRAS.364L..33M,Smartt2009ARA&A,Smartt2015PASA...32...16S,
XiangDanfeng2024ApJ...969L..15X,
LuoJingxiao2025ApJ...982L..55L,VanDyk2025Galax..13...33V}).
However, the initial masses of 
the observed RSG population in the Local Group galaxies have been discovered up to $25\,M_\odot$,
known as the RSG issue in stellar evolution (see e.g. \citealt{Smartt2009ARA&A,
Smartt2015PASA...32...16S, Rodriguez2022MNRAS.515..897R}).
Up to now, the RSG issue is still a debated topic  owing to highly 
uncertainties in late-stage evolution of massive stars and 
the small sample of SN IIP progenitors identified
(see e.g. \citealt{Davies2020MNRAS.496L.142D,Beasor2025ApJ...979..117B}).
\citet{Hiramatsu2021ApJ...913...55H} suggested that massive RSGs 
as the progenitors for short-plateau SNe can help to clarify the RSG issue.

In this paper, we present the optical photometric and spectroscopic observations of SN~2025aedz, 
an SN IIP characterized by a rapid post-peak decline and a typical short-plateau phase.
In Sect. 2, we report the discovery and data reduction of SN~2025aedz.
We present the photometric evolution in Sect. 3 and the hydrodynamical modelling for 
light curves in Sect. 4. The spectroscopic evolution is shown in Sect. 5.
A discussion is given in Sect. 6, and a summary is provided in Sect. 7.

\section{Observations and data reduction} \label{sect:observation}

\begin{figure*}
    \centering
    \begin{minipage}{0.45\textwidth}
        \centering
        \includegraphics[width=\linewidth]{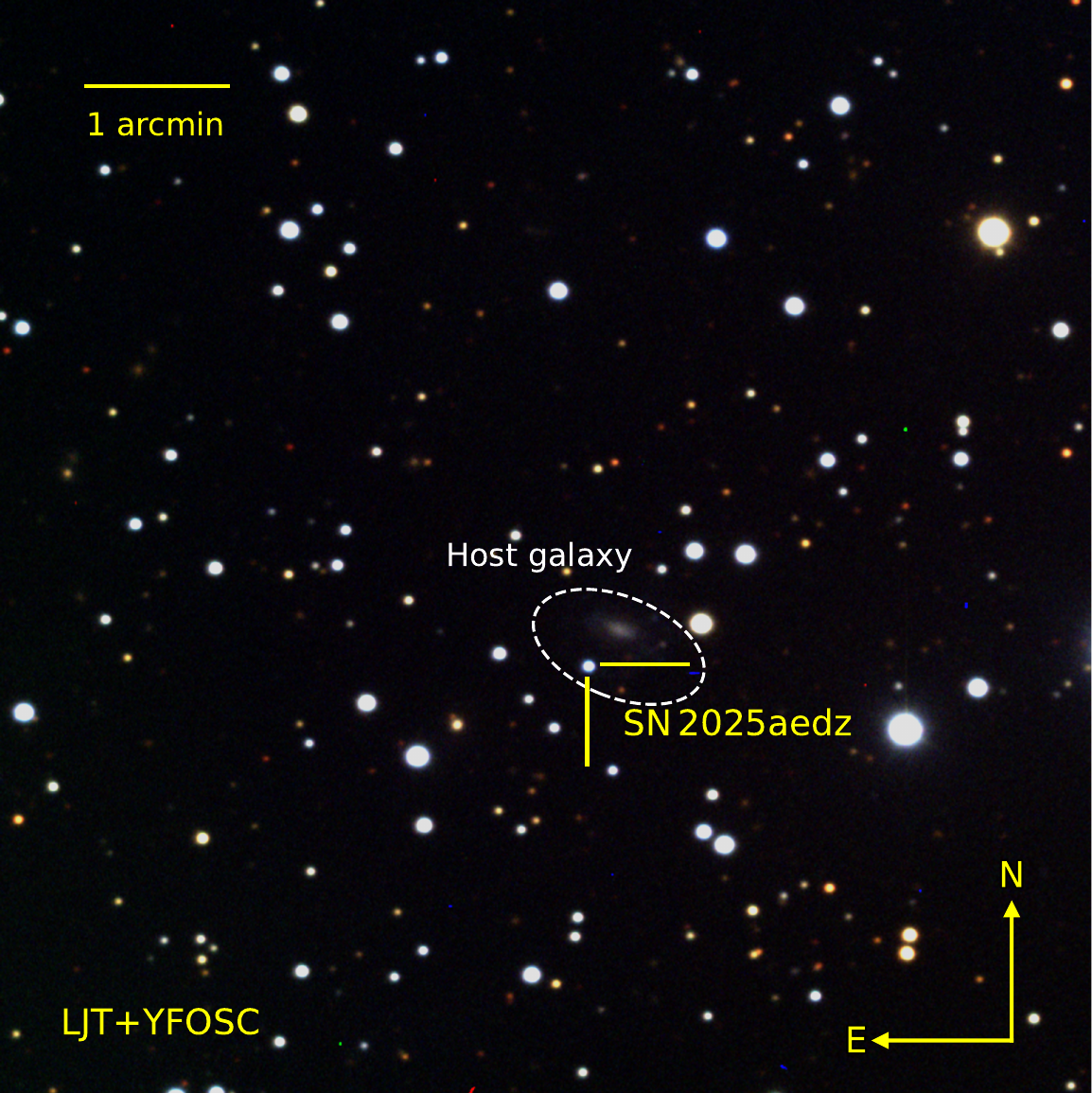}
    \end{minipage}
    \begin{minipage}{0.45\textwidth}
        \centering
        \includegraphics[width=\linewidth]{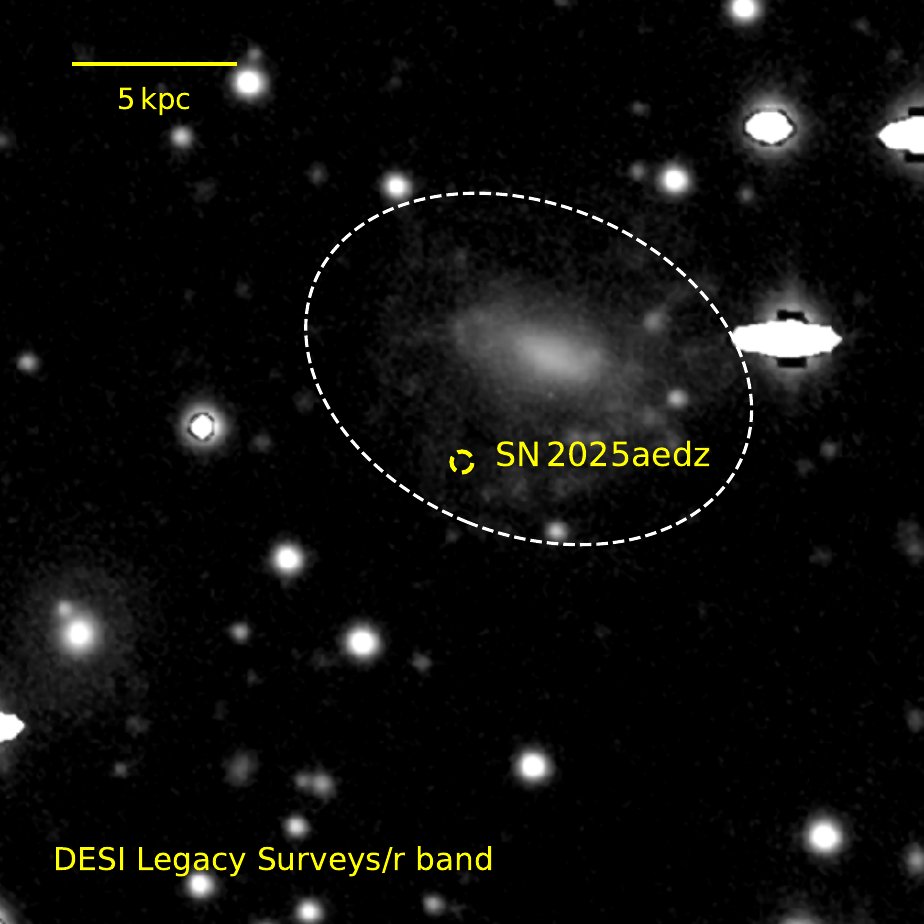}
    \end{minipage}
    \caption{
    Location of SN\,2025aedz in the host WISEA J055044.99+585507.5. 
    \textit{Left panel:} Image captured by the Lijiang 2.4\,m telescope (LJT) using the $gri$ filters, through the combination of data collected across multiple observational epochs.
    \textit{Right panel:} Zoomed-in $r$-band image from the DESI Legacy Surveys, providing a clearer view of the host galaxy structure.
    The dashed white line indicates the approximate outline of the host galaxy, while the yellow circle marks the location of SN\,2025aedz.}
   \label{Fig:location}
\end{figure*}

\subsection{Discovery}

SN\,2025aedz was discovered on 18.81 November 2025 with a clear-filter magnitude of $17.19\, \mathrm{mag}$, corresponding to MJD = 60997.81  (see \citealt{2025TNSTR4639....1S}). 
After its discovery, it was classified as an SN II with a redshift of $0.012$ (see \citealt{2025TNSCR4740....1B}).
Its J2000 coordinates are $\rm \alpha = 05^{h}50^{m}46.633^{s}$, $\rm \delta = +58^{\circ}54^{\prime}53.17^{\prime\prime}$, placing it $14.4^{\prime\prime}$ south and $24.6^{\prime\prime}$ east of the core of the host galaxy WISEA J055044.99+585507.5.
At a distance of 47.2\,Mpc, as estimated using the standard candle method (SCM) described below, the angular offset corresponds to a
projected separation of approximately 6\,kpc from the centre of the
host galaxy.
Figure~\ref{Fig:location} illustrates the location of SN\,2025aedz within the host galaxy.
In Table~\ref{tab:SN_properties}, we list the basic properties of SN\,2025aedz.

\begin{figure}
   \centering
   \includegraphics[width = \linewidth]{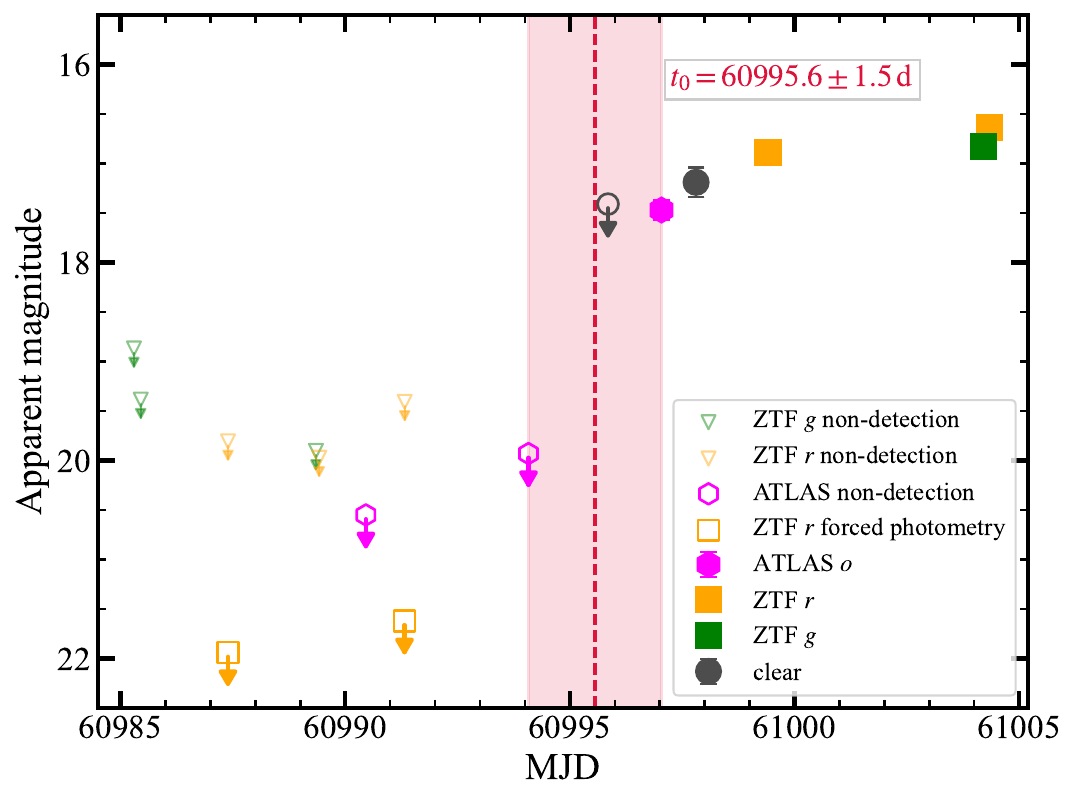}
      \caption{Early-time detections and non-detection limits of SN\,2025aedz. 
      The dashed red line marks the midpoint between the last reliable non-detection and the first detection, which is adopted as the estimated explosion epoch. 
      Note that the clear-band non-detection at MJD~=~60995.85 is not used due to its shallow limiting magnitude of 17.41 mag.
      }
     \label{Fig:explosion_epoch}
\end{figure}

Figure~\ref{Fig:explosion_epoch} presents the early-time observation from the Asteroid Terrestrial-impact Last Alert System (ATLAS) survey, Zwicky Transient Facility (ZTF) and clear-filter observations.
\citet{2025TNSTR4639....1S} reported a last non-detection with a clear-filter magnitude limit of 17.41\,mag on MJD = 60995.85. 
Since the most recent non-detection reported by \citet{2025TNSTR4639....1S} has a relatively shallow limiting magnitude of only 17.41\,mag, we instead utilized the deeper $o$-band non-detection at MJD 60994.08 with a limit magnitude of 19.93 mag as the last non-detection. 
We then took the $o$-band detection at MJD 60997.04 with $17.47$ mag as the first detection. 
We adopted the midpoint between these two epochs as the estimated explosion epoch, with half of the time interval between them taken as the uncertainty. 
This gives an explosion epoch of $t_0=60995.6\pm1.5\,$d, which is adopted throughout this work.

\begin{table}
\centering
\caption{Basic information and plateau parameters in the $r$ band of SN\,2025aedz.}
\label{tab:SN_properties}
\renewcommand{\arraystretch}{1.2}
\begin{tabularx}{0.9 \linewidth}{@{}lX@{}}
\hline\hline
RA (J2000) & 05:50:46.633 \\
DEC (J2000) & +58:54:53.17\\
Redshift $z$ & $0.012\pm0.002$\\
Distance $d$ & $47.2 \pm 7.4 \, \mathrm{Mpc}$\\
$E(B - V)_{\mathrm{Gal}}$& $0.16 \, \mathrm{mag}$\\
$E(B - V)_{\mathrm{host}}$ & $\sim0 \, \mathrm{mag}$\\
Estimated Explosion (MJD) & $60995.6\pm1.5$\\
\hline 
Plateau duration (Pd) & 50$\pm$3\,d \\
Optically thick phase duration (OPTd) & 68$\pm$3\,d \\
Plateau length ($t_{\rm PT}$) & 78.7$\pm$0.5\,d \\
\hline
\end{tabularx}
\end{table}

No accurate measurement of the host galaxy's recession velocity or redshift is available in the public databases.
Due to the large projected separation between the SN and its host galaxy centre, no prominent host-galaxy emission or absorption lines are detected in the SN spectra.
We therefore estimated the redshift via spectral-template matching using both SNID-SAGE (see \citealt{Stoppa2026MNRAS.549g1066S}) and NGSF (see \citealt{2022TNSAN.191....1G}).
SNID-SAGE yields $z=0.010$--0.014, with good matches to SN\,1999em, SN\,2015bs, and SN\,2017pn.
NGSF gives $z=0.011$--0.013, with good matches to SN\,1999em, SN\,2013fs, and SN\,2014G.
Based on the overlap between the two methods, we adopted $z=0.012\pm0.002$ for SN\,2025aedz in this work.
The quoted uncertainty should therefore be regarded as a formal statistical error, while the total uncertainty may be larger because of systematic effects inherent to template-matching methods \citep{2020ApJ...895...32F}.

In addition, we detect Galactic \Nai\ D absorption in the early spectra, but find no clear host-galaxy \Nai\ D absorption features, likely because the SN is located at a large projected distance from the centre of its host galaxy.
We adopted a Galactic reddening of $E(B-V)_{\rm Gal} = 0.16$\,mag (see \citealt{Schlafly2011ApJ...737..103S}), assuming $R_V = 3.1$ (see \citealt{Cardelli1989ApJ...345..245C}).
We therefore assumed that the host-galaxy extinction is negligible, as it is expected to be much smaller than the Galactic extinction.
Accordingly, we adopted a total extinction of $E(B-V)_{\rm total} = 0.16$\,mag for SN\,2025aedz.

\subsection{Distance}

As no direct measurement of the host-galaxy recession velocity is available, we estimated the distance using the SCM for SNe II, which relies on the empirical luminosity–velocity relation (see \citealt{Hamuy2002ApJ...566L..63H}).
For SN\,2025aedz, we calculated the apparent $V$-band magnitude at 50\,d after explosion by interpolating the $g$- and $r$-band light curves to 50\,d and converting the interpolated photometry to $V$ using the SDSS transformation relations of \citet{Lupton2005SDSSTransform}.
This method provided a $V$-band magnitude of $m_V^{50}=17.67\pm0.04$.
In addition, we convolved the +49.2\,d spectrum with the filter response curves from the \texttt{speclite.filters} package in \texttt{python} and applied a flux correction using the $gr$ photometry obtained one day after the spectroscopic observation.
We obtained a $V$-band magnitude of $17.69\pm0.06$\,mag, which is consistent with the value previously derived using the colour transformation.
The \Feii\,$\lambda5169$ velocity of $v_{\rm FeII}=3780\pm450\,\mathrm{km\,s^{-1}}$ was taken from the spectra at $+49.2$\,d.
Using the median $V$-band magnitude derived from the two methods and \Feii velocity near 50\,d, we obtain an SCM distance of $47.2\pm7.4$\,Mpc based on the calibration of \citet{Hamuy2002ApJ...566L..63H}.
The quoted uncertainty includes the propagated uncertainties in the $\Feii$ velocity, photometric magnitude, and $H_0$, as well as the statistical scatter of the SCM calibration, but does not account for systematic uncertainties related to the explosion epoch or the applicability of the SCM to this non-standard SN~II.
This SCM distance is consistent, within the uncertainties, with the Hubble-flow distance of $\sim49$\,Mpc inferred from $z=0.012$, assuming a update cosmology with $H_0=73.04\pm1.04\,\mathrm{km\,s^{-1}\,Mpc^{-1}}$ (see \citealt{2022ApJ...934L...7R}). 
We therefore adopt the SCM distance throughout the following analysis.

We note that SN\,2025aedz is located outside the locus of normal-plateau SNe~IIP in several light-curve parameter planes and shows evidence of an unusually rapid early decline. 
Its SCM distance should therefore be regarded as an indicative estimate rather than a precision distance, because the calibration of \citet{Hamuy2002ApJ...566L..63H} was established primarily from ordinary SNe~IIP. 
The consistency with the independent Hubble-flow distance suggests that the calibration is not grossly inappropriate for this object, but possible systematic deviations associated with its short plateau and early-time circumstellar material (CSM) contribution cannot be excluded.

\subsection{Data reduction}

Optical photometry of SN\,2025aedz in the \textit{ugriz} bands was obtained with the Lijiang 2.4\,m telescope (LJT) equipped with the Yunnan Faint Object Spectrograph and Camera (YFOSC; see \citealt{2015RAA....15..918F,2019RAA....19..149W}).
Additional observations in the \textit{gri} bands were carried out with the Multi-mode Autonomous Terminal for Compatible Hybrid-optics (MATCH), operated by Wuhan University \citep{2026arXiv260938858X}. 

Aperture photometry for the LJT and MATCH images was performed using the AutoPhOT pipeline (see \citealt{2022A&A...667A..62B}).
For a small subset of images, particularly those obtained in the $u$ and $z$ bands, AutoPHOT could not complete the photometric reduction as the available catalogues contained too few suitable comparison stars, or because the images lacked a sufficient number of bright, isolated point sources for reliable full width at half maximum (FWHM) estimation. 
We therefore performed aperture photometry using comparison stars with the standard \texttt{IRAF} routines (see e.g. \citealt{Tody1986SPIE..627..733T,Tody1993ASPC...52..173T}).
The resulting LJT and MATCH photometric measurements are listed in Table~\ref{tab:phot_data}.

Additionally, ZTF also observed SN\,2025aedz (ZTF25acegczg) in the \textit{g} and \textit{r} bands.
The public ZTF data were downloaded to complement our dataset \citep{2019PASP..131a8003M}\footnote{\url{https://alerce.online/object/ZTF25acegczg}}.
The \textit{c}- and \textit{o}-band photometry from the ATLAS survey was also included \citep{2018PASP..130f4505T,2020PASP..132h5002S}.
The forced-photometry light curves were retrieved from the ATLAS data-release server \citep{Shingles2021TNSAN...7....1S}\footnote{\url{https://fallingstar-data.com/forcedphot/}}.
To improve the data quality, same-filter measurements were combined into daily stacks using the public stacking tool, with sigma clipping applied to reject outliers\footnote{\url{https://github.com/thespacedoctor/plot-results-from-atlas-force-photometry-service}}.

For SN\,2025aedz, spectroscopic observations were obtained with LJT/YFOSC \citep{2019RAA....19..149W}, and the observing log is provided in Table~\ref{table:specinfo}.
The data were reduced using standard \texttt{IRAF} procedures, including bias subtraction, flat-fielding, wavelength calibration, atmospheric-extinction correction, and flux calibration based on spectrophotometric standard stars observed at similar airmasses.
Telluric absorption features were removed as part of the reduction.

\section{Photometric evolution}
\begin{figure}[htbp]
   \centering
  \sidecaption
   \includegraphics[width = \linewidth]{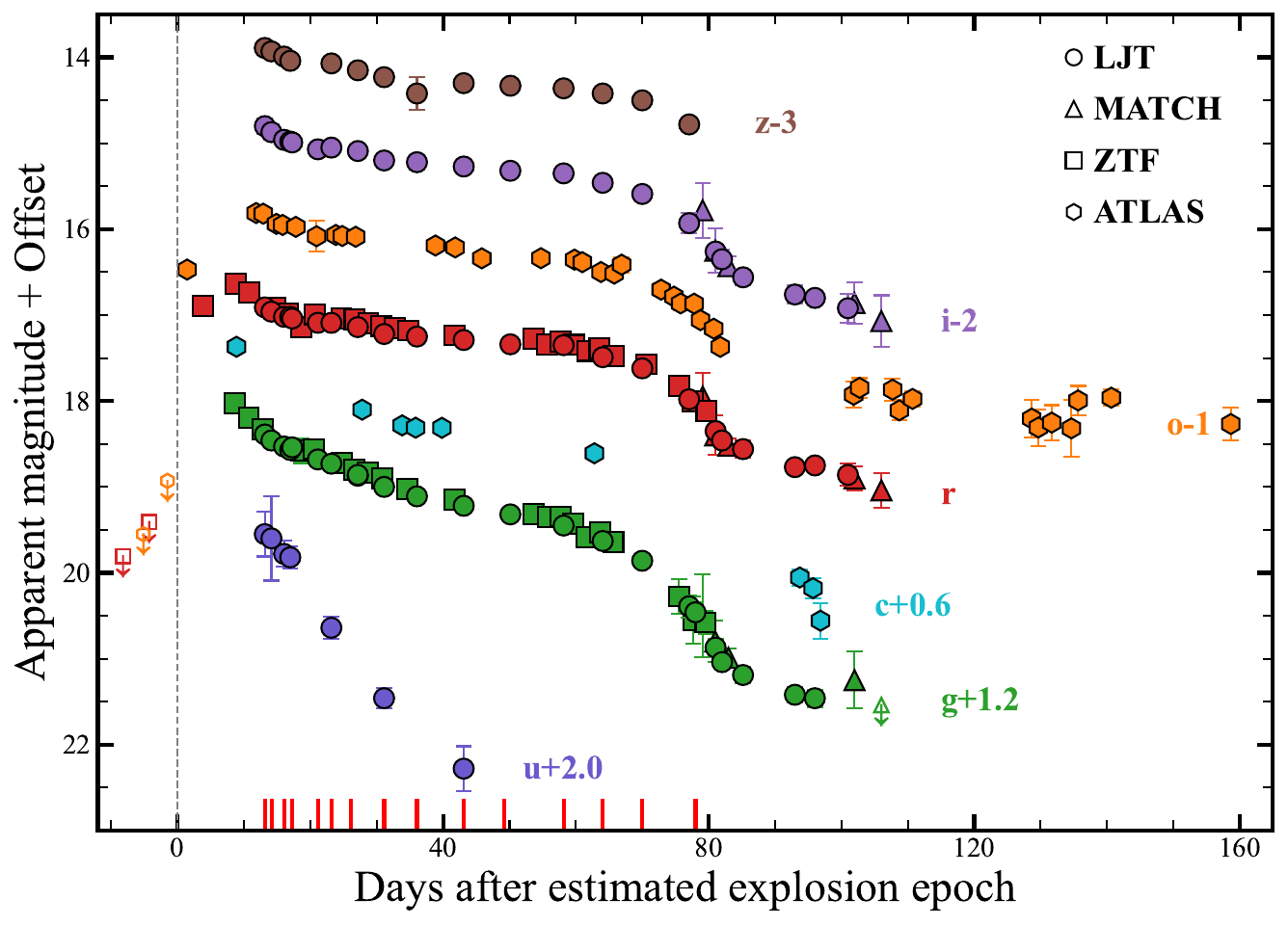}
   \caption{Multi-band light-curve evolution of SN\,2025aedz based on LJT, MATCH, ZTF and ATLAS. 
   The short red vertical ticks at the bottom mark the epochs of spectroscopic observations.}
    \label{Fig:LC}
\end{figure}  

\begin{table}
    \centering
    \caption{Three decline rates (in $\rm mag\, /100\,d$) of the multi-band light curves of SN\,2025aedz.}
    \renewcommand{\arraystretch}{1.2}
    \setlength{\tabcolsep}{8pt}
    \begin{tabular}{cccc}
        \hline \hline
        Band &  $s_1\,$(5-18\,d)  & $s_2$\,(20-65\,d)  & $s_3$\,(>90\,d)  \\
        \hline
        $u$ & \multicolumn{2}{c}{10.33$\pm$0.76} & -- \\
        $g$  &6.28$\pm$0.58 & 2.03$\pm$0.08 & 2.10$\pm$1.04 \\    
        $c$ & -- & 1.54$\pm$0.08 & -- \\
        $r$  & 4.86$\pm$0.40 & 0.83$\pm$0.07 & 1.48$\pm$1.08 \\
        $o$ & 3.85$\pm$0.30 & 0.92$\pm$0.04 & 0.40$\pm$0.23 \\
        $i$ & 4.46$\pm$2.02 & 0.88$\pm$0.16 & 3.22$\pm$0.50\\
        $z$ & -- & 0.74$\pm$0.14 & -- \\
        \hline
    \end{tabular}
    \label{tab:decline_rate}
\end{table}

\subsection{Apparent magnitude light curve} \label{Sect:Apparent_LC}

SN\,2025aedz was monitored photometrically over an extended period, and its multi-band optical light curves are presented in Figure~\ref{Fig:LC}.
The pre-maximum rising evolution is not very well observed.
We estimated the peak magnitudes by fitting the early-time light curve in $r$ band with a second-order polynomial, using a Markov chain Monte Carlo (MCMC) approach.
For the $r$ band, the fitted peak is $16.65\pm0.02$\,mag at $\mathrm{MJD}=61003.9\pm0.2$.
Taking the estimated explosion epoch of $t_0=60995.6\pm1.5$\,d, the corresponding rise time from zero flux at explosion to the peak flux is approximately $8.3$\,d, with an uncertainty of about $1.5$\,d.

To quantitatively characterize the shape of the light curve, we followed the description proposed by \citet{Anderson2014ApJ...786...67A}, who parameterized the evolution of SNe~IIP with three decline rates (in magnitudes per 100\,d), as follows:
(1) $s_1$ is the decline rate of the initial post-peak, steeper slope immediately following the peak.
(2) $s_2$ is the decline rate of the second, shallower slope, commonly referred to the plateau phase.
(3) $s_3$ is the linear decline rate of the slope reached after the transition from the plateau, commonly referred to the radioactive tail.
This parameterization has since become a standard framework for statistical analyses and comparative studies of SNe IIP \citep[see e.g.][]{Anderson2014ApJ...786...67A,Das2026PASP..138b4204D}.
We measured these slopes for each band, adopting the phase intervals of $5$--$18$\,d for $s_1$, $20$--$65$\,d for $s_2$, and $>90$\,d for $s_3$.
The resulting decline rates are listed in Table~\ref{tab:decline_rate}.
Note that for the limited $u$-band coverage the different phases are not distinguished, and a single decline rate is reported.
The relatively large uncertainty in $s_3$ for SN\,2025aedz arises from the limited and lower quality of the photometric data at late phases.
Additionally, the $o$-band light curve shows a shallower late-time decline than the $gri$ bands. 
However, the $o$-band measurements at phases later than $\sim120$\,d are close to the ATLAS detection limit and may be affected by larger photometric uncertainties \citep{2018PASP..130f4505T}.

For SNe~IIP, \citet{Anderson2014ApJ...786...67A} reported mean decline rates in the $V$ band of $2.65\pm1.50$\,mag\,(100\,d)$^{-1}$ for the $s_1$ phase, $1.27\pm0.93$\,mag\,(100\,d)$^{-1}$ for the $s_2$ (plateau) phase, and $1.47\pm0.82$\,mag\,(100\,d)$^{-1}$ for the $s_3$ (radioactive tail) phase, where the quoted uncertainties represent the standard deviations.
Using the $g$ band of SN\,2025aedz as the closest band to the $V$ band, its $s_1$ slope ($6.3\pm0.6$\,mag\,(100\,d)$^{-1}$) is markedly larger than this range.
The $r$- and $o$-band $s_1$ slopes ($4.9\pm0.4$ and $3.9\pm0.3$\,mag\,(100\,d)$^{-1}$, respectively) also exhibit a pronounced decline following the luminosity peak.
The $s_2$ and $s_3$ $g$-band slopes of SN\,2025aedz also lie close to the upper boundary of the SN~IIP distribution.
Furthermore, \citet{Das2026PASP..138b4204D} derived a median $r$-band plateau slope of $0.5^{+0.6}_{-0.4}$\,mag\,(100\,d)$^{-1}$ for SNe~IIP brighter than $-16$\,mag based on the ZTF survey, where the lower and upper limits correspond to the 16th and 84th
percentiles of the distribution, respectively.
The $r$-band plateau slope of SN\,2025aedz ($\sim0.8$\,mag\,(100\,d)$^{-1}$) is likewise on the higher side of this distribution.

\subsection{Plateau parameters}

\begin{figure}
   \centering
   \includegraphics[width = \linewidth]{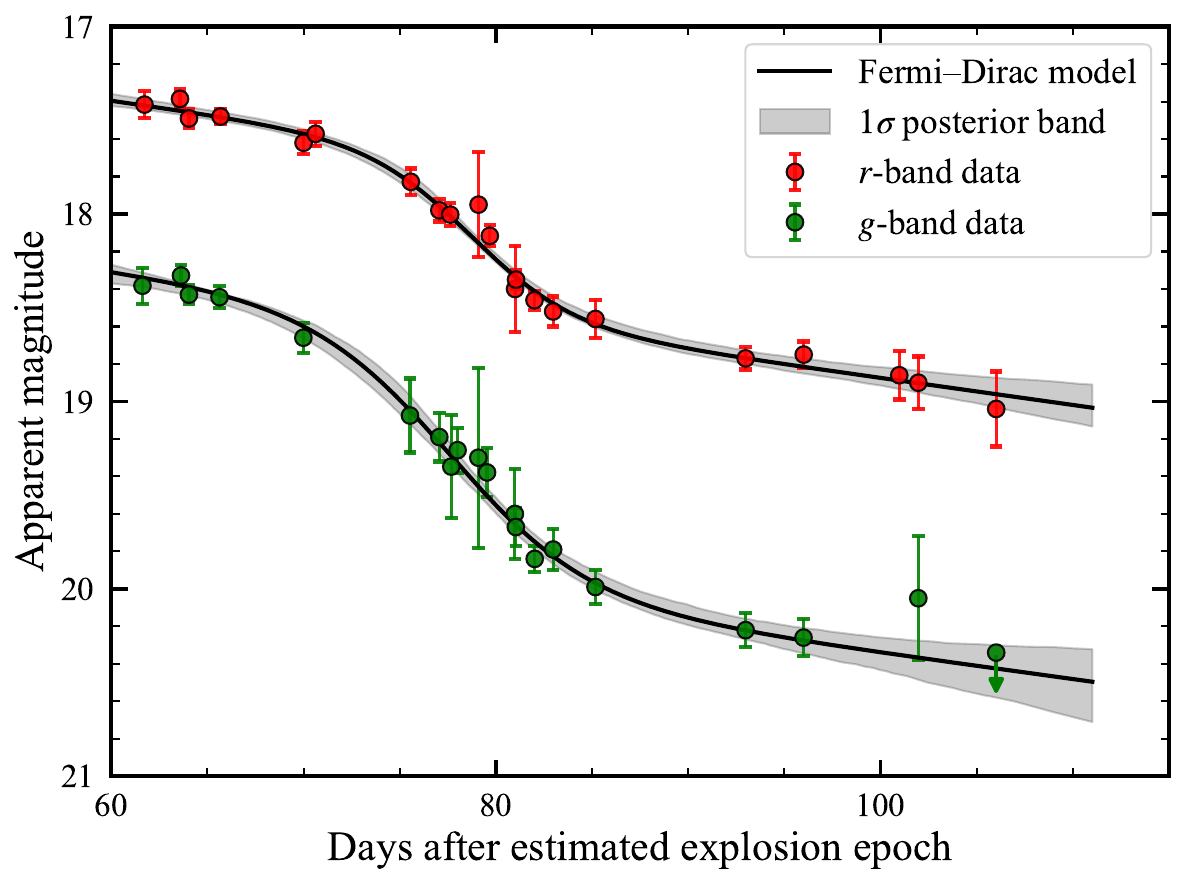}
   \caption{$gr$-band light-curve fitting of SN\,2025aedz from +60\,d to the last available epoch using a Fermi-Dirac phenomenological transition model and MCMC posterior sampling.
   The red and green points show the combined $r$- and $g$- band photometry from all instruments, respectively.
   The solid black curve marks the posterior-median model, and the  shaded grey area denotes the 1$\sigma$ posterior credible band of the model light curve.
   }
   \label{Fig:FDmodel}
\end{figure}

The transition from the end of the plateau phase to the tail can be described by the following phenomenological function:
\begin{equation}
m(t)=\frac{\Delta m}{1+e^{\left(t-t_{\mathrm{PT}}\right) / w_0}}+p_0\left(t-t_{\mathrm{PT}}\right)+m_0,
\end{equation}
where $m(t)$ is the magnitude at time $t$ \citep[see e.g.][]{2010ApJ...715..833O,2016MNRAS.459.3939V,2021ApJ...906...56D,2021MNRAS.505.1742R,Ravi2025ApJ...982...12R}.
Here, $\Delta m$ is the depth of the drop\footnote{This parameter is denoted as $-a_0$ in other works; we rewrite it as $\Delta m$, which has a similar meaning to $m_{\rm tail}-m_{\rm end}$ and is negative for fading events.}, and $w_0$ measures the slope of the drop.
The parameter $t_{\mathrm{PT}}$ is the mid-point of the transition from the plateau to the radioactive tail, while $p_0$ constrains the slope before and after the drop, and $m_0$ serves as a fitted zero-point.
The first term of the equation is a Fermi-Dirac function, which provides a description of the transition between the plateau and radioactive phases but has no direct physical meaning.

We employed MCMC sampling based on equation (1) to fit the $gr$-band evolution of SN\,2025aedz.
Figure~\ref{Fig:FDmodel} presents the fitting result, and the corresponding corner plot is shown in Figure~\ref{Fig:FDmodel_corner} in the appendix.
The Fermi-Dirac phenomenological function provides a good fit to the transition of SN\,2025aedz.
For $r$-band light curve, the posterior distributions yield $t_{\mathrm{PT}}=78.7\pm0.5$\,d, $|\Delta m|=0.9^{+0.3}_{-0.2}$\,mag, and $w_0=2.7^{+0.8}_{-0.7}$\,d.
For $g$-band light curve, the posterior distributions yield $t_{\mathrm{PT}}=77.7^{+0.7}_{-0.9}$\,d, $|\Delta m|=1.5^{+0.3}_{-0.4}$\,mag, and $w_0=3.6^{+1.0}_{-1.1}$\,d.
Compared with the statistical results of \citet{2016MNRAS.459.3939V}, in which most SNe~IIP have $t_{\mathrm{PT}}\sim80$--$120$\,d, $|\Delta m|\sim1$--$2.5$\,mag, and $w_0\sim2$--$5$\,d, SN\,2025aedz occupies the short end of the $t_{\mathrm{PT}}$ distribution, has small $|\Delta m|$ values among the sample, while its $w_0$ is similar to the sample.

\citet{Anderson2014ApJ...786...67A} characterized the plateau of SNe~IIP using two timescales: the optically thick phase duration 'OPTd', defined as the interval from the explosion to the end of the plateau ($t_{\rm end}$), and the plateau duration 'Pd', defined as the length of the $s_2$ phase.
We measured these parameters for SN\,2025aedz, obtaining $\mathrm{Pd}\approx50\pm3$\,d and $\mathrm{OPTd}\approx68\pm3$\,d.
Compared with the mean OPTd of $83.7\pm16.7$\,d reported by \citet{Anderson2014ApJ...786...67A}, SN\,2025aedz lies towards the short end of the distribution.
\citet{Das2026PASP..138b4204D} derived median values of $\mathrm{OPTd}=90^{+16}_{-20}$\,d and $\mathrm{Pd}=76^{+15}_{-16}$\,d for their $r$-band ZTF sample, where the lower and upper limits correspond to the 16th and 84th percentiles of the distribution, respectively.
Compared with these values, the plateau length of SN\,2025aedz is also notably short.

\subsection{Absolute magnitude light curve}

\begin{figure}
   \centering
   \includegraphics[width = 1\columnwidth]{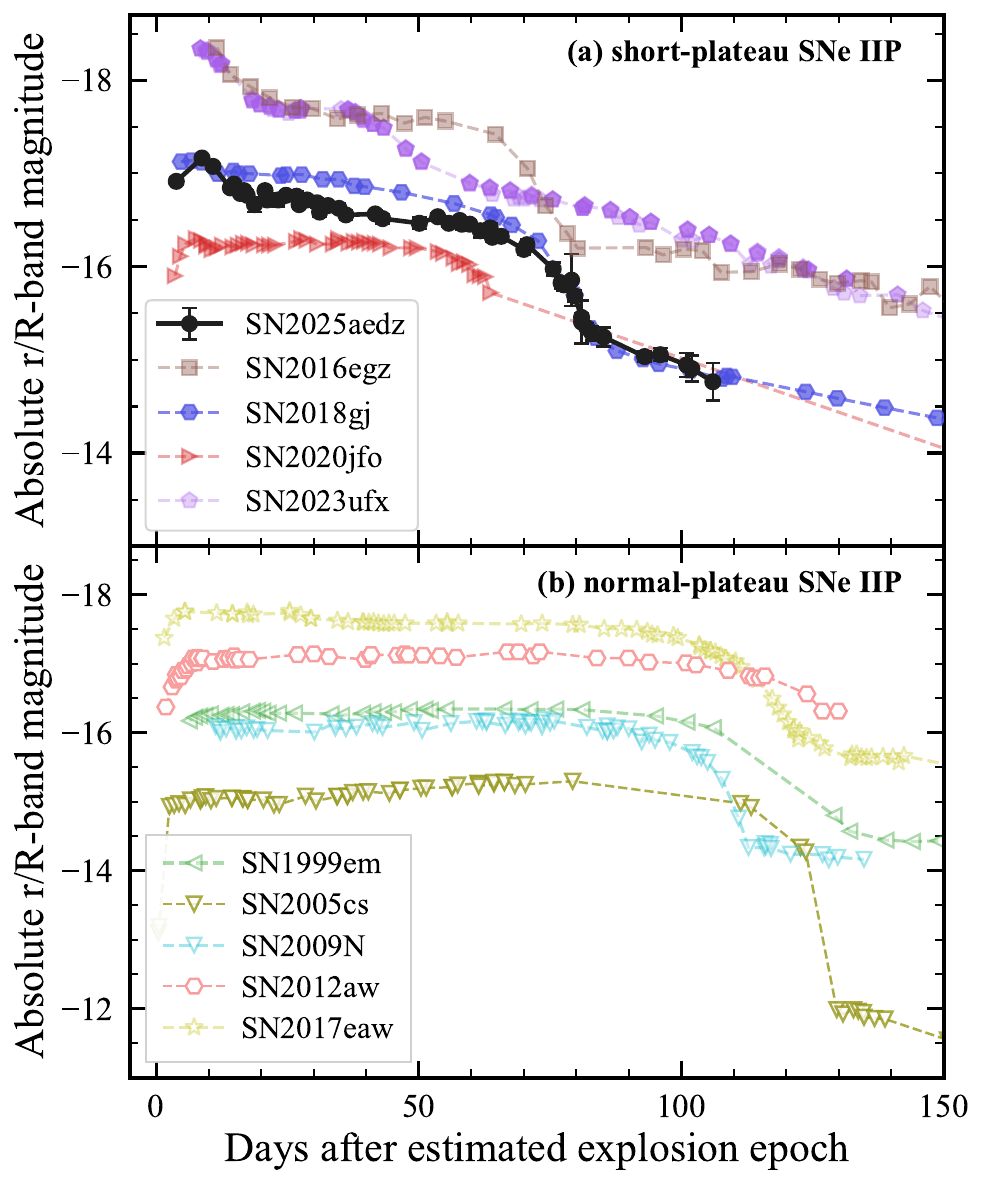}
  \caption{Comparisons between the absolute $r/R$-band light curves of SN\,2025aedz and those of well-observed SNe IIP.
    \textit{Panel~(a):} Representative short-plateau SNe IIP, shown with filled symbols.
    \textit{Panel~(b):} Typical normal-plateau SNe IIP with plateau durations of $\sim100--120$\,d, shown with open symbols.}
    \label{Fig:r_com}
\end{figure}

We constructed the absolute $r$-band light curve of SN\,2025aedz by adopting the distance and extinction listed in Table~\ref{tab:SN_properties}, as shown in Figure~\ref{Fig:r_com}.
SN\,2025aedz reaches a peak absolute $r$-band magnitude of $M_r=-17.16\pm0.03$\,mag by adopting the distance and extinction as given in Section~\ref{sect:observation}.
To place SN\,2025aedz in context, we also collected a sample of well-observed SNe~IIP for comparison, whose distances and extinction parameters are listed in Table~\ref{tab:SNe_II_info}.
To distinguish the diverse plateau evolution among SNe~IIP, we divided the comparison sample into two groups in Figure~\ref{Fig:r_com}. 
In this figure, panel (a) shows representative short-plateau SNe~IIP, while panel (b) presents typical normal-plateau SNe~IIP.

As shown in Figure~\ref{Fig:r_com}, typical SNe~IIP such as SN\,1999em end their plateau stage at $\sim100$--$130$\,d in panel (b), whereas the recent short-plateau events presented in panel~(a) all exhibit a noticeably shorter plateau.
A post-peak decline during the $s_1$ phase is also observed in some ordinary SNe IIP, although it is generally less pronounced in the comparison objects shown in panel (b).
In panel (a), SN\,2023ufx and SN\,2025aedz display particularly prominent $s_1$ declines.
By comparison, SN\,2025aedz shows a large $s_1$ slope of $\sim5$\,mag\,(100\,d)$^{-1}$, approaching the $\sim6$\,mag\,(100\,d)$^{-1}$ of SN\,2023ufx.
Its plateau length and plateau slope are very similar to those of SN\,2018gj.
The radioactive tail of SN\,2025aedz is comparable to those of SN\,2018gj and SN\,2020jfo.

\subsection{Colour evolution}

\begin{figure}
   \centering
   \includegraphics[width = 0.98\linewidth]{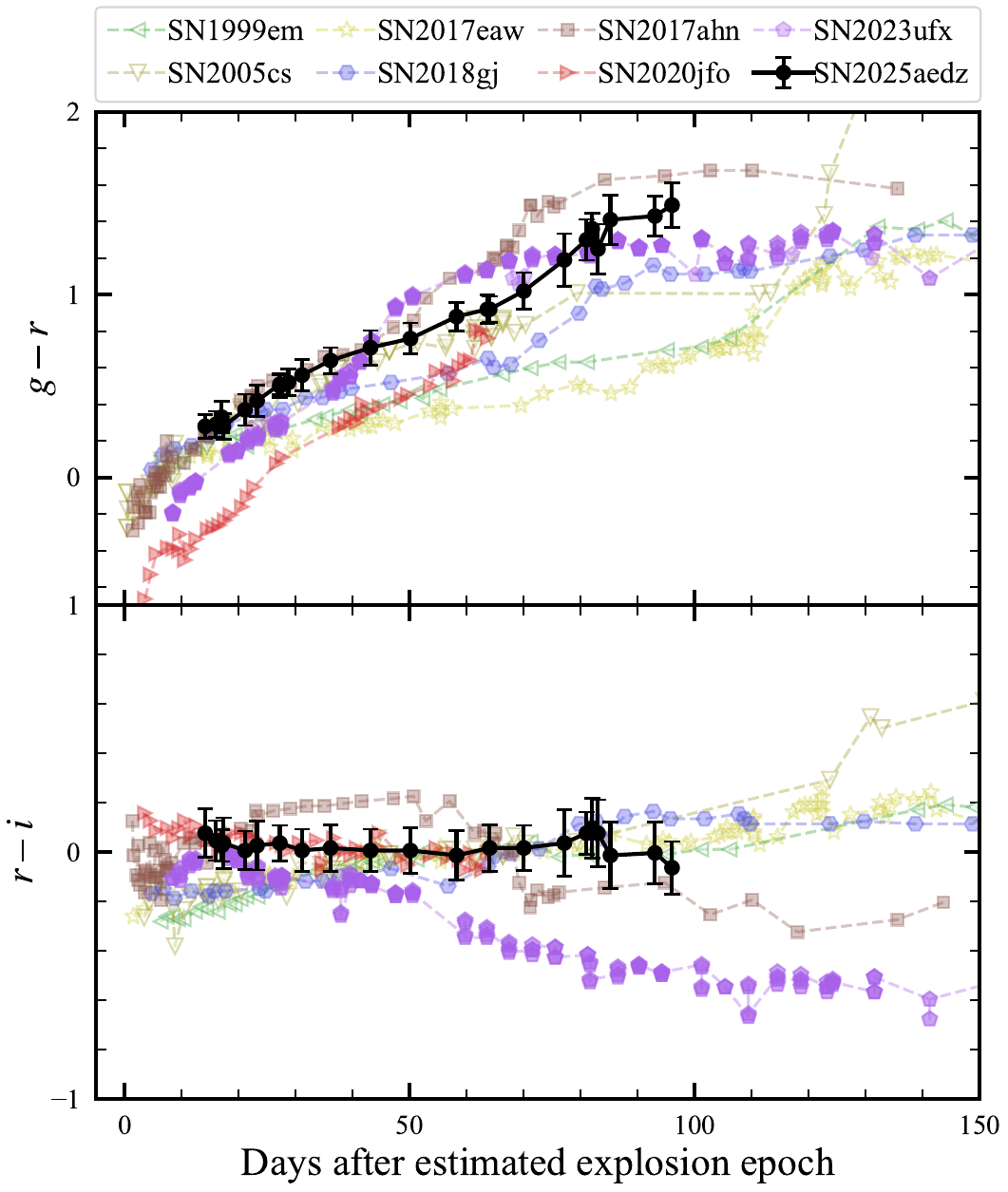}
   \caption{Colour evolution of SN\,2025aedz, alongside with that of other well-studied SNe II.
    All colours have been corrected for both Milky Way and host galaxy extinction.
   }
    \label{Fig:color}
\end{figure}

Figure~\ref{Fig:color} shows the dereddened $(g-r)$ and $(r-i)$ colour curves of SN\,2025aedz together with those of comparison SNe~II.
All colour measurements include corrections for both Milky Way foreground extinction and host-galaxy reddening.
When SDSS-band photometry was unavailable for the comparison sample, Johnson--Cousins magnitudes were transformed using the relations given by \citet{2006A&A...460..339J}.

After the explosion, the $(g-r)$ colour of SN\,2025aedz becomes progressively redder with time, following a trend similar to that of SN\,2018gj but slightly redder overall.
This monotonic reddening is a common feature of SNe~IIP and reflects the cooling of the expanding ejecta as the photosphere recedes and the hydrogen recombination front moves inward, leading to a steady decrease in the photospheric temperature during the plateau phase.
The redder $(g-r)$ colour of SN\,2025aedz may indicate a faster cooling and recombination of the hydrogen envelope.

The $(r-i)$ colour, in contrast, shows little variation throughout the evolution, again resembling that of SN\,2018gj.
This is expected since both bands are less sensitive to the temperature evolution than $(g-r)$, and the $(r-i)$ colour is more strongly affected by the strengthening of the broad \ha\ emission entering the $r$ band as the SN evolves.
In the $r-i$ panel, SN\,2017ahn and SN\,2023ufx exhibit a blueward trend after the end of the plateau, whereas this feature is not evident in the other comparison objects, including SN\,2025aedz.
This blueward trend may be related to the strengthening of emission features (e.g. \ha) in the $r$ band relative to the $i$ band during the transition to the nebular phase.

\subsection{Bolometric light curve} \label{sect:Bol}

\begin{figure}
   \centering
   \includegraphics[width = 0.98\linewidth]{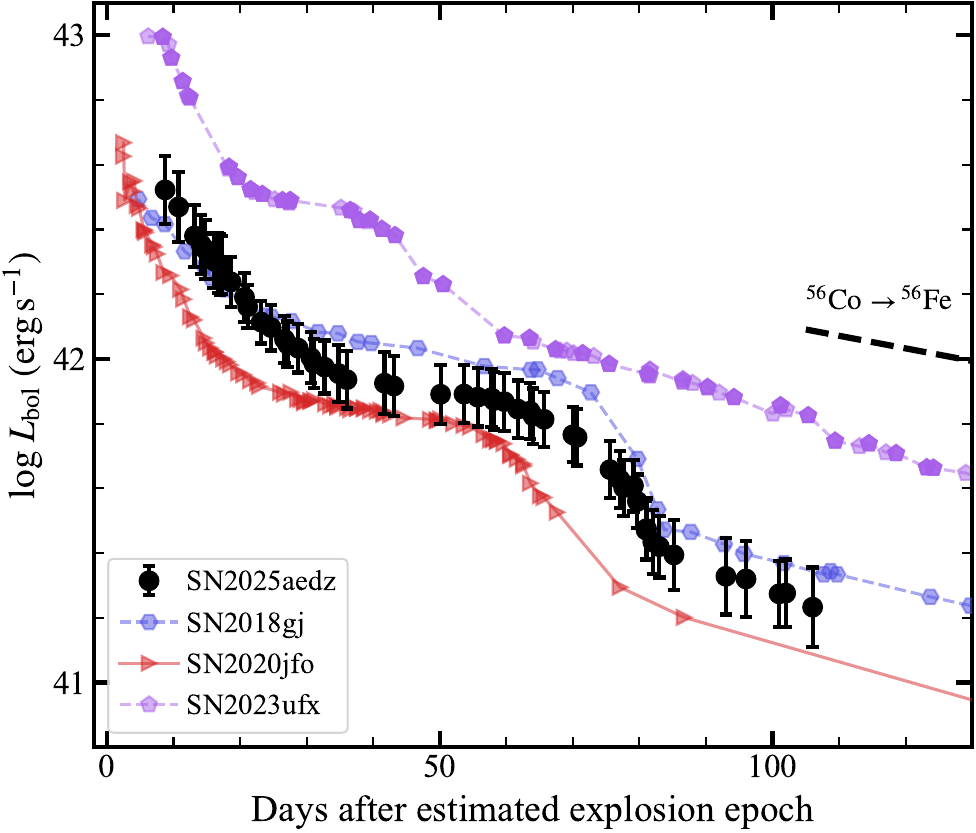}
   \caption{Bolometric luminosity evolution of SN\,2025aedz compared with other short-plateau SNe IIP.
   }
   \label{Fig:BL}
\end{figure}

\begin{figure}
  \centering
  \includegraphics[width=\linewidth]{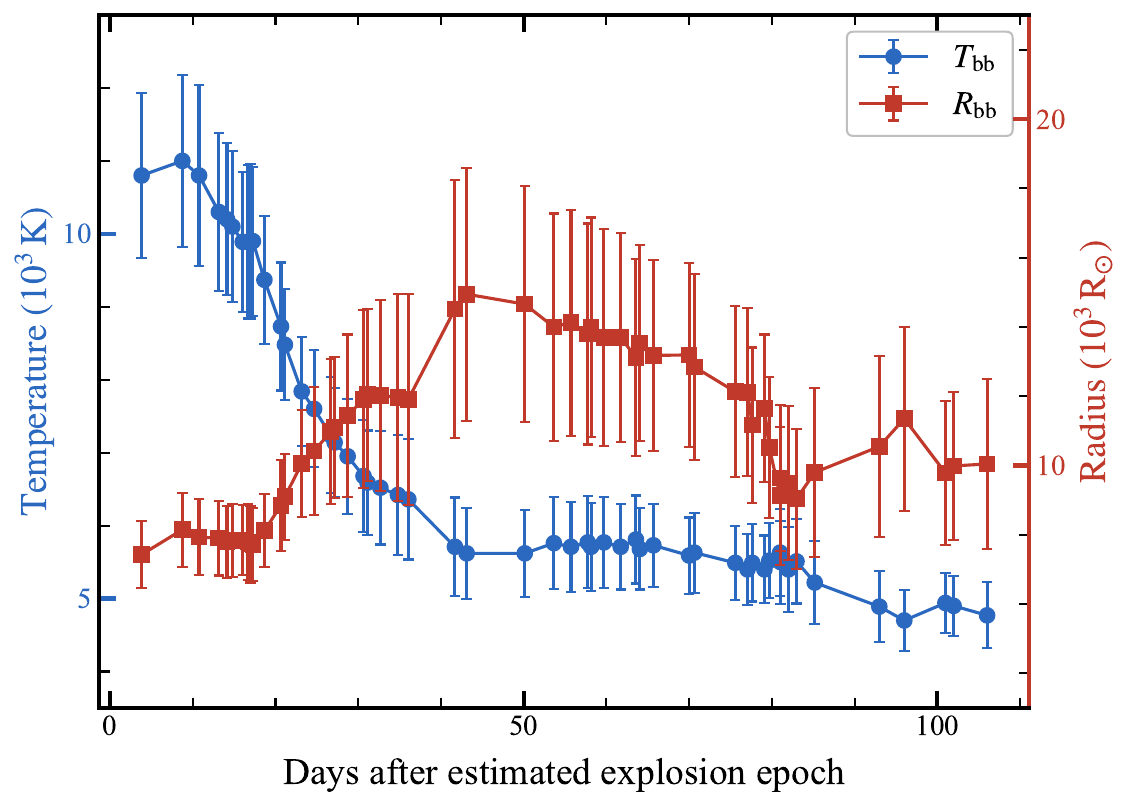}
  \caption{Temporal evolution of the blackbody temperature $T_{\rm bb}$ and photospheric radius $R_{\rm bb}$ derived from the optical photometry of SN\,2025aedz using the \texttt{Superbol} package.}
  \label{fig:bb_params}
\end{figure}

The bolometric light curve of SN\,2025aedz was constructed using the available $ugcroiz$ photometry.
For consistency, the bolometric light curves of the comparison sample were constructed using all available optical-band photometry for each SN.
The bolometric light curve was constructed with the publicly available \texttt{SuperBol} package by fitting a blackbody SED to the extinction-corrected multi-band photometry at each epoch and integrating the fitted SED over the adopted wavelength range
\citep{2018RNAAS...2..230N}. 
The integration includes the model-dependent extrapolation outside the observed photometric bands.
Reliable multi-band coverage is available up to $\sim110$\,d after
explosion.
At later epochs, only ATLAS $o$-band photometry is available.
These data are not extrapolated to construct a bolometric light curve.
The late-time $o$-band decline is relatively slower than the declines
measured in the $gri$ bands at $\sim90$--$110$\,d.
This behaviour may be partly affected by the proximity to the ATLAS detection limit, which can result in larger photometric uncertainties at late epochs \citep{2018PASP..130f4505T}. 
We therefore do not use the post-$110$\,d $o$-band data to derive the bolometric luminosity.

Adopting the reddening and distances listed in Table~\ref{tab:SNe_II_info}, we applied the same procedure to SN\,2025aedz and the comparison short-plateau SNe IIP.
Figure~\ref{Fig:BL} shows the resulting luminosity evolution, and Figure~\ref{fig:bb_params} presents the evolution of the relevant blackbody parameters.

As shown in Figure~\ref{Fig:BL}, SN\,2025aedz reaches a peak bolometric luminosity of $\sim3\times10^{42}\,\mathrm{erg\,s^{-1}}$.
Since this measurement corresponds to the epoch of the $r$-band maximum, we consider it to be an approximation of the peak bolometric luminosity.
Its luminosity evolution during the plateau and radioactive-tail phases is slightly lower than that of SN\,2018gj, suggesting that the two SNe may have similar $^{56}$Ni yields.
In Figure~\ref{fig:bb_params}, the early blackbody temperature of SN\,2025aedz is about $11{\,}000$\,K, after which it decreases rapidly.  
The blackbody temperature then remains roughly constant at $\sim5\,500$\,K during the plateau phase, before declining slightly to $\sim5\,000$\,K in the tail phase.
The blackbody radius rises continuously until $\sim50$\,d, reaching $\sim1.5\times10^{4}\,R_\odot$, and then decreases slowly thereafter.
The overall temperature and radius evolution of SN\,2025aedz also closely resembles that of SN\,2018gj (see Figure~5 of \citealt{Teja2023ApJ...954..155T}).

\subsection{The two-component model}

\begin{figure}
  \centering
  \includegraphics[width=\linewidth]{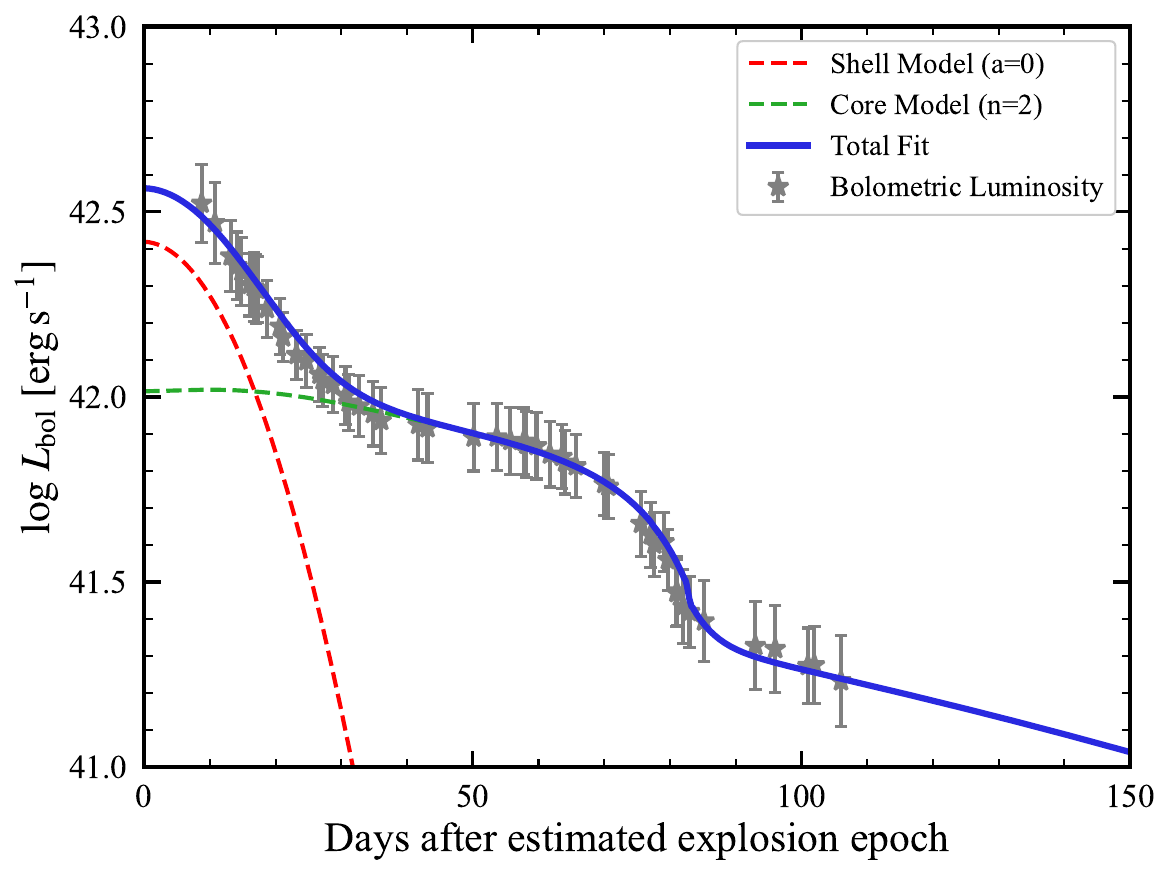}
  \caption{The two-component model fits to the SN\,2025aedz bolometric light curve on the basis of the model of \citet{2016A&A...589A..53N}.
  The contributions from the shell and the core are shown independently.}
  \label{fig:LC2}
\end{figure}

We inferred the explosion properties of SN\,2025aedz by modelling its bolometric light curve with the two-component prescription (see \citealt{2016A&A...589A..53N}).
In this framework, the ejecta are represented by a compact, He-rich inner component together with a more extended, low-mass H-rich envelope.
The resulting best-fit light curve is displayed in Figure~\ref{fig:LC2}, 
in which the fitted parameters are summarized in Table~\ref{tab:LC2_parameters}.
For the density structure and selected progenitor quantities, we adopted values consistent with the setup of \citet{2016A&A...589A..53N}.
The fit to the tail yields a synthesized $^{56}$Ni mass of $\sim0.03\pm0.01\,M_\odot$ for SN\,2025aedz.
Although the radioactive tail of SN~2018gj is very similar to that of SN\,2025aedz in Figure~\ref{Fig:BL}, our derived $^{56}$Ni mass is slightly lower than the $0.031\pm0.005\,M_\odot$ obtained for SN\,2018gj from the hydrodynamic model of \citet{Utrobin2024Ap&SS.369...49U}.
This small difference likely reflects the different modelling approaches adopted in the two studies.

It is worth noting that the parameters adopted in the two-component model are subject to parameter degeneracies, and the shell parameters are introduced primarily to reproduce the early-time bolometric light-curve evolution rather than to represent the actual physical properties of the progenitor immediately before explosion \citep{2016A&A...589A..53N,2025ApJ...993...39V}. 
In particular, the rapid post-peak decline is more likely to be associated with the effects of CSM interaction \citep{2016ApJ...829..109M,2018MNRAS.476.2840M}.

Because the early-time rise is incompletely sampled, we adopt the
midpoint between the last non-detection and the first detection as a
reference explosion epoch, $t_0=60995.6\pm1.5$.
We estimate an uncertainty of approximately 1.5\,d, which is intended to represent the allowed interval between these two observations rather than a formal statistical error.
\citet{2016A&A...589A..53N} investigated the effect of different explosion epochs on the modelling of SNe IIP using the two-component model and found that an uncertainty of about seven days in the explosion date results in moderate relative errors of approximately 5--10\% in the derived masses of the inner core ($M_{\rm core}$), outer envelope mass ($M_{\rm shell}$), and the initial radius of the core ($R_{\rm core}$).
Since the explosion-epoch uncertainty adopted here is about 1.5 d, we expect the corresponding systematic changes in these parameters to be smaller, of order a few per cent.

\begin{table}
    \centering
    \caption{The parameters of SN\,2025aedz based on the two-component model.}
    \renewcommand{\arraystretch}{1.2}
     \setlength{\tabcolsep}{12pt}
    \begin{tabular*}{0.8\columnwidth}{@{\extracolsep{\fill}}ccc}
        \hline\hline
        Parameter  & Core (a=0) & Shell (n=2) \\
        \hline
        $R_0\ (10^{12}\,\mathrm{cm})$ & 20 & 60 \\
        $M_{\mathrm{ej}}\,({M}_{\odot})$ & 8.0 & 0.5 \\
        $M_{\mathrm{Ni}}\,({M}_{\odot})$ & 0.03 & -- \\
        $E_{\mathrm{tot}}\, (10^{51}\,\mathrm{erg})$ & 2.2 & 1.3 \\
        $E_{\mathrm{kin}} / E_{\mathrm{th}}$ & 4.5 & 10 \\
        $\kappa\ (\mathrm{cm}^2\,\mathrm{g}^{-1})$ & 0.2 & 0.4 \\
        \hline
    \end{tabular*}
\tablefoot{$R_0$ is the initial ejecta radius, $M_{\mathrm{ej}}$ is the ejecta mass, $M_{\mathrm{Ni}}$ is the nickel mass, $E_{\mathrm{tot}}$ is the total energy including kinetic energy ($E_{\mathrm{kin}}$) and thermal energy ($E_{\mathrm{th}}$), and $\kappa$ is the Thomson-scattering opacity.}
    \label{tab:LC2_parameters}
\end{table}

\section{Hydrodynamical modelling}

\begin{table*}[htbp]
\centering
\caption{
Basic parameters of the hydrodynamical models for SN\,2025aedz, including CSM-interaction models (1--6) and models without CSM interaction (7--12). 
}
\begin{tabular}{cccccccc}
\hline\hline
Model & $M_{\rm ej}$ ($M_\odot$) & $M_{\rm Ni}$ ($M_\odot$) & $E_{\rm exp}$ ($10^{50}$ erg) & $\dot{M}$ ($M_\odot$ yr$^{-1}$) & $v_{\rm w}$ (km s$^{-1}$) & $t_{\rm CSM}$ (yr) & $\chi^2/N$ \\
\hline
1 & 7.45 & 0.030 & 6 & 0.1 & 10.0 & 2 & 2.59 \\
2 & 7.45 & 0.030 & 6 & 0.1 & 10.0 & 5 & 3.34 \\
3 & 7.45 & 0.030 & 6 & 0.1 & 5.0 & 5 & 0.77 \\
\textbf{4} & \textbf{7.45} & \textbf{0.032} & \textbf{6} & \textbf{0.1} & \textbf{2.5} & \textbf{5.5} & \textbf{0.14} \\
5 & 7.45 & 0.030 & 6 & 0.1 & 2.5 & 6 & 0.36 \\
6 & 7.45 & 0.030 & 6 & 0.1 & 2.5 & 10 & 4.49 \\
\hline
7 & 6.2 & 0.040 & 4 & - & - & - & 10.31  \\ 
8 & 6.2 & 0.040 & 6 & - & - & - & 6.25  \\ 
9 & 6.2 & 0.040 & 8 & - & - & - & 4.85  \\ 
10 & 8.2 & 0.040 & 4 & - & - & - & 10.73  \\ 
11 & 8.2 & 0.040 & 6 & - & - & - & 5.17  \\ 
12 & 8.2 & 0.040 & 8 & - & - & - & 2.76  \\ 
\hline
\end{tabular}
\label{tab:explosion_parameters}
\tablefoot{The models adopt different parameter combinations, where $\dot{M}$ is the wind mass-loss rate and $t_{\rm CSM}$ is the duration of wind mass loss before SN explosion.}
\end{table*}

\begin{figure*}
   \centering
   \includegraphics[width = 0.95\textwidth]{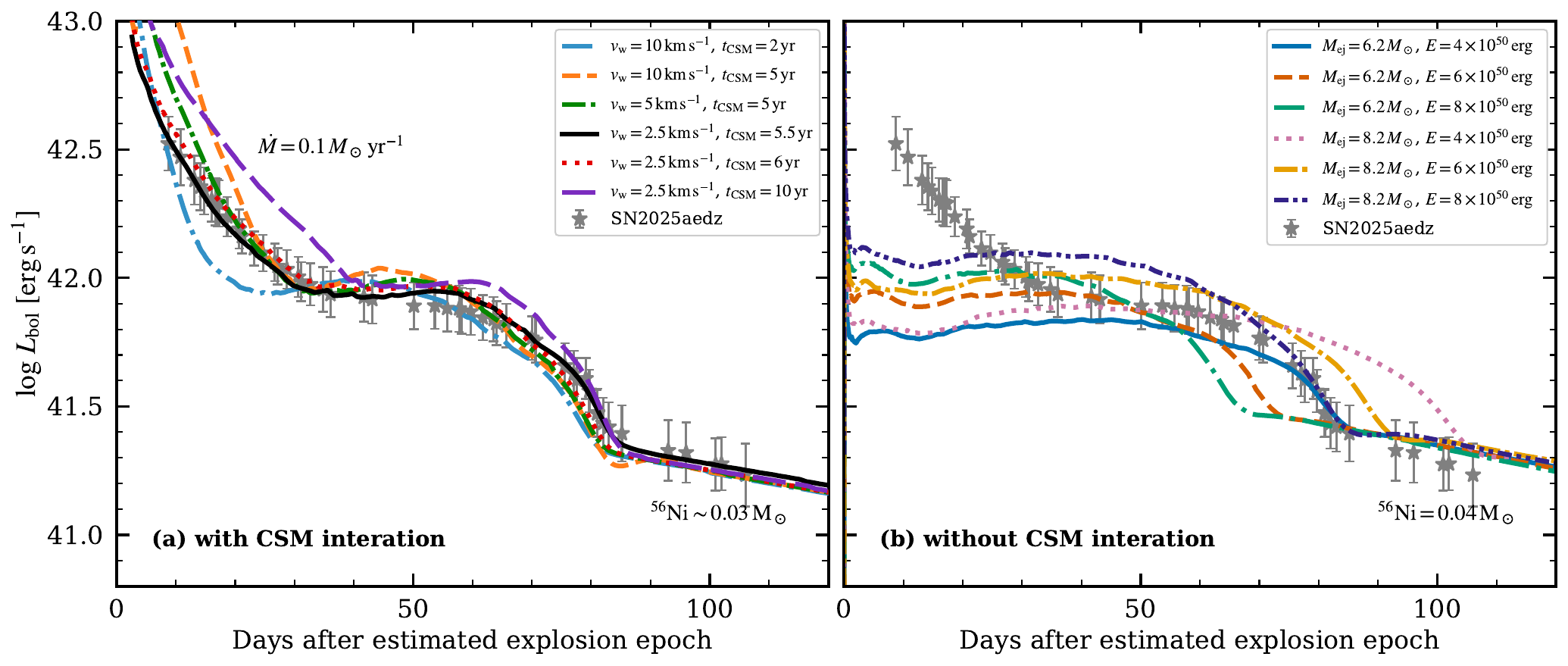}
   \caption{
  Bolometric light curves generated by the \texttt{MESA+STELLA} models. 
   Panel (a) illustrates the effects of different CSM density profiles on the light-curve evolution, while panel (b) shows the effects of different ejecta masses and explosion energies without CSM interaction.
}
    \label{Fig:stelar_fitting}
\end{figure*}

\subsection{Methods and setup}

Although the early $s_1$ phase of SN\,2025aedz is more pronounced and exhibits a steeper decline rate, comparable to that of SN\,2023ufx, such behaviour may indicate an additional contribution from early-time CSM interaction \citep{2016ApJ...829..109M,2018MNRAS.476.2840M}.
To better investigate the possible role of CSM interaction and understand the pre-explosion progenitor properties in SN\,2025aedz, we carried out the modelling using the Modules for Experiments in Stellar Astrophysics (\texttt{MESA}, version r-12778; see \citealt{Paxton2011ApJS..192....3P, Paxton2013ApJS..208....4P,Paxton2015ApJS..220...15P,Paxton2018ApJS..234...34P,Paxton2019ApJS..243...10P}),
together with the radiation hydrodynamics code \texttt{STELLA} (see \citealt{Blinnikov1993A&A...273..106B,Blinnikov2006A&A...453..229B,Blinnikov2017hsn..book..843B}).
The specific modelling process and setup are as follows:

First, we employ the \texttt{make pre CCSN IIP} example provided in \texttt{MESA} test suites to evolve an $16\,{M}_{\odot}$ zero-age main-sequence (MS) star until the development of the iron core further leading to its rapid infall. 
We consider \texttt{approx21.net} as the nuclear reaction network during the evolution.
This network includes the key isotopes involved in hydrogen, helium, carbon and advanced elements burning, such as $^{\rm 1}{\rm {H}}$, $^{\rm 3}{\rm {He}}$, $^{\rm 4}{\rm {He}}$, $^{\rm {12}}{\rm C}$, $^{\rm {16}}{\rm O}$, $^{\rm {20}}{\rm {Ne}}$, $^{\rm {24}}{\rm {Mg}}$, $^{\rm {28}}{\rm {Si}}$, $^{\rm {32}}{\rm {S}}$, $^{\rm {56}}{\rm {Fe}}$, and $^{\rm {56}}{\rm {Ni}}$, etc.
This network is commonly used in modelling type II SNe, and it is adopted as the default configuration used in a test suite for simulating the evolution of a massive star from MS to iron core-collapse.
For the wind mass loss, we adopt \texttt{de Jager} wind scheme with wind mass-loss parameter ${\alpha}_{\rm {Wind}}=1.0$ \citep{1988A&AS...72..259D}.
The wind mass-loss rate under the \texttt{de Jager} wind scheme is related to the luminosity and temperature of the star, i.e., 
\begin{equation}
    {\rm {log}}{\dot M}=1.769\times{\rm {log}}({L/{L}_{\odot}})-1.676\times{\rm {log}}({T}/{\rm K})-8.158,
\end{equation}
where ${\dot {M}}$, ${L}$ and ${T}$ are wind mass-loss rate, lumosity and temperature of the star, respectively. This wind mass-loss scheme is a default setting in simulating the progenitor evolution of iron core-collapse SNe, which is a reasonable assumption.
At the final stage of the evolution, the He core of the progenitor star is about $5.56\,{M}_{\odot}$. 

Secondly, we use this model to remove inner part of $1.8\,{M}_{\odot}$, and outer part of $4.6\,{M}_{\odot}$.
The removal of the inner mass is a numerical treatment, while the outer mass corresponds to an artificial reduction of the hydrogen-rich envelope to reproduce the short plateau duration of SN\,2025aedz.
After the removal process, the total mass of the remaining envelope is $8.2{M}_{\odot}$, consisting of a $4.44\,M_{\odot}$ hydrogen envelope. 
At this moment, the radius of the top of the hydrogen envelope is about $296\,{R}_{\odot}$. 
We then inject $6\times{10}^{50}\,{\rm {erg}}$ energy into a layer ($0.2\,{M}_{\odot}$) at the inner boundary within ${5}\,{\rm {ms}}$ to achieve the synthetic explosion.\footnote{
The adopted explosion energy is comparable to the initial value assumed in the \texttt{MESA} models of \citet{Goldberg2019ApJ...879....3G}, in which a RSG progenitor with a radius of $300\,R_{\odot}$ was assigned an explosion energy of approximately $7\times10^{50}\,{\rm erg}$.}
Since MESA does not consider nuclear reaction during the shock propagating stage, hence, we added $0.03-0.04\,{M}_{\odot}$ of $^{\rm {56}}{\rm {Ni}}$ into the He core by decreasing the corresponding oxygen mass fraction. The nickel mass is inferred from previous two-component model fit in Sect. 3.6, which is a reasonable mass for some core-collapse SNe.
We also consider different density distribution of CSM, 
as shown in Table~\ref{tab:explosion_parameters}.

Finally, we utilize the \texttt{STELLA} to calculate light curves resulted from the explosion of the progenitor star. The default settings in \texttt{STELLA} are adopted, where the discrepancies for different light curves are resulted from the difference in various progenitors properties, which is used as \texttt{STELLA} input files.
To better understand how different progenitor properties in affecting the light curves, we also calculated the light curves of exploded progenitor stars with different ejecta masses (${M}_{\rm {ej}}=6.2\,{M}_{\odot}$) and explosive energy (${E}=4.0\times{10}^{50}$ and ${8.0}\times{10}^{50}\,{\rm {erg}}$) but without considering CSM.

\subsection{Modelling results}
Figure~\ref{Fig:stelar_fitting} shows the light curves with different configurations of ejecta masses, explosive energy and various of CSM properties together with the bolometric light curve of SN\,2025aedz.
The corresponding model parameters are listed in Table~\ref{tab:explosion_parameters}. 
In Figure~\ref{Fig:stelar_fitting},
panel (b) shows how different ejecta mass and explosive energy in affecting plateau phase, in which all the models considered $0.04\,{M}_{\odot}$ of nickel but without considering CSM. 
From this panel,  we can see that the plateau duration is both affected by ejecta mass and explosive energy. For a fitted ejecta mass, a higher explosive energy will result in a shorter and brighter plateau phase. 
For a fitted explosive energy, increasing ejecta mass yields a corresponding rise in both plateau duration and plateau-phase luminosity.
We also found that all models in panel (b) produce slightly brighter radioactive tails than that of SN\,2025aedz at late times, although the differences remain within the observational uncertainties.
This suggests that the adopted $^{56}$Ni mass of $0.04\,M_\odot$ is close to the upper end of its allowed range.

In contrast, different CSM configurations (panel a) lead to significant differences in the early-time light-curve evolution. 
Under the same explosion energy and ejecta mass, the CSM properties are the dominant factor governing the early-time decline of the light curve.
To quantitatively evaluate the model performance, we calculated the chi-square per data point ($\chi^2/N$) between the model bolometric light curves and the observed bolometric light curve of SN\,2025aedz using 51 data points. 
Among the six models in Table~\ref{tab:explosion_parameters}, Model~4 provides the best fit, yielding the minimum $\chi^2/N$ value.
The best-fitting parameters of the CSM property are as follows:
the wind moss-loss rate is $0.1{M}_{\odot}\,{\rm {yr}^{-1}}$, wind velocity $v_{\rm w}$ is 2.5$\, \mathrm{ km\,s^{-1} }$, and the wind lasts for ${5.5}\,{\rm yr}$.
The models shown in panel (a) adopt an $^{56}$Ni mass of $0.03\,M_\odot$, while Model~4 uses a slightly higher value of $0.032\,M_\odot$. 
Considering the fit to the radioactive tail and the uncertainties in the bolometric light curve, we estimate an $^{56}$Ni mass of approximately $0.032\pm0.010\,M_\odot$. 
This estimate is consistent with the $^{56}$Ni mass derived from the two-component model.
Enhanced or episodic mass loss during the final years to decades before core collapse has been suggested for massive stars and may produce dense CSM \citep{2014ARA&A..52..487S,2017NatPh..13..510Y,2018MNRAS.476.1497B}. 
The observed pre-explosion variability of the progenitor of SN\,2024abfo further suggests that massive stars can undergo significant changes during the final decades before final explosion \citep{2025ApJ...987L..10N}.
These results provide a possible physical context for the CSM adopted in our modelling.
However, the CSM configuration derived here should not be interpreted as a direct measurement of the pre-SN mass-loss history.
Rather, it aims to illustrate how CSM interaction may drive the observed early-time light-curve evolution.

It is worth noting that this CSM configuration is intended only to illustrate a possible effect of CSM interaction.
Detailed single or binary star evolutionary models are needed in the future to better understand the progenitor channel of the exploding SN.
The difference of the physical parameters derived from the two-component model and hydrodynamical model is likely related to the strong correlations among the parameters involved in the light-curve modelling \citep{2016A&A...589A..53N}.
The parameter degeneracies allow different combinations of physical parameters to reproduce similar light-curve evolution \citep{2013ApJ...773...76C}.
Consequently, the differences between the parameter sets primarily reflect their different definitions and modelling assumptions.
For more studies on the light-curve modelling of SNe, see, for example, 
\citet{WuChengyuan2024ApJ...967L..45W},
\citet{LiuLiang-Duan2025ApJ...992...20L},
\citet{Sarin2025MNRAS.544.2653S},
\citet{Sarin2026MNRAS.552g1640S},
\citet{Sengupta2026ApJ...996...18S},
\citet{ZhangYu-Hao2026ApJ...999..186Z},
and \citet{ZhangZhengyang2026PhRvD.114b3020Z}.

\begin{figure*}
   \centering
   \includegraphics[width = 0.85\linewidth]{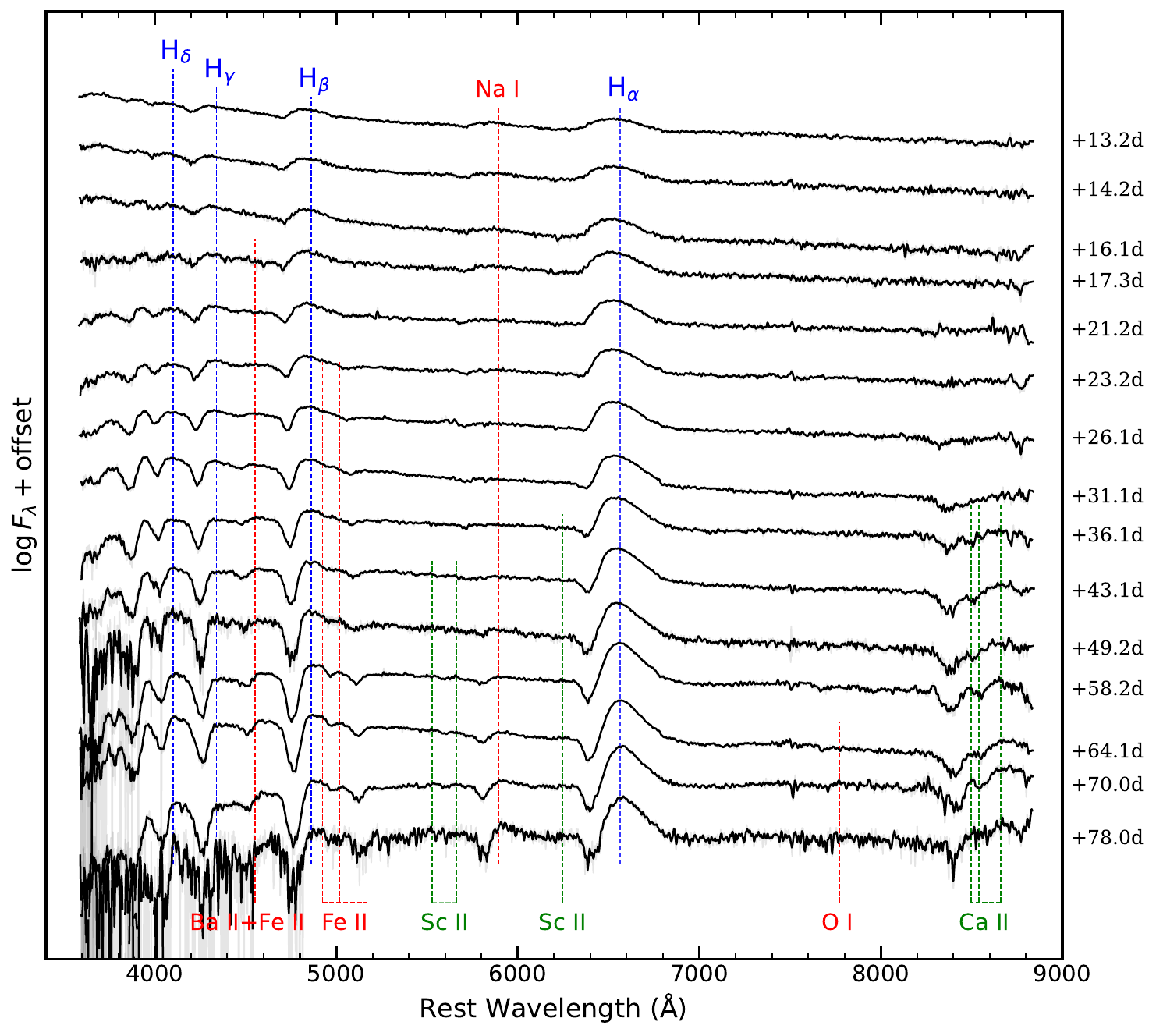}
   \caption{
   Spectral evolution sequences of SN\,2025aedz, in which vertical dashed lines mark the main transition positions of H, He, and other ions, with the corresponding rest wavelengths indicated for the labelled transitions. Phases relative to the derived explosion epoch (MJD=60995.6) are presented on the right. All spectra have been shifted to the rest frame and corrected for extinction. Spectra with low signal-to-noise ratios have been smoothed with a Savitzky–Golay filter.}
    \label{Fig:spec}
\end{figure*}

\section{Spectroscopic evolution}

\subsection{Spectral evolution sequence}

Figure~\ref{Fig:spec} presents the spectral evolution of SN\,2025aedz.
Owing to the lack of observations at very early phases, our first spectrum at $+13.2$\,d already corresponds to the plateau phase.
This spectrum is characterized by broad P-Cygni Balmer lines (\ha, \hb, \hg, \hd) superimposed on a blue continuum, a typical signature of the photospheric phase of SNe~IIP, in which the line profiles trace the rapidly expanding hydrogen-rich ejecta.
A blackbody fit to this spectrum gives a temperature of $\sim8000$\,K, which is slightly lower than the value derived from the photometry in Figure~\ref{fig:bb_params}, but is consistent within the uncertainties.

As the SN evolves, the Balmer lines gradually strengthen in emission and their absorption components deepen, reflecting the recession of the photosphere and the cooling of the ejecta.
The spectrum at $+23.2$\,d shows a weak \Feii\,$\lambda5169$ profile, marking the emergence of metal lines as the photospheric temperature decreases and the line-forming region moves into deeper, slower-expanding layers.
By $+31.1$\,d, the \Caii~NIR triplet begins to appear, becoming one of the prominent features at later phases.
The \Scii\ lines, which are commonly used as a temperature and metallicity indicator in SNe~IIP, remain inconspicuous throughout the entire evolution \citep{2014MNRAS.440.1856D}.

The last spectrum, obtained at $+78.0$\,d, corresponds to the transition from the plateau to the radioactive-tail phase, with a blackbody temperature of $\sim4000$\,K.
At this epoch the spectrum does not exhibit significant changes in line profiles, nor the emergence of forbidden lines, indicating that SN\,2025aedz has not yet entered the nebular phase by the end of our spectroscopic coverage.

\subsection{Comparison with other SNe~IIP}

\begin{figure*}
   \centering 
   \includegraphics[width = 0.95 \linewidth]{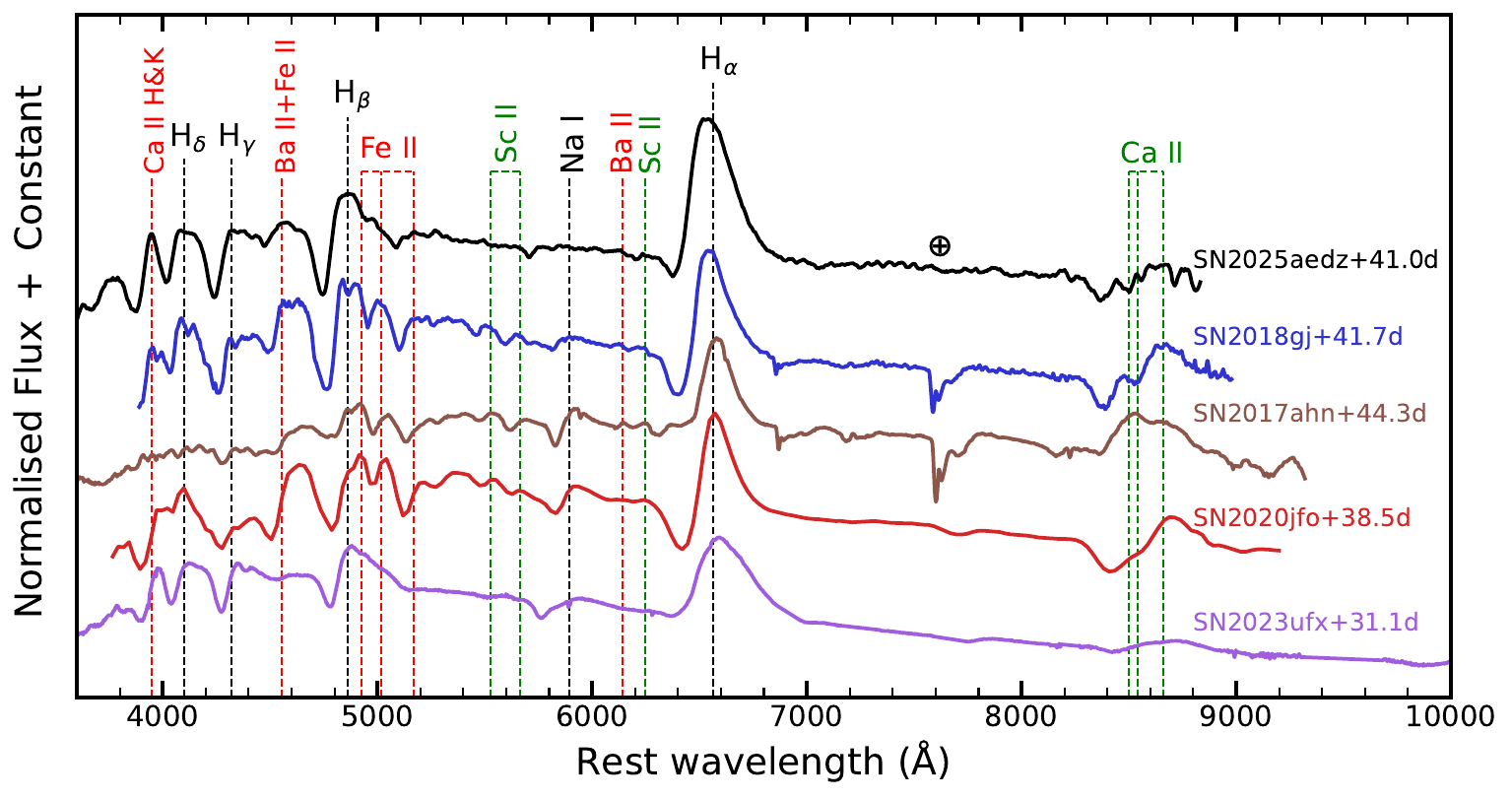}
   \caption{Intermediate-phase spectral comparison between SN\,2025aedz and other short-plateau SNe~IIP. 
   The phase of each spectrum, measured relative to its explosion epoch, is labelled on the right, and all spectra are plotted in the rest frame. 
   Prominent spectral transitions are indicated at their corresponding rest wavelengths. The Earth symbols mark telluric absorption features.
}
    \label{Fig:spec_com}
\end{figure*}

Figure~\ref{Fig:spec_com} compares the mid-plateau spectra of SN\,2025aedz with those of other short-plateau SNe~IIP at similar phases.
Overall, SN\,2025aedz remains most similar to SN\,2018gj, consistent with the resemblance already seen in their light curve and colour evolution.
A closer inspection, however, reveals some differences in the line profiles: the hydrogen Balmer lines of SN\,2025aedz show broader emission components, whereas the absorption components of SN\,2018gj are deeper.
To quantify the \ha\ profile differences, we measured the FWHM of the emission component at comparable phases.
The \ha\ emission component of SN\,2025aedz has an FWHM of $\sim204\,\AA$ ($\rm \sim9300\,km\,s^{-1}$), which is broader than those measured for SN\,2018gj ($\sim159\,\AA$), SN\,2017ahn ($\sim150\,\AA$), and SN\,2020jfo ($\sim148\,\AA$), although narrower than that of SN\,2023ufx ($\sim260\,\AA$).
The broader \ha\ emission in SN\,2025aedz indicates a larger velocity extent of the line-emitting hydrogen-rich ejecta at this phase.

In addition, the metal lines, such as \Feii, \Scii, and \Baii, are more prominent in SN\,2018gj, SN\,2017ahn and SN\,2020jfo, while in SN\,2023ufx these metal features remain weak owing to its short plateau and correspondingly higher photospheric temperature at this epoch.
By comparison, the metal lines of SN\,2025aedz are notably weaker: apart from \Feii, only a faint \Scii\ feature can be discerned.
The relatively weak metal lines of SN\,2025aedz persist even in the last spectrum at $+78.0$\,d, when the photosphere has already cooled to $\sim4000$\,K.
\citet{2014MNRAS.440.1856D} showed that the progenitors with higher metallicity generally produce stronger \Feii\ and \Naid\ absorption features, corresponding to larger absolute values of the line-absorption equivalent widths.
The weak \Feii\ and \Naid\ absorption features observed in SN\,2025aedz are therefore qualitatively consistent with a relatively low metallicity, although a quantitative metallicity constraint would require equivalent-width measurements and detailed spectral modelling.

\subsection{Spectral line velocity evolution}

\begin{figure}
   \centering
   \includegraphics[width = 1 \linewidth]{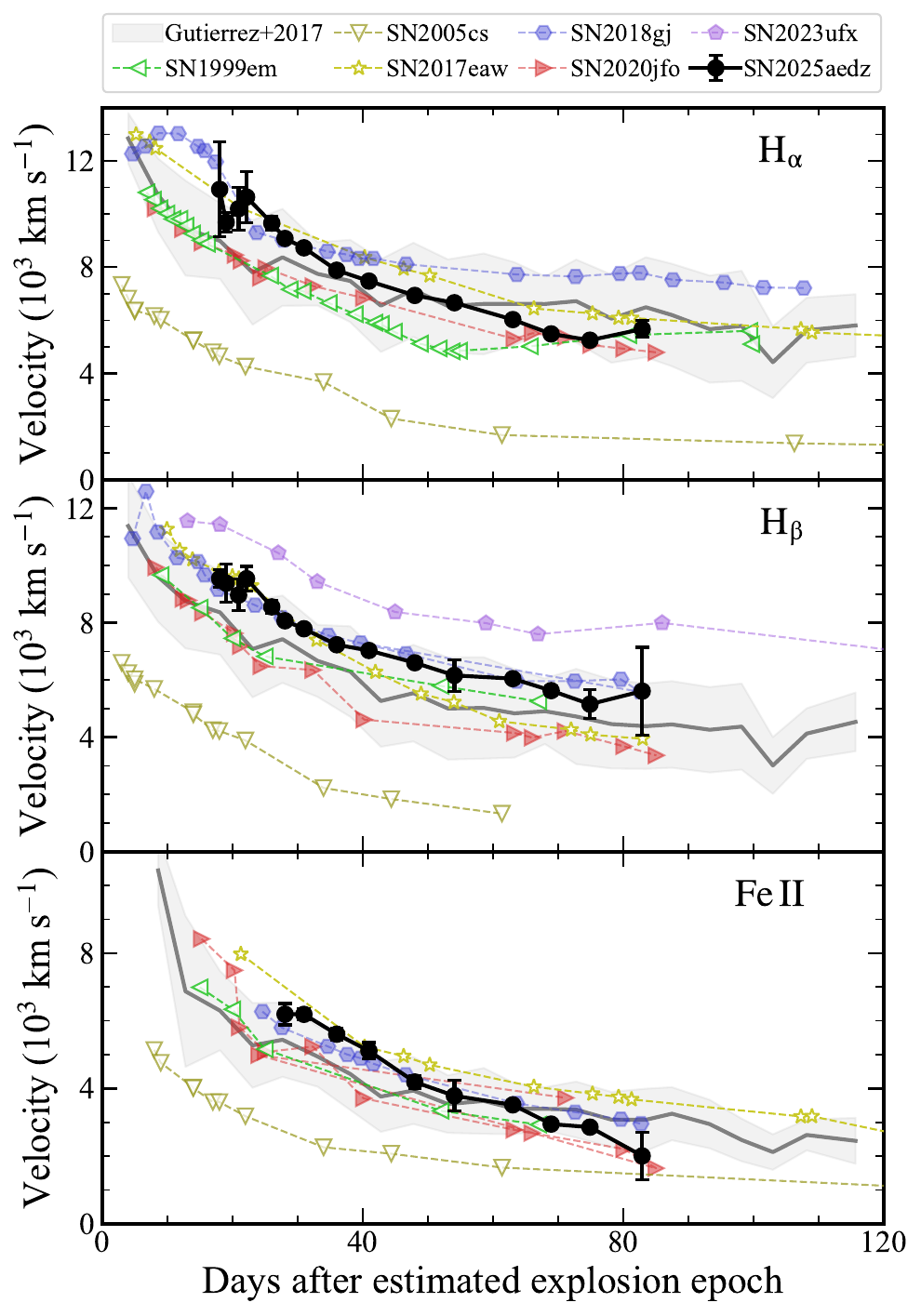}
   \caption{Evolution of the ejecta expanding velocities of \ha, \hb\ and \Feii\,$\lambda\,5169$ for SN\,2025aedz, compared with those of representative SNe II.
   Here all velocities are measured from the minima of the P-Cygni absorption components. 
   The shaded grey region shows the average velocity evolution of the SN II sample from \citet{2017ApJ...850...89G}.
   }
    \label{Fig:line_velocity}
\end{figure}

Figure~\ref{Fig:line_velocity} shows the velocity evolution of \ha, \hb\  and \Feii~$\lambda5169$.
For each rest-frame spectrum, we measured the line velocity from the absorption minimum of the corresponding P-Cygni profile.
The local line profile at each epoch was modelled with a two-component Gaussian function in an MCMC framework, and the posterior median was taken as the representative velocity.
During the photospheric phase, the absorption minimum of \Feii~$\lambda5169$ is commonly used as a proxy for the
photospheric velocity, while the \ha\ and \hb\ lines generally sample higher-velocity ejecta above the photosphere and typically yield larger absorption-minimum \citep{2005A&A...439..671D,2012MNRAS.419.2783T}.
For reference, we overplot the mean trend and dispersion of the SN~II sample from \citet{2017ApJ...850...89G}, together with several comparison SNe~IIP in Figure~\ref{Fig:line_velocity}.

The \ha\ velocity of SN\,2025aedz declines more rapidly overall.
At early phases it is comparable to that of SN\,2018gj, but after $\sim30$\,d it drops quickly and falls below the comparison objects.
The \hb\ velocity evolution is similar to that of SN\,2018gj, becoming slightly lower than SN\,2018gj at later phases.
The \Feii~$\lambda5169$ velocity also decreases faster than in the comparison sample.
In general, the line velocities of SN\,2025aedz are higher than those of the short-plateau SN\,2020jfo and the mean trend of \citet{2017ApJ...850...89G}, yet they appear to decline more steeply with time.

This combination of relatively high expansion velocities and a rapid velocity decline is broadly consistent with the relatively luminous early-time evolution and short plateau of SN\,2025aedz.
The rapid decline of the \Feii~$\lambda5169$ velocity suggests a fast recession of the photosphere through the ejecta, consistent with a steep density gradient and a relatively low-mass hydrogen envelope \citep{2011MNRAS.410.1739D,2012MNRAS.419.2783T}.
Such an envelope would recombine more quickly, leading to the short-plateau duration observed in SN\,2025aedz.

\section{Discussion}  \label{sect:discussion}

\begin{figure}
    \centering
    \includegraphics[width=1\linewidth]{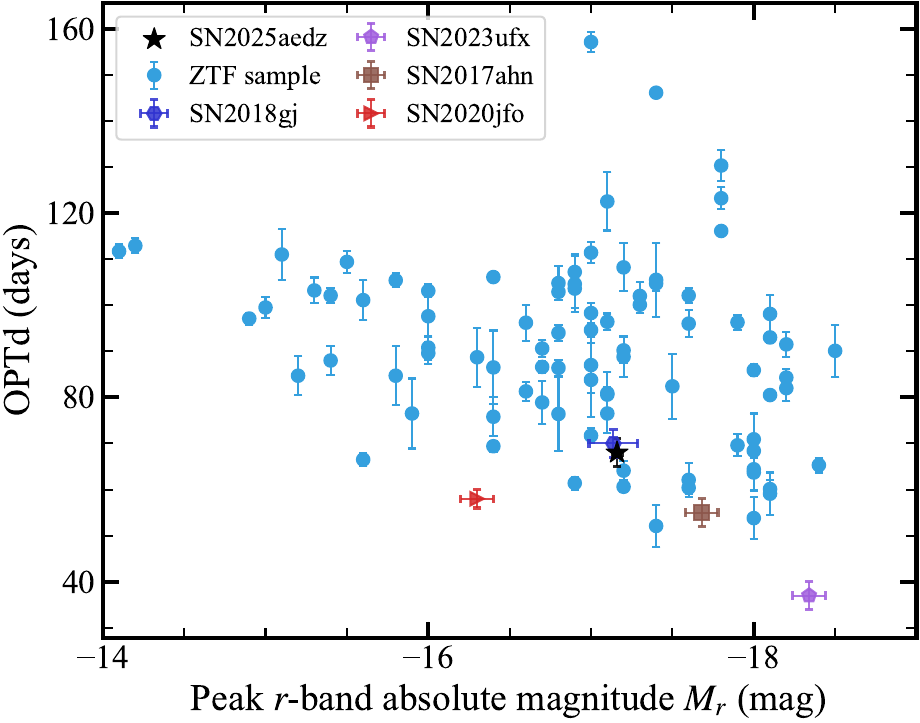}
    \caption{Comparison of OPTd versus peak absolute $r$-band magnitude ($M_r$). 
    Light blue points show the ZTF sample from \citet{Das2026PASP..138b4204D}.
    Sources with OPTd uncertainties larger than 8\,d were excluded. Coloured symbols mark representative short-plateau SNe. 
    For SN without $r$-band coverage, nearby $R$-band photometry was adopted as proxies.}
    \label{fig:Mr-OPTd}
\end{figure}

\subsection{Light-curve parameter comparisons}

To place SN\,2025aedz in the broader population of SNe~IIP, we compare its light-curve parameters with two diagnostic diagrams.
Figure~\ref{fig:Mr-OPTd} shows the relation between the duration of the optically thick phase (OPTd) and the peak absolute $r$-band magnitude ($M_r$).
The ZTF reference sample from \citet{Das2026PASP..138b4204D} defines the locus of typical SNe IIP in this plane, after excluding sources with OPTd uncertainties larger than 8\,d.
Representative short-plateau SNe are highlighted with coloured symbols for direct comparison, including SN\,2018gj, SN\,2023ufx and SN\,2020jfo.
SN\,2025aedz appears in this diagram alongside these reference objects and is most similar to SN\,2018gj. 
The two SNe have comparable OPTd and peak magnitude $M_r$.
Overall, the ZTF sample spans OPTd $\sim70$--$110$\,d, whereas the shortest known luminous short-plateau event, SN\,2023ufx, has an estimated OPTd of $\sim37$\,d based on its published $r$-band light curve and lies at the extreme edge of the population \cite{Ravi2025ApJ...982...12R}.
The location of SN\,2025aedz bridges SN\,2023ufx and the bulk of normal plateau SNe IIP.

\begin{figure}
    \centering
    \includegraphics[width=1\linewidth]{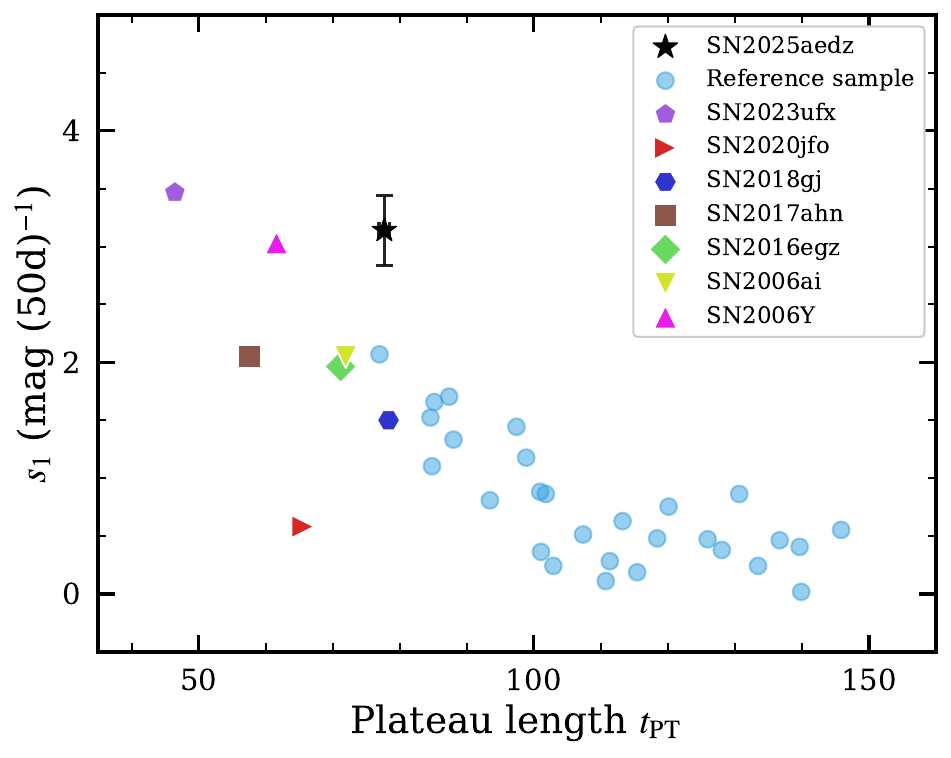}
    \caption{
    Comparison of plateau length ($t_{\rm PT}$) versus the early decline rate ($s_1$). 
    Note that the plateau length and decline rate shown in this figure are based on $V$-band measurements. 
    For SN\,2025aedz, the nearby $g$-band observations were used as a proxy owing to the lack of suitable $V$-band data.
    The reference sample is adopted from \citet{Anderson2014ApJ...786...67A,Anderson2024A&A...692A..95A}, \citet{2016MNRAS.459.3939V} and \citet{2019MNRAS.490.2799D}.
    }
    \label{fig:tPT-s1}
\end{figure}

Figure~\ref{fig:tPT-s1} examines the relation between plateau length, $t_{\rm PT}$, and the early post-peak decline rate, $s_1$, using $V$-band measurements from the compilations of some previous studies (see e.g. \citealt{Anderson2014ApJ...786...67A,Anderson2024A&A...692A..95A,2016MNRAS.459.3939V,2019MNRAS.490.2799D}).
For SN\,2025aedz, nearby $g$-band observations were used as a proxy because suitable $V$-band data were not available.
Given the similar wavelength and the early-time focus of $s_1$, this substitution is unlikely to change the qualitative classification, although small systematic offsets relative to the $V$-band reference sample cannot be excluded.

In Figure~\ref{fig:tPT-s1}, short-plateau SNe generally occupy the region of short $t_{\rm PT}$ and a large range of $s_1$.
SN\,2025aedz has an early $g$-band decline rate of $s_1 \approx 3.1 \pm 0.3$\,mag\,(50\,d)$^{-1}$, comparable to SN\,2023ufx ($\sim3.5$\,mag\,(50\,d)$^{-1}$) and SN\,2006Y ($\sim3$\,mag\,(50\,d)$^{-1}$).
These values are substantially higher than those of normal SNe~IIP in the reference sample ($\sim0--2$\,mag\,(50\,d)$^{-1}$), placing SN\,2025aedz among the most rapidly declining events in this comparison.
A similar trend is seen in the $r/R$-band comparison of Figure~\ref{Fig:r_com}.
SN\,2025aedz and SN\,2023ufx exhibit closely matched early decline rates that exceed those of the other comparison SNe, although the peak of SN\,2025aedz occurs later and the onset of the $s_1$ segment is correspondingly delayed.
By contrast, SN\,2018gj--which shows otherwise very similar photometric and spectroscopic evolution--does not display a comparably prominent early $s_1$ phase, with $s_1 \approx 1.5$\,mag\,(50\,d)$^{-1}$.
Together, these comparisons suggest that the early peak brightness and the steep post-peak $s_1$ decline in SN\,2025aedz, SN\,2023ufx, and SN\,2006Y may reflect enhanced emission during the first weeks after explosion, plausibly driven by CSM interaction.
The more gradual early decline of SN\,2018gj may indicate a weaker or absent CSM component despite its overall similarity to SN\,2025aedz in later phases.

\subsection{Progenitor and formation channel}

The plateau evolution of SN\,2025aedz is similar to that of SN\,2018gj, and the velocity evolution inferred from \hb\ and \Feii\ also follows a comparable trend. 
Although the early $s_1$ phase of SN\,2025aedz is more pronounced and shows a steeper decline rate, close to that observed in SN\,2023ufx, this may be attributed to a stronger contribution from early-time CSM interaction \citep{2016ApJ...829..109M,2018MNRAS.476.2840M}. 
However, due to the lack of spectroscopic observations shortly after explosion, this scenario cannot be directly confirmed from the spectra.

Short-plateau SNe IIP are often associated with progenitors retaining relatively low-mass hydrogen envelopes, since the plateau duration is expected to increase with the mass of the recombining H-rich envelope \citep[see e.g.][]{Popov1993ApJ...414..712P,Kasen2009ApJ...703.2205K,Goldberg2019ApJ...879....3G,Martinez2022A&A...660A..42M}.
This interpretation has been invoked for several short-plateau events.
For example, the hydrogen-envelope mass for the progenitor of SN\,2023ufx was estimated to be $\sim 1.2~M_\odot$ \citep{Ravi2025ApJ...982...12R}, while the best-fitting model for SN\,2018gj yielded a hydrogen-envelope mass of $3.26~M_\odot$ \citep{Teja2022ApJ...930...34T}. 
For zero-age MS stars with initial masses above $8~M_\odot$, binary evolution provides a natural channel for producing such low-mass hydrogen envelopes, although robust observational evidence remains limited. 
Note that SN\,2018gj is located far from the centre of its host galaxy and resides in a relatively old environment, suggesting that it may have undergone binary-driven mass loss followed by a final binary merger before final core collapse, giving rise to a short-plateau event \citep[see][]{NiuZexi2026SciBu..71.1023N}.  
In fact, SN\,2025aedz is also located at a relatively large projected distance from the centre of its host galaxy.
The possible presence of early-time CSM interaction further supports the possibility of a binary-evolution origin.
However, this interpretation remains tentative without more direct constraints on the progenitor environment, CSM properties, or binary history.
For more studies on the RSG mass loss and/or binary interaction implications for the origin of SNe, see, for example, 
\citet{vanLoonJacco2025Galax..13...72V}, 
\citet{vanLoonJacco2026MNRAS.546ag012V}, and \citet{Munoz-Sanchez2026NatAs..10..702M}.

\section{Summary}

Overall, SN\,2025aedz is a typical short-plateau SN IIP, with a rapidly declining post-peak light curve.
Its peak absolute magnitude in the $r$ band is relatively bright, at about $-17.1$\,mag.
Its $r$-band plateau is short, with $\mathrm{Pd}\approx50\pm3$\,d, $\mathrm{OPTd}\approx68\pm3$\,d, and $t_{\rm PT}=78.7\pm0.5$\,d, whereas the radioactive tail indicates a $^{56}$Ni mass of $\sim0.03\pm0.01\,M_\odot$.
We performed radiation hydrodynamic modelling of a $16\,M_{\odot}$ progenitor using \texttt{MESA+STELLA} with different CSM configurations. 
The preferred model requires a wind mass-loss rate of $0.1\,M_{\odot}\,{\rm yr}^{-1}$ sustained for $\sim5.5\,{\rm yr}$ prior to explosion to reproduce the steep early decline of the bolometric light curve.
However, the model parameters are subject to degeneracies, and the inferred CSM configuration should not necessarily be interpreted as a direct representation of the actual pre-SN mass-loss history. 
Rather, it demonstrates that CSM interaction offers a viable mechanism for shaping the observed early-time light-curve evolution.

It is worth noting that the spectra of SN\,2025aedz are broadly consistent with those of normal SNe~IIP, but show weak metal lines and rapidly declining expansion velocities. 
Its overall evolution is closest to SN\,2018gj, whereas the steep early post-peak decline is reminiscent of SN\,2023ufx and may point to an early contribution from circumstellar interaction. 
The short plateau and fast photospheric recession suggest a relatively low-mass hydrogen envelope, possibly formed through rapid mass loss or binary evolution before explosion.
The current small number of short-plateau events is insufficient to constrain their properties. 
To improve our understanding of this kind of SN, larger samples and more theoretical modelling are needed.

\section*{Data availability}
The photometric data are presented in the tables in the Appendix. 
The spectra are available through the Weizmann Interactive Supernova Data Repository \citep[WISeREP; ][]{2012PASP..124..668Y} at \url{https://www.wiserep.org/object/29424}.

\begin{acknowledgements}
We acknowledge the support of the staff of the LJT and the Wuhan University 1\,m telescope.
The LJT is supported by the Chinese Academy of Sciences (CAS) and the People's Government of Yunnan Province, and operated and managed by Yunnan Observatories, CAS.

This study is supported by 
the National Natural Science Foundation of China (Nos 12225304, 12288102, 12303054, 12473032), 
the CAS Project for Young Scientists in Basic Research (YSBR-148), 
the Yunnan Revitalization Talent Support Program (Yunling Scholar Project, Young Talent Project), the Yunnan Science and Technology Program (Nos 202401AU070063, 202501AS070005, 202501AW070001, 202605AS350010 and 202601BC070011), the New Cornerstone Science Foundation through the XPLORER PRIZE, and the International Centre of Supernovae (ICESUN), Yunnan Key Laboratory of Supernova Research (No. 202505AV340004). J.Z. is supported by the Yunnan Science and Technology Program (Nos 202501AV070012 and 202401BC070007) and the National Natural Science Foundation of China (Nos 12173082 and 12333008). Y.-Z. Cai acknowledges financial support from the SOXS project (PI S. Campana). 
\end{acknowledgements}

\bibliographystyle{aa}
\bibliography{aa62144-26.bib}

@ARTICLE{2026arXiv260938858X,
       author = {{Xu}, Sai-En and {You}, Bei and {He}, Han and {Yang}, Shuai-Kang and {Fan}, Xiao and {Qu}, Yi-Fei and {Hu}, Rui-Xiang and {Fei}, Hui-Bo and {Chen}, Bo-An and {Zhu}, Zong-Hong and {Li}, Zheng-Yang and {Li}, Xiao-Yan and {Han}, Zi-Jian and {Cong}, Jia-Nan and {Chen}, Chao and {Chen}, Jia-Li and {Zheng}, Kai-Wen and {Yang}, Yi-Qiao and {Li}, Qing-Shan and {Sun}, Zhen-Guang and {Zhou}, Tong and {Zhang}, Kai and {Wu}, Jiajia and {Yin}, Xuhang and {Yuan}, Liang},
        title = "{MATCH: Performance Characterization and Scientific Observations of the Wuhan University 1-m Optical Telescope}",
      journal = {arXiv e-prints},
         year = 2026,
        month = sep,
          eid = {arXiv:2609.38858},
        pages = {arXiv:2609.38858},
          doi = {10.48550/arXiv.2609.38858},
archivePrefix = {arXiv},
       eprint = {2609.38858},
 primaryClass = {astro-ph.IM},
       adsurl = {https://ui.adsabs.harvard.edu/abs/2026arXiv260938858X}
}

@ARTICLE{2024ApJ...976..178T,
       author = {{Tucker}, Michael A. and {Hinkle}, Jason and {Angus}, Charlotte R. and {Auchettl}, Katie and {Hoogendam}, Willem B. and {Shappee}, Benjamin and {Kochanek}, Christopher S. and {Ashall}, Chris and {de Boer}, Thomas and {Chambers}, Kenneth C. and {Desai}, Dhvanil D. and {Do}, Aaron and {Fulton}, Michael D. and {Gao}, Hua and {Herman}, Joanna and {Huber}, Mark and {Lidman}, Chris and {Lin}, Chien-Cheng and {Lowe}, Thomas B. and {Magnier}, Eugene A. and {Martin}, Bailey and {M{\'\i}nguez}, Paloma and {Nicholl}, Matt and {Pursiainen}, Miika and {Smartt}, S.~J. and {Smith}, Ken W. and {Srivastav}, Shubham and {Tucker}, Brad E. and {Wainscoat}, Richard J.},
        title = "{The Extremely Metal-poor SN 2023ufx: A Local Analog to High-redshift Type II Supernovae}",
      journal = {\apj},
         year = 2024,
        month = dec,
       volume = {976},
       number = {2},
          eid = {178},
        pages = {178},
          doi = {10.3847/1538-4357/ad8448},
archivePrefix = {arXiv},
       eprint = {2405.00113},
 primaryClass = {astro-ph.SR},
       adsurl = {https://ui.adsabs.harvard.edu/abs/2024ApJ...976..178T}
}

@ARTICLE{Munoz-Sanchez2026NatAs..10..702M,
       author = {{Mu{\~n}oz-Sanchez}, Gonzalo and {Kalitsounaki}, Maria and {de Wit}, Stephan and {Antoniadis}, Konstantinos and {Bonanos}, Alceste Zoe and {Zapartas}, Emmanouil and {Boutsia}, Konstantina and {Christodoulou}, Evangelia and {Maravelias}, Grigoris and {Soszy{\'n}ski}, Igor and {Udalski}, Andrzej},
        title = "{The dramatic transition of the extreme red supergiant WOH G64 to a yellow hypergiant}",
      journal = {Nature Astronomy},
         year = 2026,
        month = may,
       volume = {10},
        pages = {702-713},
          doi = {10.1038/s41550-026-02789-7},
archivePrefix = {arXiv},
       eprint = {2411.19329},
 primaryClass = {astro-ph.SR},
       adsurl = {https://ui.adsabs.harvard.edu/abs/2026NatAs..10..702M}
}

@ARTICLE{ZhangZhengyang2026PhRvD.114b3020Z,
       author = {{Zhang}, Zhengyang and {Zha}, Shuai and {Sarin}, Nikhil and {Moriya}, Takashi J. and {Wu}, Chengyuan and {Wang}, Bo},
        title = "{Surrogate models for type II supernovae: Probing low-energy explosions and interaction-free regimes}",
      journal = {\prd},
         year = 2026,
        month = jul,
       volume = {114},
       number = {2},
          eid = {023020},
        pages = {023020},
          doi = {10.1103/hsms-7v8x},
archivePrefix = {arXiv},
       eprint = {2607.17488},
 primaryClass = {astro-ph.SR},
       adsurl = {https://ui.adsabs.harvard.edu/abs/2026PhRvD.114b3020Z}
}

@ARTICLE{Sarin2026MNRAS.552g1640S,
       author = {{Sarin}, Nikhil and {Hirai}, Ryosuke},
        title = "{From mass-loss histories to light curves: a generalized framework for interaction-powered transients}",
      journal = {\mnras},
         year = 2026,
        month = oct,
       volume = {552},
       number = {1},
          eid = {stag1640},
        pages = {stag1640},
          doi = {10.1093/mnras/stag1640},
archivePrefix = {arXiv},
       eprint = {2605.19571},
 primaryClass = {astro-ph.HE},
       adsurl = {https://ui.adsabs.harvard.edu/abs/2026MNRAS.552g1640S}
}

@ARTICLE{Sarin2025MNRAS.544.2653S,
       author = {{Sarin}, Nikhil and {Moriya}, Takashi J. and {Singh}, Avinash and {Gangopadhyay}, Anjasha and {Hinds}, K.-Ryan and {Schulze}, Steve and {Omand}, Conor M.~B. and {Das}, Kaustav K.},
        title = "{Surrogate models for light curves and photosphere properties of Type II supernovae}",
      journal = {\mnras},
         year = 2025,
        month = dec,
       volume = {544},
       number = {2},
        pages = {2653-2663},
          doi = {10.1093/mnras/staf1808},
archivePrefix = {arXiv},
       eprint = {2506.02107},
 primaryClass = {astro-ph.HE},
       adsurl = {https://ui.adsabs.harvard.edu/abs/2025MNRAS.544.2653S}
}

@ARTICLE{vanLoonJacco2026MNRAS.546ag012V,
       author = {{van Loon}, Jacco Th and {Ohnaka}, Keiichi},
        title = "{A phoenix rises from the ashes: WOH G64 is still a red supergiant, for now}",
      journal = {\mnras},
         year = 2026,
        month = feb,
       volume = {546},
       number = {2},
          eid = {stag012},
        pages = {stag012},
          doi = {10.1093/mnras/stag012},
archivePrefix = {arXiv},
       eprint = {2601.02057},
 primaryClass = {astro-ph.SR},
       adsurl = {https://ui.adsabs.harvard.edu/abs/2026MNRAS.546ag012V}
}

@ARTICLE{vanLoonJacco2025Galax..13...72V,
       author = {{van Loon}, Jacco Th.},
        title = "{Red Supergiant Mass Loss and Mass-Loss Rates}",
      journal = {Galaxies},
         year = 2025,
        month = jun,
       volume = {13},
       number = {4},
          eid = {72},
        pages = {72},
          doi = {10.3390/galaxies13040072},
archivePrefix = {arXiv},
       eprint = {2507.15971},
 primaryClass = {astro-ph.SR},
       adsurl = {https://ui.adsabs.harvard.edu/abs/2025Galax..13...72V}
}

@ARTICLE{Sengupta2026ApJ...996...18S,
       author = {{Sengupta}, Sutirtha and {Sujit}, Das and {Sarangi}, Arkaprabha},
        title = "{Dance to Demise{\textemdash}How Massive Stars May Form Dense Circumstellar Shells before Explosion}",
      journal = {\apj},
         year = 2026,
        month = jan,
       volume = {996},
       number = {1},
          eid = {18},
        pages = {18},
          doi = {10.3847/1538-4357/ae129c},
archivePrefix = {arXiv},
       eprint = {2508.04497},
 primaryClass = {astro-ph.SR},
       adsurl = {https://ui.adsabs.harvard.edu/abs/2026ApJ...996...18S}
}

@ARTICLE{1988A&AS...72..259D,
       author = {{de Jager}, C. and {Nieuwenhuijzen}, H. and {van der Hucht}, K.~A.},
        title = "{Mass loss rates in the Hertzsprung-Russell diagram.}",
      journal = {\aaps},
         year = 1988,
        month = feb,
       volume = {72},
        pages = {259-289},
       adsurl = {https://ui.adsabs.harvard.edu/abs/1988A&AS...72..259D}
}

@ARTICLE{2017NatPh..13..510Y,
       author = {{Yaron}, O. and {Perley}, D.~A. and {Gal-Yam}, A. and {Groh}, J.~H. and {Horesh}, A. and {Ofek}, E.~O. and {Kulkarni}, S.~R. and {Sollerman}, J. and {Fransson}, C. and {Rubin}, A. and {Szabo}, P. and {Sapir}, N. and {Taddia}, F. and {Cenko}, S.~B. and {Valenti}, S. and {Arcavi}, I. and {Howell}, D.~A. and {Kasliwal}, M.~M. and {Vreeswijk}, P.~M. and {Khazov}, D. and {Fox}, O.~D. and {Cao}, Y. and {Gnat}, O. and {Kelly}, P.~L. and {Nugent}, P.~E. and {Filippenko}, A.~V. and {Laher}, R.~R. and {Wozniak}, P.~R. and {Lee}, W.~H. and {Rebbapragada}, U.~D. and {Maguire}, K. and {Sullivan}, M. and {Soumagnac}, M.~T.},
        title = "{Confined dense circumstellar material surrounding a regular type II supernova}",
      journal = {Nature Physics},
         year = 2017,
        month = feb,
       volume = {13},
       number = {5},
        pages = {510-517},
          doi = {10.1038/nphys4025},
archivePrefix = {arXiv},
       eprint = {1701.02596},
 primaryClass = {astro-ph.HE},
       adsurl = {https://ui.adsabs.harvard.edu/abs/2017NatPh..13..510Y}
}

@ARTICLE{2025ApJ...987L..10N,
       author = {{Niu}, Zexi and {Sun}, Ning-Chen and {Maund}, Justyn R. and {Guo}, Zhen and {Li}, Wenxiong and {Sun}, Meng and {Liu}, Jifeng},
        title = "{Discovery of a Variable Yellow Supergiant Progenitor for the Type IIb SN 2024abfo}",
      journal = {\apjl},
         year = 2025,
        month = jul,
       volume = {987},
       number = {1},
          eid = {L10},
        pages = {L10},
          doi = {10.3847/2041-8213/ade4cd},
archivePrefix = {arXiv},
       eprint = {2504.20407},
 primaryClass = {astro-ph.HE},
       adsurl = {https://ui.adsabs.harvard.edu/abs/2025ApJ...987L..10N}
}

@ARTICLE{2014ARA&A..52..487S,
       author = {{Smith}, Nathan},
        title = "{Mass Loss: Its Effect on the Evolution and Fate of High-Mass Stars}",
      journal = {\araa},
         year = 2014,
        month = aug,
       volume = {52},
        pages = {487-528},
          doi = {10.1146/annurev-astro-081913-040025},
archivePrefix = {arXiv},
       eprint = {1402.1237},
 primaryClass = {astro-ph.SR},
       adsurl = {https://ui.adsabs.harvard.edu/abs/2014ARA&A..52..487S}
}

@ARTICLE{2018MNRAS.476.1497B,
       author = {{Bullivant}, Christopher and {Smith}, Nathan and {Williams}, G. Grant and {Mauerhan}, Jon C. and {Andrews}, Jennifer E. and {Fong}, Wen-Fai and {Bilinski}, Christopher and {Kilpatrick}, Charles D. and {Milne}, Peter A. and {Fox}, Ori D. and {Cenko}, S. Bradley and {Filippenko}, Alexei V. and {Zheng}, WeiKang and {Kelly}, Patrick L. and {Clubb}, Kelsey I.},
        title = "{SN 2013fs and SN 2013fr: exploring the circumstellar-material diversity in Type II supernovae}",
      journal = {\mnras},
         year = 2018,
        month = may,
       volume = {476},
       number = {2},
        pages = {1497-1518},
          doi = {10.1093/mnras/sty045},
archivePrefix = {arXiv},
       eprint = {1801.01532},
 primaryClass = {astro-ph.HE},
       adsurl = {https://ui.adsabs.harvard.edu/abs/2018MNRAS.476.1497B}
}

@ARTICLE{2011MNRAS.410.1739D,
       author = {{Dessart}, Luc and {Hillier}, D. John},
        title = "{Non-LTE time-dependent spectroscopic modelling of Type II-plateau supernovae from the photospheric to the nebular phase: case study for 15 and 25 M$_{☉}$ progenitor stars}",
      journal = {\mnras},
         year = 2011,
        month = jan,
       volume = {410},
       number = {3},
        pages = {1739-1760},
          doi = {10.1111/j.1365-2966.2010.17557.x},
archivePrefix = {arXiv},
       eprint = {1008.3238},
 primaryClass = {astro-ph.SR},
       adsurl = {https://ui.adsabs.harvard.edu/abs/2011MNRAS.410.1739D}
}

@ARTICLE{2005A&A...439..671D,
       author = {{Dessart}, L. and {Hillier}, D.~J.},
        title = "{Distance determinations using type II supernovae and the expanding photosphere method}",
      journal = {\aap},
         year = 2005,
        month = aug,
       volume = {439},
       number = {2},
        pages = {671-685},
          doi = {10.1051/0004-6361:20053217},
archivePrefix = {arXiv},
       eprint = {astro-ph/0505465},
 primaryClass = {astro-ph},
       adsurl = {https://ui.adsabs.harvard.edu/abs/2005A&A...439..671D}
}

@ARTICLE{2012MNRAS.419.2783T,
       author = {{Tak{\'a}ts}, K. and {Vink{\'o}}, J.},
        title = "{Measuring expansion velocities in Type II-P supernovae}",
      journal = {\mnras},
         year = 2012,
        month = feb,
       volume = {419},
       number = {4},
        pages = {2783-2796},
          doi = {10.1111/j.1365-2966.2011.19921.x},
archivePrefix = {arXiv},
       eprint = {1109.5873},
 primaryClass = {astro-ph.SR},
       adsurl = {https://ui.adsabs.harvard.edu/abs/2012MNRAS.419.2783T}
}

@ARTICLE{2014MNRAS.440.1856D,
       author = {{Dessart}, L. and {Gutierrez}, C.~P. and {Hamuy}, M. and {Hillier}, D.~J. and {Lanz}, T. and {Anderson}, J.~P. and {Folatelli}, G. and {Freedman}, W.~L. and {Ley}, F. and {Morrell}, N. and {Persson}, S.~E. and {Phillips}, M.~M. and {Stritzinger}, M. and {Suntzeff}, N.~B.},
        title = "{Type II Plateau supernovae as metallicity probes of the Universe}",
      journal = {\mnras},
         year = 2014,
        month = may,
       volume = {440},
       number = {2},
        pages = {1856-1864},
          doi = {10.1093/mnras/stu417},
archivePrefix = {arXiv},
       eprint = {1403.1167},
 primaryClass = {astro-ph.SR},
       adsurl = {https://ui.adsabs.harvard.edu/abs/2014MNRAS.440.1856D}
}

@ARTICLE{2020ApJ...895...32F,
       author = {{Fremling}, C. and {Miller}, A.~A. and {Sharma}, Y. and {Dugas}, A. and {Perley}, D.~A. and {Taggart}, K. and {Sollerman}, J. and {Goobar}, A. and {Graham}, M.~L. and {Neill}, J.~D. and {Nordin}, J. and {Rigault}, M. and {Walters}, R. and {Andreoni}, I. and {Bagdasaryan}, A. and {Belicki}, J. and {Cannella}, C. and {Bellm}, E.~C. and {Cenko}, S.~B. and {De}, K. and {Dekany}, R. and {Frederick}, S. and {Golkhou}, V.~Z. and {Graham}, M.~J. and {Helou}, G. and {Ho}, A.~Y.~Q. and {Kasliwal}, M.~M. and {Kupfer}, T. and {Laher}, R.~R. and {Mahabal}, A. and {Masci}, F.~J. and {Riddle}, R. and {Rusholme}, B. and {Schulze}, S. and {Shupe}, D.~L. and {Smith}, R.~M. and {van Velzen}, S. and {Yan}, Lin and {Yao}, Y. and {Zhuang}, Z. and {Kulkarni}, S.~R.},
        title = "{The Zwicky Transient Facility Bright Transient Survey. I. Spectroscopic Classification and the Redshift Completeness of Local Galaxy Catalogs}",
      journal = {\apj},
         year = 2020,
        month = may,
       volume = {895},
       number = {1},
          eid = {32},
        pages = {32},
          doi = {10.3847/1538-4357/ab8943},
archivePrefix = {arXiv},
       eprint = {1910.12973},
 primaryClass = {astro-ph.HE},
       adsurl = {https://ui.adsabs.harvard.edu/abs/2020ApJ...895...32F}
}

@ARTICLE{2022ApJ...934L...7R,
       author = {{Riess}, Adam G. and {Yuan}, Wenlong and {Macri}, Lucas M. and {Scolnic}, Dan and {Brout}, Dillon and {Casertano}, Stefano and {Jones}, David O. and {Murakami}, Yukei and {Anand}, Gagandeep S. and {Breuval}, Louise and {Brink}, Thomas G. and {Filippenko}, Alexei V. and {Hoffmann}, Samantha and {Jha}, Saurabh W. and {D'arcy Kenworthy}, W. and {Mackenty}, John and {Stahl}, Benjamin E. and {Zheng}, WeiKang},
        title = "{A Comprehensive Measurement of the Local Value of the Hubble Constant with 1 km s$^{-1}$ Mpc$^{-1}$ Uncertainty from the Hubble Space Telescope and the SH0ES Team}",
      journal = {\apjl},
         year = 2022,
        month = jul,
       volume = {934},
       number = {1},
          eid = {L7},
        pages = {L7},
          doi = {10.3847/2041-8213/ac5c5b},
archivePrefix = {arXiv},
       eprint = {2112.04510},
 primaryClass = {astro-ph.CO},
       adsurl = {https://ui.adsabs.harvard.edu/abs/2022ApJ...934L...7R}
}

@ARTICLE{2013ApJ...773...76C,
       author = {{Chatzopoulos}, E. and {Wheeler}, J. Craig and {Vinko}, J. and {Horvath}, Z.~L. and {Nagy}, A.},
        title = "{Analytical Light Curve Models of Superluminous Supernovae: {\ensuremath{\chi}}$^{2}$-minimization of Parameter Fits}",
      journal = {\apj},
         year = 2013,
        month = aug,
       volume = {773},
       number = {1},
          eid = {76},
        pages = {76},
          doi = {10.1088/0004-637X/773/1/76},
archivePrefix = {arXiv},
       eprint = {1306.3447},
 primaryClass = {astro-ph.HE},
       adsurl = {https://ui.adsabs.harvard.edu/abs/2013ApJ...773...76C}
}

@ARTICLE{2025ApJ...993...39V,
       author = {{Vink{\'o}}, J{\'o}zsef and {Bodola}, Zs{\'o}fia R{\'e}ka and {G{\H{o}}d{\'e}ny}, {\'A}kos and {Cs{\'a}k}, Szelina Fruzsina and {K{\"o}nyves-T{\'o}th}, R{\'e}ka and {Nagy}, Andrea P. and {Szalai}, Tam{\'a}s and {B{\'a}nhidi}, Dominik and {B{\'\i}r{\'o}}, Imre Barna and {B{\'o}di}, Attila and {Bora}, Zs{\'o}fia and {Cs{\'a}nyi}, Istv{\'a}n and {Cseh}, Borb{\'a}la and {Heged{\"u}s}, Tibor and {Horti-D{\'a}vid}, {\'A}goston and {Jo{\'o}}, Andr{\'a}s P{\'e}ter and {Kalup}, Csilla and {Kriskovics}, Levente and {Mochn{\'a}cs}, Erika and {P{\'a}l}, Andr{\'a}s and {Reg{\'a}ly}, Zsolt and {Seli}, B{\'a}lint and {S{\'o}dor}, {\'A}d{\'a}m and {Szab{\'o}}, Norton Oliv{\'e}r and {Szak{\'a}ts}, R{\'o}bert and {Sz{\'e}kely}, P{\'e}ter and {Varga}, V{\'a}zsony and {Vida}, Kriszti{\'a}n},
        title = "{SN 2023ixf in M101: Physical Parameters from Bolometric Light Curve Modeling}",
      journal = {\apj},
         year = 2025,
        month = nov,
       volume = {993},
       number = {1},
          eid = {39},
        pages = {39},
          doi = {10.3847/1538-4357/ae0614},
archivePrefix = {arXiv},
       eprint = {2508.06654},
 primaryClass = {astro-ph.HE},
       adsurl = {https://ui.adsabs.harvard.edu/abs/2025ApJ...993...39V}
}

@ARTICLE{2016ApJ...829..109M,
       author = {{Morozova}, Viktoriya and {Piro}, Anthony L. and {Renzo}, Mathieu and {Ott}, Christian D.},
        title = "{Numerical Modeling of the Early Light Curves of Type IIP Supernovae}",
      journal = {\apj},
         year = 2016,
        month = oct,
       volume = {829},
       number = {2},
          eid = {109},
        pages = {109},
          doi = {10.3847/0004-637X/829/2/109},
archivePrefix = {arXiv},
       eprint = {1603.08530},
 primaryClass = {astro-ph.HE},
       adsurl = {https://ui.adsabs.harvard.edu/abs/2016ApJ...829..109M}
}

@ARTICLE{2018MNRAS.476.2840M,
       author = {{Moriya}, Takashi J. and {F{\"o}rster}, Francisco and {Yoon}, Sung-Chul and {Gr{\"a}fener}, G{\"o}tz and {Blinnikov}, Sergei I.},
        title = "{Type IIP supernova light curves affected by the acceleration of red supergiant winds}",
      journal = {\mnras},
         year = 2018,
        month = may,
       volume = {476},
       number = {2},
        pages = {2840-2851},
          doi = {10.1093/mnras/sty475},
archivePrefix = {arXiv},
       eprint = {1802.07752},
 primaryClass = {astro-ph.HE},
       adsurl = {https://ui.adsabs.harvard.edu/abs/2018MNRAS.476.2840M}
}

@ARTICLE{Martinez2022A&A...660A..42M,
       author = {{Martinez}, L. and {Anderson}, J.~P. and {Bersten}, M.~C. and {Hamuy}, M. and {Gonz{\'a}lez-Gait{\'a}n}, S. and {Orellana}, M. and {Stritzinger}, M. and {Phillips}, M.~M. and {Guti{\'e}rrez}, C.~P. and {Burns}, C. and {de Jaeger}, T. and {Ertini}, K. and {Folatelli}, G. and {F{\"o}rster}, F. and {Galbany}, L. and {Hoeflich}, P. and {Hsiao}, E.~Y. and {Morrell}, N. and {Pessi}, P.~J. and {Suntzeff}, N.~B.},
        title = "{Type II supernovae from the Carnegie Supernova Project-I. III. Understanding SN II diversity through correlations between physical and observed properties}",
      journal = {\aap},
         year = 2022,
        month = apr,
       volume = {660},
          eid = {A42},
        pages = {A42},
          doi = {10.1051/0004-6361/202142555},
archivePrefix = {arXiv},
       eprint = {2202.11220},
 primaryClass = {astro-ph.SR},
       adsurl = {https://ui.adsabs.harvard.edu/abs/2022A&A...660A..42M}
}

@ARTICLE{Teja2026ApJ..1005....4T,
       author = {{Teja}, Rishabh Singh and {Sahu}, D.~K. and {Anupama}, G.~C. and {Singh}, Avinash and {Dutta}, Amrit and {Rameshan}, Gitika and {Das}, Hrishav and {Kawabata}, Koji S. and {Singh}, Mridweeka and {Bhalerao}, Varun},
        title = "{Subluminous Type IIP SN 2024abfl as a Result of a Significantly Low-energy Fe-core Collapse}",
      journal = {\apj},
         year = 2026,
        month = jul,
       volume = {1005},
       number = {1},
          eid = {4},
        pages = {4},
          doi = {10.3847/1538-4357/ae75dd},
       adsurl = {https://ui.adsabs.harvard.edu/abs/2026ApJ..1005....4T}
}

@ARTICLE{Goldberg2019ApJ...879....3G,
       author = {{Goldberg}, Jared A. and {Bildsten}, Lars and {Paxton}, Bill},
        title = "{Inferring Explosion Properties from Type II-Plateau Supernova Light Curves}",
      journal = {\apj},
         year = 2019,
        month = jul,
       volume = {879},
       number = {1},
          eid = {3},
        pages = {3},
          doi = {10.3847/1538-4357/ab22b6},
archivePrefix = {arXiv},
       eprint = {1903.09114},
 primaryClass = {astro-ph.SR},
       adsurl = {https://ui.adsabs.harvard.edu/abs/2019ApJ...879....3G}
}

@ARTICLE{Kasen2009ApJ...703.2205K,
       author = {{Kasen}, Daniel and {Woosley}, S.~E.},
        title = "{Type II Supernovae: Model Light Curves and Standard Candle Relationships}",
      journal = {\apj},
         year = 2009,
        month = oct,
       volume = {703},
       number = {2},
        pages = {2205-2216},
          doi = {10.1088/0004-637X/703/2/2205},
archivePrefix = {arXiv},
       eprint = {0910.1590},
 primaryClass = {astro-ph.CO},
       adsurl = {https://ui.adsabs.harvard.edu/abs/2009ApJ...703.2205K}
}

@ARTICLE{2019MNRAS.490.2799D,
       author = {{de Jaeger}, T. and {Zheng}, W. and {Stahl}, B.~E. and {Filippenko}, A.~V. and {Brink}, T.~G. and {Bigley}, A. and {Blanchard}, K. and {Blanchard}, P.~K. and {Bradley}, J. and {Cargill}, S.~K. and {Casper}, C. and {Cenko}, S.~B. and {Channa}, S. and {Choi}, B.~Y. and {Clubb}, K.~I. and {Cobb}, B.~E. and {Cohen}, D. and {de Kouchkovsky}, M. and {Ellison}, M. and {Falcon}, E. and {Fox}, O.~D. and {Fuller}, K. and {Ganeshalingam}, M. and {Gould}, C. and {Graham}, M.~L. and {Halevi}, G. and {Hayakawa}, K.~T. and {Hestenes}, J. and {Hyland}, M.~P. and {Jeffers}, B. and {Joubert}, N. and {Kandrashoff}, M.~T. and {Kelly}, P.~L. and {Kim}, H. and {Kim}, M. and {Kumar}, S. and {Leonard}, E.~J. and {Li}, G.~Z. and {Lowe}, T.~B. and {Lu}, P. and {Mason}, M. and {McAllister}, K.~J. and {Mauerhan}, J.~C. and {Modjaz}, M. and {Molloy}, J. and {Perley}, D.~A. and {Pina}, K. and {Poznanski}, D. and {Ross}, T.~W. and {Shivvers}, I. and {Silverman}, J.~M. and {Soler}, C. and {Stegman}, S. and {Taylor}, S. and {Tang}, K. and {Wilkins}, A. and {Wang}, Xiaofeng and {Wang}, Xianggao and {Yuk}, H. and {Yunus}, S. and {Zhang}, K.~D.},
        title = "{The Berkeley sample of Type II supernovae: BVRI light curves and spectroscopy of 55 SNe II}",
      journal = {\mnras},
         year = 2019,
        month = dec,
       volume = {490},
       number = {2},
        pages = {2799-2821},
          doi = {10.1093/mnras/stz2714},
archivePrefix = {arXiv},
       eprint = {1909.13813},
 primaryClass = {astro-ph.HE},
       adsurl = {https://ui.adsabs.harvard.edu/abs/2019MNRAS.490.2799D}
}

@ARTICLE{Das2026PASP..138b4204D,
       author = {{Das}, Kaustav K. and {Kasliwal}, Mansi M. and {Sollerman}, Jesper and {Fremling}, Christoffer and {Moriya}, Takashi J. and {Hinds}, K.-Ryan and {Perley}, Daniel A. and {Bellm}, Eric C. and {Chen}, Tracy X. and {O'Connor}, Evan P. and {Coughlin}, Michael W. and {Jacobson-Gal{\'a}n}, W.~V. and {Gangopadhyay}, Anjasha and {Graham}, Matthew and {Kulkarni}, S.~R. and {Purdum}, Josiah and {Sarin}, Nikhil and {Schulze}, Steve and {Singh}, Avinash and {Tsuna}, Daichi and {Wold}, Avery},
        title = "{Low-luminosity Type IIP Supernovae from the Zwicky Transient Facility Census of the Local Universe. II. Lightcurve Analysis}",
      journal = {\pasp},
         year = 2026,
        month = feb,
       volume = {138},
       number = {2},
          eid = {024204},
        pages = {024204},
          doi = {10.1088/1538-3873/ae33f5},
archivePrefix = {arXiv},
       eprint = {2506.20068},
 primaryClass = {astro-ph.HE},
       adsurl = {https://ui.adsabs.harvard.edu/abs/2026PASP..138b4204D}
}

@INPROCEEDINGS{Tody1993ASPC...52..173T,
   author = {{Tody}, D.},
    title = "{IRAF in the Nineties}",
booktitle = {Astronomical Data Analysis Software and Systems II},
     year = 1993,
   series = {Astronomical Society of the Pacific Conference Series},
   volume = 52,
   editor = {{Hanisch}, R.~J. and {Brissenden}, R.~J.~V. and {Barnes}, J.
	},
    month = jan,
    pages = {173},
   adsurl = {http://adsabs.harvard.edu/abs/1993ASPC...52..173T}
}

@INPROCEEDINGS{Tody1986SPIE..627..733T,
   author = {{Tody}, D.},
    title = "{The IRAF Data Reduction and Analysis System}",
booktitle = {Instrumentation in astronomy VI},
     year = 1986,
   series = {\procspie},
   volume = 627,
   editor = {{Crawford}, D.~L.},
    month = jan,
    pages = {733},
      doi = {10.1117/12.968154},
   adsurl = {http://adsabs.harvard.edu/abs/1986SPIE..627..733T}
}

@ARTICLE{Sollerman2021A&A...655A.105S,
       author = {{Sollerman}, J. and {Yang}, S. and {Schulze}, S. and {Strotjohann}, N.~L. and {Jerkstrand}, A. and {Van Dyk}, S.~D. and {Kool}, E.~C. and {Barbarino}, C. and {Brink}, T.~G. and {Bruch}, R. and {De}, K. and {Filippenko}, A.~V. and {Fremling}, C. and {Patra}, K.~C. and {Perley}, D. and {Yan}, L. and {Yang}, Y. and {Andreoni}, I. and {Campbell}, R. and {Coughlin}, M. and {Kasliwal}, M. and {Kim}, Y.-L. and {Rigault}, M. and {Shin}, K. and {Tzanidakis}, A. and {Ashley}, M.~C.~B. and {Moore}, A.~M. and {Travouillon}, T.},
        title = "{The Type II supernova SN 2020jfo in M 61, implications for progenitor system, and explosion dynamics}",
      journal = {\aap},
         year = 2021,
        month = nov,
       volume = {655},
          eid = {A105},
        pages = {A105},
          doi = {10.1051/0004-6361/202141374},
archivePrefix = {arXiv},
       eprint = {2107.14503},
 primaryClass = {astro-ph.HE},
       adsurl = {https://ui.adsabs.harvard.edu/abs/2021A&A...655A.105S}
}

@ARTICLE{Ailawadhi2023MNRAS.519..248A,
       author = {{Ailawadhi}, B. and {Dastidar}, R. and {Misra}, K. and {Roy}, R. and {Hiramatsu}, D. and {Howell}, D.~A. and {Brink}, T.~G. and {Zheng}, W. and {Galbany}, L. and {Shahbandeh}, M. and {Arcavi}, I. and {Ashall}, C. and {Bostroem}, K.~A. and {Burke}, J. and {Chapman}, T. and {Dimple} and {Filippenko}, A.~V. and {Gangopadhyay}, A. and {Ghosh}, A. and {Hoffman}, A.~M. and {Hosseinzadeh}, G. and {Jennings}, C. and {Jha}, V.~K. and {Kumar}, A. and {Karamehmetoglu}, E. and {McCully}, C. and {McGinness}, E. and {M{\"u}ller-Bravo}, T.~E. and {Murakami}, Y.~S. and {Pandey}, S.~B. and {Pellegrino}, C. and {Piscarreta}, L. and {Rho}, J. and {Stritzinger}, M. and {Sunseri}, J. and {Van Dyk}, S.~D. and {Yadav}, L.},
        title = "{Photometric and spectroscopic analysis of the Type II SN 2020jfo with a short plateau}",
      journal = {\mnras},
         year = 2023,
        month = feb,
       volume = {519},
       number = {1},
        pages = {248-270},
          doi = {10.1093/mnras/stac3234},
archivePrefix = {arXiv},
       eprint = {2211.02823},
 primaryClass = {astro-ph.HE},
       adsurl = {https://ui.adsabs.harvard.edu/abs/2023MNRAS.519..248A}
}

@ARTICLE{Teja2022ApJ...930...34T,
       author = {{Teja}, Rishabh Singh and {Singh}, Avinash and {Sahu}, D.~K. and {Anupama}, G.~C. and {Kumar}, Brajesh and {Nayana}, A.~J.},
        title = "{SN 2020jfo: A Short-plateau Type II Supernova from a Low-mass Progenitor}",
      journal = {\apj},
         year = 2022,
        month = may,
       volume = {930},
       number = {1},
          eid = {34},
        pages = {34},
          doi = {10.3847/1538-4357/ac610b},
archivePrefix = {arXiv},
       eprint = {2202.09412},
 primaryClass = {astro-ph.HE},
       adsurl = {https://ui.adsabs.harvard.edu/abs/2022ApJ...930...34T}
}

@ARTICLE{Kilpatrick2023MNRAS.524.2161K,
       author = {{Kilpatrick}, Charles D. and {Izzo}, Luca and {Bentley}, Rory O. and {Chambers}, Kenneth C. and {Coulter}, David A. and {Drout}, Maria R. and {de Boer}, Thomas and {Foley}, Ryan J. and {Gall}, Christa and {Halford}, Melissa R. and {Jones}, David O. and {Langeroodi}, Danial and {Lin}, Chien-Cheng and {Magnier}, Eugene A. and {McGill}, Peter and {O'Grady}, Anna J.~G. and {Pan}, Yen-Chen and {Ramirez-Ruiz}, Enrico and {Rest}, Armin and {Swift}, Jonathan J. and {Tinyanont}, Samaporn and {Villar}, V. Ashley and {Wainscoat}, Richard J. and {Wasserman}, Amanda Rose and {Yadavalli}, S. Karthik and {Yang}, Grace},
        title = "{Type II-P supernova progenitor star initial masses and SN 2020jfo: direct detection, light-curve properties, nebular spectroscopy, and local environment}",
      journal = {\mnras},
         year = 2023,
        month = sep,
       volume = {524},
       number = {2},
        pages = {2161-2185},
          doi = {10.1093/mnras/stad1954},
archivePrefix = {arXiv},
       eprint = {2307.00550},
 primaryClass = {astro-ph.SR},
       adsurl = {https://ui.adsabs.harvard.edu/abs/2023MNRAS.524.2161K}
}

@ARTICLE{Sahu2006MNRAS.372.1315S,
       author = {{Sahu}, D.~K. and {Anupama}, G.~C. and {Srividya}, S. and {Muneer}, S.},
        title = "{Photometric and spectroscopic evolution of the Type IIP supernova SN 2004et}",
      journal = {\mnras},
         year = 2006,
        month = nov,
       volume = {372},
       number = {3},
        pages = {1315-1324},
          doi = {10.1111/j.1365-2966.2006.10937.x},
archivePrefix = {arXiv},
       eprint = {astro-ph/0608432},
 primaryClass = {astro-ph},
       adsurl = {https://ui.adsabs.harvard.edu/abs/2006MNRAS.372.1315S}
}

@ARTICLE{Tartaglia2021ApJ...907...52T,
       author = {{Tartaglia}, L. and {Sand}, D.~J. and {Groh}, J.~H. and {Valenti}, S. and {Wyatt}, S.~D. and {Bostroem}, K.~A. and {Brown}, P.~J. and {Yang}, S. and {Burke}, J. and {Chen}, T.-W. and {Davis}, S. and {F{\"o}rster}, F. and {Galbany}, L. and {Haislip}, J. and {Hiramatsu}, D. and {Hosseinzadeh}, G. and {Howell}, D.~A. and {Hsiao}, E.~Y. and {Jha}, S.~W. and {Kouprianov}, V. and {Kuncarayakti}, H. and {Lyman}, J.~D. and {McCully}, C. and {Phillips}, M.~M. and {Rau}, A. and {Reichart}, D.~E. and {Shahbandeh}, M. and {Strader}, J.},
        title = "{The Early Discovery of SN 2017ahn: Signatures of Persistent Interaction in a Fast-declining Type II Supernova}",
      journal = {\apj},
         year = 2021,
        month = jan,
       volume = {907},
       number = {1},
          eid = {52},
        pages = {52},
          doi = {10.3847/1538-4357/abca8a},
archivePrefix = {arXiv},
       eprint = {2008.06515},
 primaryClass = {astro-ph.HE},
       adsurl = {https://ui.adsabs.harvard.edu/abs/2021ApJ...907...52T}
}

@misc{Lupton2005SDSSTransform,
  author       = {{Lupton}, Robert H.},
  title        = {{SDSS} Magnitudes to Physical Fluxes and the {$gri$}--to--{$UBVR_{\mathrm c}I_{\mathrm c}$} Transformations},
  year         = {2005},
  howpublished = {\url{https://www.sdss3.org/dr8/algorithms/sdssUBVRITransform.php}}
}

@ARTICLE{2022TNSAN.191....1G,
       author = {{Goldwasser}, S. and {Yaron}, O. and {Sass}, A. and {Irani}, I. and {Gal-Yam}, A. and {Howell}, D.~A.},
        title = "{The Next Generation SuperFit (NGSF) tool is now available for online execution on WISeREP}",
      journal = {Transient Name Server AstroNote},
         year = 2022,
        month = sep,
       volume = {191},
        pages = {1},
       adsurl = {https://ui.adsabs.harvard.edu/abs/2022TNSAN.191....1G}
}

@ARTICLE{Stoppa2026MNRAS.549g1066S,
       author = {{Stoppa}, Fiorenzo and {Smartt}, Stephen J.},
        title = "{SNID─SAGE: a modern framework for interactive supernova classification and spectral analysis}",
      journal = {\mnras},
         year = 2026,
        month = jul,
       volume = {549},
       number = {4},
          eid = {stag1066},
        pages = {stag1066},
          doi = {10.1093/mnras/stag1066},
archivePrefix = {arXiv},
       eprint = {2603.28741},
 primaryClass = {astro-ph.IM},
       adsurl = {https://ui.adsabs.harvard.edu/abs/2026MNRAS.549g1066S}
}

@ARTICLE{2025TNSCR4740....1B,
       author = {{Balcon}, C.},
        title = "{Transient Classification Report for 2025-11-27}",
      journal = {Transient Name Server Classification Report},
         year = 2025,
        month = nov,
       volume = {2025-4740},
        pages = {1},
       adsurl = {https://ui.adsabs.harvard.edu/abs/2025TNSCR4740....1B}
}

@ARTICLE{2025TNSTR4639....1S,
       author = {{Sasaoka}, T. and {Koshi}, R. and {Sako}, S. and {Tominaga}, N. and {Taguchi}, K.},
        title = "{Tomo-e Gozen Transient Discovery Report for 2025-11-18}",
      journal = {Transient Name Server Discovery Report},
         year = 2025,
        month = nov,
       volume = {2025-4639},
        pages = {1},
       adsurl = {https://ui.adsabs.harvard.edu/abs/2025TNSTR4639....1S}
}

@ARTICLE{2019PASP..131a8003M,
       author = {{Masci}, Frank J. and {Laher}, Russ R. and {Rusholme}, Ben and {Shupe}, David L. and {Groom}, Steven and {Surace}, Jason and {Jackson}, Edward and {Monkewitz}, Serge and {Beck}, Ron and {Flynn}, David and {Terek}, Scott and {Landry}, Walter and {Hacopians}, Eugean and {Desai}, Vandana and {Howell}, Justin and {Brooke}, Tim and {Imel}, David and {Wachter}, Stefanie and {Ye}, Quan-Zhi and {Lin}, Hsing-Wen and {Cenko}, S. Bradley and {Cunningham}, Virginia and {Rebbapragada}, Umaa and {Bue}, Brian and {Miller}, Adam A. and {Mahabal}, Ashish and {Bellm}, Eric C. and {Patterson}, Maria T. and {Juri{\'c}}, Mario and {Golkhou}, V. Zach and {Ofek}, Eran O. and {Walters}, Richard and {Graham}, Matthew and {Kasliwal}, Mansi M. and {Dekany}, Richard G. and {Kupfer}, Thomas and {Burdge}, Kevin and {Cannella}, Christopher B. and {Barlow}, Tom and {Van Sistine}, Angela and {Giomi}, Matteo and {Fremling}, Christoffer and {Blagorodnova}, Nadejda and {Levitan}, David and {Riddle}, Reed and {Smith}, Roger M. and {Helou}, George and {Prince}, Thomas A. and {Kulkarni}, Shrinivas R.},
        title = "{The Zwicky Transient Facility: Data Processing, Products, and Archive}",
      journal = {\pasp},
         year = 2019,
        month = jan,
       volume = {131},
       number = {995},
        pages = {018003},
          doi = {10.1088/1538-3873/aae8ac},
archivePrefix = {arXiv},
       eprint = {1902.01872},
 primaryClass = {astro-ph.IM},
       adsurl = {https://ui.adsabs.harvard.edu/abs/2019PASP..131a8003M}
}

@ARTICLE{2020PASP..132h5002S,
       author = {{Smith}, K.~W. and {Smartt}, S.~J. and {Young}, D.~R. and {Tonry}, J.~L. and {Denneau}, L. and {Flewelling}, H. and {Heinze}, A.~N. and {Weiland}, H.~J. and {Stalder}, B. and {Rest}, A. and {Stubbs}, C.~W. and {Anderson}, J.~P. and {Chen}, T.-W. and {Clark}, P. and {Do}, A. and {F{\"o}rster}, F. and {Fulton}, M. and {Gillanders}, J. and {McBrien}, O.~R. and {O'Neill}, D. and {Srivastav}, S. and {Wright}, D.~E.},
        title = "{Design and Operation of the ATLAS Transient Science Server}",
      journal = {\pasp},
         year = 2020,
        month = aug,
       volume = {132},
       number = {1014},
          eid = {085002},
        pages = {085002},
          doi = {10.1088/1538-3873/ab936e},
archivePrefix = {arXiv},
       eprint = {2003.09052},
 primaryClass = {astro-ph.IM},
       adsurl = {https://ui.adsabs.harvard.edu/abs/2020PASP..132h5002S}
}

@ARTICLE{2018PASP..130f4505T,
       author = {{Tonry}, J.~L. and {Denneau}, L. and {Heinze}, A.~N. and {Stalder}, B. and {Smith}, K.~W. and {Smartt}, S.~J. and {Stubbs}, C.~W. and {Weiland}, H.~J. and {Rest}, A.},
        title = "{ATLAS: A High-cadence All-sky Survey System}",
      journal = {\pasp},
         year = 2018,
        month = jun,
       volume = {130},
       number = {988},
        pages = {064505},
          doi = {10.1088/1538-3873/aabadf},
archivePrefix = {arXiv},
       eprint = {1802.00879},
 primaryClass = {astro-ph.IM},
       adsurl = {https://ui.adsabs.harvard.edu/abs/2018PASP..130f4505T}
}

@ARTICLE{Anderson2024A&A...692A..95A,
       author = {{Anderson}, J.~P. and {Contreras}, C. and {Stritzinger}, M.~D. and {Hamuy}, M. and {Phillips}, M.~M. and {Suntzeff}, N.~B. and {Morrell}, N. and {Gonz{\'a}lez-Gait{\'a}n}, S. and {Guti{\'e}rrez}, C.~P. and {Burns}, C.~R. and {Hsiao}, E.~Y. and {Anais}, J. and {Ashall}, C. and {Baltay}, C. and {Baron}, E. and {Bersten}, M. and {Busta}, L. and {Castell{\'o}n}, S. and {de Jaeger}, T. and {DePoy}, D. and {Filippenko}, A.~V. and {Folatelli}, G. and {F{\"o}rster}, F. and {Galbany}, L. and {Gall}, C. and {Goobar}, A. and {Gonzalez}, C. and {Hadjiyska}, E. and {Hoeflich}, P. and {Krisciunas}, K. and {Krzemi{\'n}ski}, W. and {Li}, W. and {Madore}, B. and {Marshall}, J. and {Martinez}, L. and {Nugent}, P. and {Pessi}, P.~J. and {Piro}, A.~L. and {Rheault}, J.-P. and {Ryder}, S. and {Ser{\'o}n}, J. and {Shappee}, B.~J. and {Taddia}, F. and {Torres}, S. and {Thomas-Osip}, J. and {Uddin}, S.},
        title = "{Optical and near-infrared photometry of 94 type II supernovae from the Carnegie Supernova Project}",
      journal = {\aap},
         year = 2024,
        month = dec,
       volume = {692},
          eid = {A95},
        pages = {A95},
          doi = {10.1051/0004-6361/202244401},
archivePrefix = {arXiv},
       eprint = {2410.06738},
 primaryClass = {astro-ph.CO},
       adsurl = {https://ui.adsabs.harvard.edu/abs/2024A&A...692A..95A}
}

@ARTICLE{XiangDanfeng2024ApJ...969L..15X,
       author = {{Xiang}, Danfeng and {Mo}, Jun and {Wang}, Xiaofeng and {Wang}, Lingzhi and {Zhang}, Jujia and {Lin}, Han and {Chen}, Liyang and {Song}, Cuiying and {Liu}, Liang-Duan and {Wang}, Zhenyu and {Li}, Gaici},
        title = "{The Red Supergiant Progenitor of Type II Supernova 2024ggi}",
      journal = {\apjl},
         year = 2024,
        month = jul,
       volume = {969},
       number = {1},
          eid = {L15},
        pages = {L15},
          doi = {10.3847/2041-8213/ad54b3},
archivePrefix = {arXiv},
       eprint = {2405.07699},
 primaryClass = {astro-ph.HE},
       adsurl = {https://ui.adsabs.harvard.edu/abs/2024ApJ...969L..15X}
}

@ARTICLE{VanDyk2025Galax..13...33V,
       author = {{Van Dyk}, Schuyler D.},
        title = "{Red Supergiants as Supernova Progenitors}",
      journal = {Galaxies},
         year = 2025,
        month = apr,
       volume = {13},
       number = {2},
          eid = {33},
        pages = {33},
          doi = {10.3390/galaxies13020033},
archivePrefix = {arXiv},
       eprint = {2507.15973},
 primaryClass = {astro-ph.SR},
       adsurl = {https://ui.adsabs.harvard.edu/abs/2025Galax..13...33V}
}

@ARTICLE{Galbany2016AJ....151...33G,
       author = {{Galbany}, Llu{\'\i}s and {Hamuy}, Mario and {Phillips}, Mark M. and {Suntzeff}, Nicholas B. and {Maza}, Jos{\'e} and {de Jaeger}, Thomas and {Moraga}, Tania and {Gonz{\'a}lez-Gait{\'a}n}, Santiago and {Krisciunas}, Kevin and {Morrell}, Nidia I. and {Thomas-Osip}, Joanna and {Krzeminski}, Wojtek and {Gonz{\'a}lez}, Luis and {Antezana}, Roberto and {Wishnjewski}, Marina and {McCarthy}, Patrick and {Anderson}, Joseph P. and {Guti{\'e}rrez}, Claudia P. and {Stritzinger}, Maximilian and {Folatelli}, Gast{\'o}n and {Anguita}, Claudio and {Galaz}, Gaspar and {Green}, Elisabeth M. and {Impey}, Chris and {Kim}, Yong-Cheol and {Kirhakos}, Sofia and {Malkan}, Mathew A. and {Mulchaey}, John S. and {Phillips}, Andrew C. and {Pizzella}, Alessandro and {Prosser}, Charles F. and {Schmidt}, Brian P. and {Schommer}, Robert A. and {Sherry}, William and {Strolger}, Louis-Gregory and {Wells}, Lisa A. and {Williger}, Gerard M.},
        title = "{UBVRIz Light Curves of 51 Type II Supernovae}",
      journal = {\aj},
         year = 2016,
        month = feb,
       volume = {151},
       number = {2},
          eid = {33},
        pages = {33},
          doi = {10.3847/0004-6256/151/2/33},
archivePrefix = {arXiv},
       eprint = {1511.08402},
 primaryClass = {astro-ph.SR},
       adsurl = {https://ui.adsabs.harvard.edu/abs/2016AJ....151...33G}
}

@ARTICLE{Maund2005MNRAS.364L..33M,
       author = {{Maund}, Justyn R. and {Smartt}, Stephen J. and {Danziger}, I. John},
        title = "{The progenitor of SN 2005cs in the Whirlpool Galaxy}",
      journal = {\mnras},
         year = 2005,
        month = nov,
       volume = {364},
       number = {1},
        pages = {L33-L37},
          doi = {10.1111/j.1745-3933.2005.00100.x},
archivePrefix = {arXiv},
       eprint = {astro-ph/0507502},
 primaryClass = {astro-ph},
       adsurl = {https://ui.adsabs.harvard.edu/abs/2005MNRAS.364L..33M}
}

@ARTICLE{Spiro2014MNRAS.439.2873S,
       author = {{Spiro}, S. and {Pastorello}, A. and {Pumo}, M.~L. and {Zampieri}, L. and {Turatto}, M. and {Smartt}, S.~J. and {Benetti}, S. and {Cappellaro}, E. and {Valenti}, S. and {Agnoletto}, I. and {Altavilla}, G. and {Aoki}, T. and {Brocato}, E. and {Corsini}, E.~M. and {Di Cianno}, A. and {Elias-Rosa}, N. and {Hamuy}, M. and {Enya}, K. and {Fiaschi}, M. and {Folatelli}, G. and {Desidera}, S. and {Harutyunyan}, A. and {Howell}, D.~A. and {Kawka}, A. and {Kobayashi}, Y. and {Leibundgut}, B. and {Minezaki}, T. and {Navasardyan}, H. and {Nomoto}, K. and {Mattila}, S. and {Pietrinferni}, A. and {Pignata}, G. and {Raimondo}, G. and {Salvo}, M. and {Schmidt}, B.~P. and {Sollerman}, J. and {Spyromilio}, J. and {Taubenberger}, S. and {Valentini}, G. and {Vennes}, S. and {Yoshii}, Y.},
        title = "{Low luminosity Type II supernovae - II. Pointing towards moderate mass precursors}",
      journal = {\mnras},
         year = 2014,
        month = apr,
       volume = {439},
       number = {3},
        pages = {2873-2892},
          doi = {10.1093/mnras/stu156},
archivePrefix = {arXiv},
       eprint = {1401.5426},
 primaryClass = {astro-ph.SR},
       adsurl = {https://ui.adsabs.harvard.edu/abs/2014MNRAS.439.2873S}
}

@ARTICLE{DasKaustav2025PASP..137d4203D,
       author = {{Das}, Kaustav K. and {Kasliwal}, Mansi M. and {Fremling}, Christoffer and {Sollerman}, Jesper and {Perley}, Daniel A. and {De}, Kishalay and {Tzanidakis}, Anastasios and {Sit}, Tawny and {Adams}, Scott and {Anand}, Shreya and {Ahumuda}, Tomas and {Andreoni}, Igor and {Brennan}, Se{\'a}n and {Brink}, Thomas and {Bruch}, Rachel J. and {Chen}, Ping and {Chu}, Matthew R. and {Cook}, David O. and {Covarrubias}, Sofia and {Dahiwale}, Aishwarya and {Earley}, Nicholas and {Ho}, Anna Y.~Q. and {Gal-Yam}, Avishay and {Gangopadhyay}, Anjasha and {Hammerstein}, Erica and {Hinds}, K.-Ryan and {Karambelkar}, Viraj and {Kong}, Yihan and {Kulkarni}, S.~R. and {Jegou du Laz}, Theophile and {Liu}, Chang and {Meynardie}, William and {Miller}, Adam A. and {Nir}, Guy and {Patra}, Kishore C. and {Pessi}, Priscila J. and {Rich}, R. Michael and {Rehemtulla}, Nabeel and {Rose}, Sam and {Rusholme}, Ben and {Schulze}, Steve and {Sharma}, Yashvi and {Singh}, Avinash and {Smith}, Roger and {Stein}, Robert and {Mandigo-Stoba}, Milan Sharma and {Strotjohann}, Nora L. and {Qin}, Yu-Jing and {Wise}, Jacob and {Wold}, Avery and {Yan}, Lin and {Yang}, Yi and {Yao}, Yuhan and {Zimmerman}, Erez},
        title = "{Low-luminosity Type IIP Supernovae from the Zwicky Transient Facility Census of the Local Universe. I. Luminosity Function, Volumetric Rate}",
      journal = {\pasp},
         year = 2025,
        month = apr,
       volume = {137},
       number = {4},
          eid = {044203},
        pages = {044203},
          doi = {10.1088/1538-3873/adcaeb},
archivePrefix = {arXiv},
       eprint = {2502.19493},
 primaryClass = {astro-ph.HE},
       adsurl = {https://ui.adsabs.harvard.edu/abs/2025PASP..137d4203D}
}

@ARTICLE{Popov1993ApJ...414..712P,
       author = {{Popov}, D.~V.},
        title = "{An Analytical Model for the Plateau Stage of Type II Supernovae}",
      journal = {\apj},
         year = 1993,
        month = sep,
       volume = {414},
        pages = {712},
          doi = {10.1086/173117},
       adsurl = {https://ui.adsabs.harvard.edu/abs/1993ApJ...414..712P}
}

@ARTICLE{Filippenko1997ARA&A..35..309F,
       author = {{Filippenko}, Alexei V.},
        title = "{Optical Spectra of Supernovae}",
      journal = {\araa},
         year = 1997,
        month = jan,
       volume = {35},
        pages = {309-355},
          doi = {10.1146/annurev.astro.35.1.309},
       adsurl = {https://ui.adsabs.harvard.edu/abs/1997ARA&A..35..309F}
}

@ARTICLE{Heger2003ApJ...591..288H,
       author = {{Heger}, A. and {Fryer}, C.~L. and {Woosley}, S.~E. and {Langer}, N. and {Hartmann}, D.~H.},
        title = "{How Massive Single Stars End Their Life}",
      journal = {\apj},
         year = 2003,
        month = jul,
       volume = {591},
       number = {1},
        pages = {288-300},
          doi = {10.1086/375341},
archivePrefix = {arXiv},
       eprint = {astro-ph/0212469},
 primaryClass = {astro-ph},
       adsurl = {https://ui.adsabs.harvard.edu/abs/2003ApJ...591..288H}
}

@ARTICLE{Anderson2014ApJ...786...67A,
       author = {{Anderson}, Joseph P. and {Gonz{\'a}lez-Gait{\'a}n}, Santiago and {Hamuy}, Mario and {Guti{\'e}rrez}, Claudia P. and {Stritzinger}, Maximilian D. and {Olivares E.}, Felipe and {Phillips}, Mark M. and {Schulze}, Steve and {Antezana}, Roberto and {Bolt}, Luis and {Campillay}, Abdo and {Castell{\'o}n}, Sergio and {Contreras}, Carlos and {de Jaeger}, Thomas and {Folatelli}, Gast{\'o}n and {F{\"o}rster}, Francisco and {Freedman}, Wendy L. and {Gonz{\'a}lez}, Luis and {Hsiao}, Eric and {Krzemi{\'n}ski}, Wojtek and {Krisciunas}, Kevin and {Maza}, Jos{\'e} and {McCarthy}, Patrick and {Morrell}, Nidia I. and {Persson}, Sven E. and {Roth}, Miguel and {Salgado}, Francisco and {Suntzeff}, Nicholas B. and {Thomas-Osip}, Joanna},
        title = "{Characterizing the V-band Light-curves of Hydrogen-rich Type II Supernovae}",
      journal = {\apj},
         year = 2014,
        month = may,
       volume = {786},
       number = {1},
          eid = {67},
        pages = {67},
          doi = {10.1088/0004-637X/786/1/67},
archivePrefix = {arXiv},
       eprint = {1403.7091},
 primaryClass = {astro-ph.HE},
       adsurl = {https://ui.adsabs.harvard.edu/abs/2014ApJ...786...67A}
}

@ARTICLE{2016A&A...589A..53N,
       author = {{Nagy}, A.~P. and {Vink{\'o}}, J.},
        title = "{A two-component model for fitting light curves of core-collapse supernovae}",
      journal = {\aap},
         year = 2016,
        month = may,
       volume = {589},
          eid = {A53},
        pages = {A53},
          doi = {10.1051/0004-6361/201527931},
archivePrefix = {arXiv},
       eprint = {1602.04001},
 primaryClass = {astro-ph.IM},
       adsurl = {https://ui.adsabs.harvard.edu/abs/2016A&A...589A..53N}
}

@ARTICLE{2018RNAAS...2..230N,
       author = {{Nicholl}, Matt},
        title = "{SuperBol: A User-friendly Python Routine for Bolometric Light Curves}",
      journal = {Research Notes of the American Astronomical Society},
         year = 2018,
        month = dec,
       volume = {2},
       number = {4},
          eid = {230},
        pages = {230},
          doi = {10.3847/2515-5172/aaf799},
       adsurl = {https://ui.adsabs.harvard.edu/abs/2018RNAAS...2..230N}
}

@ARTICLE{Sanders2015ApJ...799..208S,
       author = {{Sanders}, N.~E. and {Soderberg}, A.~M. and {Gezari}, S. and {Betancourt}, M. and {Chornock}, R. and {Berger}, E. and {Foley}, R.~J. and {Challis}, P. and {Drout}, M. and {Kirshner}, R.~P. and {Lunnan}, R. and {Marion}, G.~H. and {Margutti}, R. and {McKinnon}, R. and {Milisavljevic}, D. and {Narayan}, G. and {Rest}, A. and {Kankare}, E. and {Mattila}, S. and {Smartt}, S.~J. and {Huber}, M.~E. and {Burgett}, W.~S. and {Draper}, P.~W. and {Hodapp}, K.~W. and {Kaiser}, N. and {Kudritzki}, R.~P. and {Magnier}, E.~A. and {Metcalfe}, N. and {Morgan}, J.~S. and {Price}, P.~A. and {Tonry}, J.~L. and {Wainscoat}, R.~J. and {Waters}, C.},
        title = "{Toward Characterization of the Type IIP Supernova Progenitor Population: A Statistical Sample of Light Curves from Pan-STARRS1}",
      journal = {\apj},
         year = 2015,
        month = feb,
       volume = {799},
       number = {2},
          eid = {208},
        pages = {208},
          doi = {10.1088/0004-637X/799/2/208},
archivePrefix = {arXiv},
       eprint = {1404.2004},
 primaryClass = {astro-ph.HE},
       adsurl = {https://ui.adsabs.harvard.edu/abs/2015ApJ...799..208S}
}

@ARTICLE{2017ApJ...850...89G,
       author = {{Guti{\'e}rrez}, Claudia P. and {Anderson}, Joseph P. and {Hamuy}, Mario and {Morrell}, Nidia and {Gonz{\'a}lez-Gaitan}, Santiago and {Stritzinger}, Maximilian D. and {Phillips}, Mark M. and {Galbany}, Lluis and {Folatelli}, Gast{\'o}n and {Dessart}, Luc and {Contreras}, Carlos and {Della Valle}, Massimo and {Freedman}, Wendy L. and {Hsiao}, Eric Y. and {Krisciunas}, Kevin and {Madore}, Barry F. and {Maza}, Jos{\'e} and {Suntzeff}, Nicholas B. and {Prieto}, Jose Luis and {Gonz{\'a}lez}, Luis and {Cappellaro}, Enrico and {Navarrete}, Mauricio and {Pizzella}, Alessandro and {Ruiz}, Maria T. and {Smith}, R. Chris and {Turatto}, Massimo},
        title = "{Type II Supernova Spectral Diversity. I. Observations, Sample Characterization, and Spectral Line Evolution}",
      journal = {\apj},
         year = 2017,
        month = nov,
       volume = {850},
       number = {1},
          eid = {89},
        pages = {89},
          doi = {10.3847/1538-4357/aa8f52},
archivePrefix = {arXiv},
       eprint = {1709.02487},
 primaryClass = {astro-ph.HE},
       adsurl = {https://ui.adsabs.harvard.edu/abs/2017ApJ...850...89G}
}

@ARTICLE{VanDyk2019ApJ...875..136V,
       author = {{Van Dyk}, Schuyler D. and {Zheng}, WeiKang and {Maund}, Justyn R. and {Brink}, Thomas G. and {Srinivasan}, Sundar and {Andrews}, Jennifer E. and {Smith}, Nathan and {Leonard}, Douglas C. and {Morozova}, Viktoriya and {Filippenko}, Alexei V. and {Conner}, Brody and {Milisavljevic}, Dan and {de Jaeger}, Thomas and {Long}, Knox S. and {Isaacson}, Howard and {Crossfield}, Ian J.~M. and {Kosiarek}, Molly R. and {Howard}, Andrew W. and {Fox}, Ori D. and {Kelly}, Patrick L. and {Piro}, Anthony L. and {Littlefair}, Stuart P. and {Dhillon}, Vik S. and {Wilson}, Richard and {Butterley}, Timothy and {Yunus}, Sameen and {Channa}, Sanyum and {Jeffers}, Benjamin T. and {Falcon}, Edward and {Ross}, Timothy W. and {Hestenes}, Julia C. and {Stegman}, Samantha M. and {Zhang}, Keto and {Kumar}, Sahana},
        title = "{The Type II-plateau Supernova 2017eaw in NGC 6946 and Its Red Supergiant Progenitor}",
      journal = {\apj},
         year = 2019,
        month = apr,
       volume = {875},
       number = {2},
          eid = {136},
        pages = {136},
          doi = {10.3847/1538-4357/ab1136},
archivePrefix = {arXiv},
       eprint = {1903.03872},
 primaryClass = {astro-ph.HE},
       adsurl = {https://ui.adsabs.harvard.edu/abs/2019ApJ...875..136V}
}

@ARTICLE{Szalai2019ApJ...876...19S,
       author = {{Szalai}, Tam{\'a}s and {Vink{\'o}}, J{\'o}zsef and {K{\"o}nyves-T{\'o}th}, R{\'e}ka and {Nagy}, Andrea P. and {Bostroem}, K. Azalee and {S{\'a}rneczky}, Kriszti{\'a}n and {Brown}, Peter J. and {Pejcha}, Ondrej and {B{\'o}di}, Attila and {Cseh}, Borb{\'a}la and {Cs{\"o}rnyei}, G{\'e}za and {Dencs}, Zolt{\'a}n and {Hanyecz}, Ott{\'o} and {Ign{\'a}cz}, Bernadett and {Kalup}, Csilla and {Kriskovics}, Levente and {Ordasi}, Andr{\'a}s and {P{\'a}l}, Andr{\'a}s and {Seli}, B{\'a}lint and {S{\'o}dor}, {\'A}d{\'a}m and {Szak{\'a}ts}, R{\'o}bert and {Vida}, Kriszti{\'a}n and {Zsidi}, Gabriella and {Konkoly Team} and {Arcavi}, Iair and {Ashall}, Chris and {Burke}, Jamison and {Galbany}, Llu{\'\i}s and {Hiramatsu}, Daichi and {Hosseinzadeh}, Griffin and {Hsiao}, Eric Y. and {Howell}, D. Andrew and {McCully}, Curtis and {Moran}, Shane and {Rho}, Jeonghee and {Sand}, David J. and {Shahbandeh}, Melissa and {Valenti}, Stefano and {Wang}, Xiaofeng and {Wheeler}, J. Craig and {Supernova Project}, Global},
        title = "{The Type II-P Supernova 2017eaw: From Explosion to the Nebular Phase}",
      journal = {\apj},
         year = 2019,
        month = may,
       volume = {876},
       number = {1},
          eid = {19},
        pages = {19},
          doi = {10.3847/1538-4357/ab12d0},
archivePrefix = {arXiv},
       eprint = {1903.09048},
 primaryClass = {astro-ph.HE},
       adsurl = {https://ui.adsabs.harvard.edu/abs/2019ApJ...876...19S}
}

@ARTICLE{Takats2014MNRAS.438..368T,
       author = {{Tak{\'a}ts}, K. and {Pumo}, M.~L. and {Elias-Rosa}, N. and {Pastorello}, A. and {Pignata}, G. and {Paillas}, E. and {Zampieri}, L. and {Anderson}, J.~P. and {Vink{\'o}}, J. and {Benetti}, S. and {Botticella}, M.-T. and {Bufano}, F. and {Campillay}, A. and {Cartier}, R. and {Ergon}, M. and {Folatelli}, G. and {Foley}, R.~J. and {F{\"o}rster}, F. and {Hamuy}, M. and {Hentunen}, V.-P. and {Kankare}, E. and {Leloudas}, G. and {Morrell}, N. and {Nissinen}, M. and {Phillips}, M.~M. and {Smartt}, S.~J. and {Stritzinger}, M. and {Taubenberger}, S. and {Valenti}, S. and {Van Dyk}, S.~D. and {Haislip}, J.~B. and {LaCluyze}, A.~P. and {Moore}, J.~P. and {Reichart}, D.},
        title = "{SN 2009N: linking normal and subluminous Type II-P SNe}",
      journal = {\mnras},
         year = 2014,
        month = feb,
       volume = {438},
       number = {1},
        pages = {368-387},
          doi = {10.1093/mnras/stt2203},
archivePrefix = {arXiv},
       eprint = {1311.2525},
 primaryClass = {astro-ph.SR},
       adsurl = {https://ui.adsabs.harvard.edu/abs/2014MNRAS.438..368T}
}

@ARTICLE{Elmhamdi2003MNRAS.338..939E,
       author = {{Elmhamdi}, Abouazza and {Danziger}, I.~J. and {Chugai}, N. and {Pastorello}, A. and {Turatto}, M. and {Cappellaro}, E. and {Altavilla}, G. and {Benetti}, S. and {Patat}, F. and {Salvo}, M.},
        title = "{Photometry and spectroscopy of the Type IIP SN 1999em from outburst to dust formation}",
      journal = {\mnras},
         year = 2003,
        month = feb,
       volume = {338},
       number = {4},
        pages = {939-956},
          doi = {10.1046/j.1365-8711.2003.06150.x},
archivePrefix = {arXiv},
       eprint = {astro-ph/0209623},
 primaryClass = {astro-ph},
       adsurl = {https://ui.adsabs.harvard.edu/abs/2003MNRAS.338..939E}
}

@ARTICLE{Leonard2002PASP..114...35L,
       author = {{Leonard}, Douglas C. and {Filippenko}, Alexei V. and {Gates}, Elinor L. and {Li}, Weidong and {Eastman}, Ronald G. and {Barth}, Aaron J. and {Bus}, Schelte J. and {Chornock}, Ryan and {Coil}, Alison L. and {Frink}, Sabine and {Grady}, Carol A. and {Harris}, Alan W. and {Malkan}, Matthew A. and {Matheson}, Thomas and {Quirrenbach}, Andreas and {Treffers}, Richard R.},
        title = "{The Distance to SN 1999em in NGC 1637 from the Expanding Photosphere Method}",
      journal = {\pasp},
         year = 2002,
        month = jan,
       volume = {114},
       number = {791},
        pages = {35-64},
          doi = {10.1086/324785},
archivePrefix = {arXiv},
       eprint = {astro-ph/0109535},
 primaryClass = {astro-ph},
       adsurl = {https://ui.adsabs.harvard.edu/abs/2002PASP..114...35L}
}

@ARTICLE{Hamuy2002ApJ...566L..63H,
       author = {{Hamuy}, Mario and {Pinto}, Philip A.},
        title = "{Type II Supernovae as Standardized Candles}",
      journal = {\apjl},
         year = 2002,
        month = feb,
       volume = {566},
       number = {2},
        pages = {L63-L65},
          doi = {10.1086/339676},
archivePrefix = {arXiv},
       eprint = {astro-ph/0201279},
 primaryClass = {astro-ph},
       adsurl = {https://ui.adsabs.harvard.edu/abs/2002ApJ...566L..63H}
}

@ARTICLE{Kirshner1974ApJ...193...27K,
       author = {{Kirshner}, R.~P. and {Kwan}, J.},
        title = "{Distances to extragalactic supernovae.}",
      journal = {\apj},
         year = 1974,
        month = oct,
       volume = {193},
        pages = {27-36},
          doi = {10.1086/153123},
       adsurl = {https://ui.adsabs.harvard.edu/abs/1974ApJ...193...27K}
}

@ARTICLE{Cardelli1989ApJ...345..245C,
       author = {{Cardelli}, Jason A. and {Clayton}, Geoffrey C. and {Mathis}, John S.},
        title = "{The Relationship between Infrared, Optical, and Ultraviolet Extinction}",
      journal = {\apj},
         year = 1989,
        month = oct,
       volume = {345},
        pages = {245},
          doi = {10.1086/167900},
       adsurl = {https://ui.adsabs.harvard.edu/abs/1989ApJ...345..245C}
}

@ARTICLE{Schlafly2011ApJ...737..103S,
   author = {{Schlafly}, E.~F. and {Finkbeiner}, D.~P.},
    title = "{Measuring Reddening with Sloan Digital Sky Survey Stellar Spectra and Recalibrating SFD}",
  journal = {\apj},
archivePrefix = "arXiv",
   eprint = {1012.4804},
 primaryClass = "astro-ph.GA",
     year = 2011,
    month = aug,
   volume = 737,
      eid = {103},
    pages = {103},
      doi = {10.1088/0004-637X/737/2/103},
   adsurl = {http://adsabs.harvard.edu/abs/2011ApJ...737..103S}
}

@ARTICLE{2021ApJ...906...56D,
       author = {{Dong}, Yize and {Valenti}, S. and {Bostroem}, K.~A. and {Sand}, D.~J. and {Andrews}, Jennifer E. and {Galbany}, L. and {Jha}, Saurabh W. and {Eweis}, Youssef and {Kwok}, Lindsey and {Hsiao}, E.~Y. and {Davis}, Scott and {Brown}, Peter J. and {Kuncarayakti}, H. and {Maeda}, Keiichi and {Rho}, Jeonghee and {Amaro}, R.~C. and {Anderson}, J.~P. and {Arcavi}, Iair and {Burke}, Jamison and {Dastidar}, Raya and {Folatelli}, Gast{\'o}n and {Haislip}, Joshua and {Hiramatsu}, Daichi and {Hosseinzadeh}, Griffin and {Howell}, D. Andrew and {Jencson}, J. and {Kouprianov}, Vladimir and {Lundquist}, M. and {Lyman}, J.~D. and {McCully}, Curtis and {Misra}, Kuntal and {Reichart}, Daniel E. and {S{\'a}nchez}, S.~F. and {Smith}, Nathan and {Wang}, Xiaofeng and {Wang}, Lingzhi and {Wyatt}, S.},
        title = "{Supernova 2018cuf: A Type IIP Supernova with a Slow Fall from Plateau}",
      journal = {\apj},
         year = 2021,
        month = jan,
       volume = {906},
       number = {1},
          eid = {56},
        pages = {56},
          doi = {10.3847/1538-4357/abc417},
archivePrefix = {arXiv},
       eprint = {2010.09764},
 primaryClass = {astro-ph.HE},
       adsurl = {https://ui.adsabs.harvard.edu/abs/2021ApJ...906...56D}
}

@ARTICLE{2016MNRAS.459.3939V,
       author = {{Valenti}, S. and {Howell}, D.~A. and {Stritzinger}, M.~D. and {Graham}, M.~L. and {Hosseinzadeh}, G. and {Arcavi}, I. and {Bildsten}, L. and {Jerkstrand}, A. and {McCully}, C. and {Pastorello}, A. and {Piro}, A.~L. and {Sand}, D. and {Smartt}, S.~J. and {Terreran}, G. and {Baltay}, C. and {Benetti}, S. and {Brown}, P. and {Filippenko}, A.~V. and {Fraser}, M. and {Rabinowitz}, D. and {Sullivan}, M. and {Yuan}, F.},
        title = "{The diversity of Type II supernova versus the similarity in their progenitors}",
      journal = {\mnras},
         year = 2016,
        month = jul,
       volume = {459},
       number = {4},
        pages = {3939-3962},
          doi = {10.1093/mnras/stw870},
archivePrefix = {arXiv},
       eprint = {1603.08953},
 primaryClass = {astro-ph.SR},
       adsurl = {https://ui.adsabs.harvard.edu/abs/2016MNRAS.459.3939V}
}

@ARTICLE{2006A&A...460..339J,
       author = {{Jordi}, K. and {Grebel}, E.~K. and {Ammon}, K.},
        title = "{Empirical color transformations between SDSS photometry and other photometric systems}",
      journal = {\aap},
         year = 2006,
        month = dec,
       volume = {460},
       number = {1},
        pages = {339-347},
          doi = {10.1051/0004-6361:20066082},
archivePrefix = {arXiv},
       eprint = {astro-ph/0609121},
 primaryClass = {astro-ph},
       adsurl = {https://ui.adsabs.harvard.edu/abs/2006A&A...460..339J}
}

@ARTICLE{LiLuhan2026A&A...714A.430L,
       author = {{Li}, Luhan and {Wang}, Bo and {Zhang}, Jujia and {Zhang}, Zhengyang and {Luo}, Xinjie and {Dong}, Shiyang and {Xu}, Saien and {Liu}, Zhengwei and {Han}, Zhanwen},
        title = "{A prolonged plateau-to-tail transition in the Type II supernova SN 2025abyc}",
      journal = {\aap},
         year = 2026,
        month = oct,
       volume = {714},
          eid = {A430},
        pages = {A430},
          doi = {10.1051/0004-6361/202661397},
       adsurl = {https://ui.adsabs.harvard.edu/abs/2026A&A...714A.430L}
}

@ARTICLE{2010ApJ...715..833O,
       author = {{Olivares E.}, Felipe and {Hamuy}, Mario and {Pignata}, Giuliano and {Maza}, Jos{\'e} and {Bersten}, Melina and {Phillips}, Mark M. and {Suntzeff}, Nicholas B. and {Filippenko}, Alexei V. and {Morrel}, Nidia I. and {Kirshner}, Robert P. and {Matheson}, Thomas},
        title = "{The Standardized Candle Method for Type II Plateau Supernovae}",
      journal = {\apj},
         year = 2010,
        month = jun,
       volume = {715},
       number = {2},
        pages = {833-853},
          doi = {10.1088/0004-637X/715/2/833},
archivePrefix = {arXiv},
       eprint = {1004.2534},
 primaryClass = {astro-ph.CO},
       adsurl = {https://ui.adsabs.harvard.edu/abs/2010ApJ...715..833O}
}

@ARTICLE{LuoJingxiao2025ApJ...982L..55L,
       author = {{Luo}, Jingxiao and {Zhang}, Lifu and {Chen}, Bing-Qiu and {Cheng}, Qiyuan and {Guo}, Boyang and {Li}, Jiao and {Guo}, Yanjun and {Xiong}, Jianping and {Meng}, Xiangcun and {Chen}, Xuefei and {Liu}, Zhengwei and {Han}, Zhanwen},
        title = "{The Red Supergiant Progenitor of the Type II Supernova 2024abfl}",
      journal = {\apjl},
         year = 2025,
        month = apr,
       volume = {982},
       number = {2},
          eid = {L55},
        pages = {L55},
          doi = {10.3847/2041-8213/adbf0b},
archivePrefix = {arXiv},
       eprint = {2412.13166},
 primaryClass = {astro-ph.SR},
       adsurl = {https://ui.adsabs.harvard.edu/abs/2025ApJ...982L..55L}
}

@ARTICLE{Teja2023ApJ...954..155T,
       author = {{Teja}, Rishabh Singh and {Singh}, Avinash and {Sahu}, D.~K. and {Anupama}, G.~C. and {Kumar}, Brajesh and {Nakaoka}, Tatsuya and {Kawabata}, Koji S. and {Yamanaka}, Masayuki and {Takey}, Ali and {Kawabata}, Miho},
        title = "{SN 2018gj: A Short Plateau Type II Supernova with Persistent Blueshifted Ha Emission}",
      journal = {\apj},
         year = 2023,
        month = sep,
       volume = {954},
       number = {2},
          eid = {155},
        pages = {155},
          doi = {10.3847/1538-4357/acdf5e},
archivePrefix = {arXiv},
       eprint = {2306.10136},
 primaryClass = {astro-ph.HE},
       adsurl = {https://ui.adsabs.harvard.edu/abs/2023ApJ...954..155T}
}

@ARTICLE{2021MNRAS.505.1742R,
       author = {{Rodr{\'\i}guez}, {\'O}. and {Meza}, N. and {Pineda-Garc{\'\i}a}, J. and {Ramirez}, M.},
        title = "{The iron yield of normal Type II supernovae}",
      journal = {\mnras},
         year = 2021,
        month = aug,
       volume = {505},
       number = {2},
        pages = {1742-1774},
          doi = {10.1093/mnras/stab1335},
archivePrefix = {arXiv},
       eprint = {2105.03268},
 primaryClass = {astro-ph.SR},
       adsurl = {https://ui.adsabs.harvard.edu/abs/2021MNRAS.505.1742R}
}

@ARTICLE{LiLuhan2026ApJ..1002...68L,
       author = {{Li}, Luhan and {Zhang}, Jujia and {Zhao}, Zeyi and {Li}, Liping and {Wang}, Xiaofeng and {Chen}, Liyang and {Wang}, Zeyi and {Luo}, Jingxiao and {Liu}, Zhengwei and {Han}, Zhanwen and {Wang}, Bo},
        title = "{SN 2024abfl: A Low-luminosity Type IIP Supernova at the Low-mass End of Core Collapse}",
      journal = {\apj},
         year = 2026,
        month = may,
       volume = {1002},
       number = {1},
          eid = {68},
        pages = {68},
          doi = {10.3847/1538-4357/ae5a9e},
archivePrefix = {arXiv},
       eprint = {2604.01806},
 primaryClass = {astro-ph.SR},
       adsurl = {https://ui.adsabs.harvard.edu/abs/2026ApJ..1002...68L}
}

@ARTICLE{Ravi2025ApJ...982...12R,
       author = {{Ravi}, Aravind P. and {Valenti}, Stefano and {Dong}, Yize and {Hiramatsu}, Daichi and {Barmentloo}, Stan and {Jerkstrand}, Anders and {Bostroem}, K. Azalee and {Pearson}, Jeniveve and {Shrestha}, Manisha and {Andrews}, Jennifer E. and {Sand}, David J. and {Hosseinzadeh}, Griffin and {Lundquist}, Michael and {Hoang}, Emily and {Mehta}, Darshana and {Meza Retamal}, Nicol{\'a}s and {Martas}, Aidan and {Jha}, Saurabh W. and {Janzen}, Daryl and {Subrayan}, Bhagya and {Howell}, D. Andrew and {McCully}, Curtis and {Farah}, Joseph and {Newsome}, Megan and {Padilla Gonzalez}, Estefania and {Terreran}, Giacomo and {Andrews}, Moira and {Filippenko}, Alexei V. and {Brink}, Thomas G. and {Zheng}, Weikang and {Yang}, Yi and {Vink{\'o}}, Jozsef and {Wheeler}, J. Craig and {Smith}, Nathan and {Rho}, Jeonghee and {K{\"o}nyves-T{\'o}th}, R{\'e}ka and {Guti{\'e}rrez}, Claudia P.},
        title = "{Luminous Type II Short-plateau SN 2023ufx: Asymmetric Explosion of a Partially Stripped Massive Progenitor}",
      journal = {\apj},
         year = 2025,
        month = mar,
       volume = {982},
       number = {1},
          eid = {12},
        pages = {12},
          doi = {10.3847/1538-4357/adb0bb},
archivePrefix = {arXiv},
       eprint = {2411.02493},
 primaryClass = {astro-ph.HE},
       adsurl = {https://ui.adsabs.harvard.edu/abs/2025ApJ...982...12R}
}

@ARTICLE{LinHan2025MNRAS.540.2591L,
       author = {{Lin}, Han and {Zhang}, Jujia and {Wang}, Xiaofeng and {Hu}, Maokai and {Zha}, Shuai and {Xiang}, Danfeng and {Li}, Liping and {Reguitti}, Andrea and {Zhang}, Xinghan and {Cai}, Yongzhi and {Wang}, Zhenyu and {Zhao}, Zeyi and {Zhai}, Qian and {Huang}, Fang and {Lin}, Weili and {Bai}, Jinming},
        title = "{SN 2022acko: a low-luminosity SNe IIP with signs of early circumstellar interaction}",
      journal = {\mnras},
         year = 2025,
        month = jul,
       volume = {540},
       number = {3},
        pages = {2591-2611},
          doi = {10.1093/mnras/staf893},
archivePrefix = {arXiv},
       eprint = {2512.23267},
 primaryClass = {astro-ph.HE},
       adsurl = {https://ui.adsabs.harvard.edu/abs/2025MNRAS.540.2591L}
}

@ARTICLE{Utrobin2024Ap&SS.369...49U,
       author = {{Utrobin}, V.~P. and {Chugai}, N.~N.},
        title = "{Revisiting short-plateau SN 2018gj}",
      journal = {\apss},
         year = 2024,
        month = may,
       volume = {369},
       number = {5},
          eid = {49},
        pages = {49},
          doi = {10.1007/s10509-024-04311-9},
archivePrefix = {arXiv},
       eprint = {2405.12867},
 primaryClass = {astro-ph.HE},
       adsurl = {https://ui.adsabs.harvard.edu/abs/2024Ap&SS.369...49U}
}

@ARTICLE{NiuZexi2026SciBu..71.1023N,
       author = {{Niu}, Zexi and {Sun}, Ning-Chen and {Zapartas}, Emmanouil and {Souropanis}, Dimitris and {Cui}, Yingzhen and {Maund}, Justyn R. and {Andrews}, Jeff J. and {Briel}, Max M. and {Fraser}, Morgan and {Gossage}, Seth and {Kruckow}, Matthias U. and {Liotine}, Camille and {Liu}, Zhengwei and {Podsiadlowski}, Philipp and {Srivastava}, Philipp M. and {Teng}, Elizabeth and {Wang}, Xiaofeng and {Yang}, Yi and {Liu}, Jifeng},
        title = "{A binary merger product as the direct progenitor of a Type II-P supernova}",
      journal = {Science Bulletin},
         year = 2026,
        month = mar,
       volume = {71},
       number = {5},
        pages = {1023-1033},
          doi = {10.1016/j.scib.2026.01.036},
       adsurl = {https://ui.adsabs.harvard.edu/abs/2026SciBu..71.1023N}
}

@ARTICLE{2012PASP..124..668Y,
       author = {{Yaron}, Ofer and {Gal-Yam}, Avishay},
        title = "{WISeREP{\textemdash}An Interactive Supernova Data Repository}",
      journal = {\pasp},
         year = 2012,
        month = jul,
       volume = {124},
       number = {917},
        pages = {668},
          doi = {10.1086/666656},
archivePrefix = {arXiv},
       eprint = {1204.1891},
 primaryClass = {astro-ph.IM},
       adsurl = {https://ui.adsabs.harvard.edu/abs/2012PASP..124..668Y}
}

@ARTICLE{Shingles2021TNSAN...7....1S,
       author = {{Shingles}, L. and {Smith}, K.~W. and {Young}, D.~R. and {Smartt}, S.~J. and {Tonry}, J. and {Denneau}, L. and {Heinze}, A. and {Weiland}, H. and {Flewelling}, H. and {Stalder}, B. and {Clocchiatti}, A. and {F{\"o}rster}, F. and {Pignata}, G. and {Rest}, A. and {Anderson}, J. and {Stubbs}, C. and {Erasmus}, N.},
        title = "{Release of the ATLAS Forced Photometry server for public use}",
      journal = {Transient Name Server AstroNote},
         year = 2021,
        month = jan,
       volume = {7},
        pages = {1-7},
       adsurl = {https://ui.adsabs.harvard.edu/abs/2021TNSAN...7....1S}
}

@ARTICLE{2015RAA....15..918F,
	author = {{Fan}, Yu-Feng and {Bai}, Jin-Ming and {Zhang}, Ju-Jia and {Wang}, Chuan-Jun and {Chang}, Liang and {Xin}, Yu-Xin and {Zhang}, Rui-Long},
	title = "{Rapid instrument exchanging system for the Cassegrain focus of the Lijiang 2.4-m Telescope}",
	journal = {Research in Astronomy and Astrophysics},
	year = 2015,
	month = jun,
	volume = {15},
	number = {6},
	eid = {918},
	pages = {918},
	doi = {10.1088/1674-4527/15/6/014},
	adsurl = {https://ui.adsabs.harvard.edu/abs/2015RAA....15..918F}
}

@ARTICLE{2019RAA....19..149W,
	author = {{Wang}, Chuan-Jun and {Bai}, Jin-Ming and {Fan}, Yu-Feng and
	{Mao}, Ji-Rong and {Chang}, Liang and {Xin}, Yu-Xin and
	{Zhang}, Ju-Jia and {Lun}, Bao-Li and {Wang}, Jian-Guo and
	{Zhang}, Xi-Liang and {Ying}, Mei and {Lu}, Kai-Xing and
	{Wang}, Xiao-Li and {Ji}, Kai-Fan and {Xiong}, Ding-Rong and
	{Yu}, Xiao-Guang and {Ding}, Xu and {Ye}, Kai and {Xing}, Li-Feng and
	{Yi}, Wei-Min and {Xu}, Liang and {Zheng}, Xiang-Ming and
	{Feng}, Yuan-Jie and {He}, Shou-Sheng and {Wang}, Xue-Li and
	{Liu}, Zhong and {Chen}, Dong and {Xu}, Jun and {Qin}, Song-Nian and
	{Zhang}, Rui-Long and {Tan}, Hui-Song and {Li}, Zhi and {Lou}, Ke and
	{Li}, Jian and {Liu}, Wei-Wei},
	title = "{Lijiang 2.4-meter Telescope and its instruments}",
	journal = {Research in Astronomy and Astrophysics},
	year = 2019,
	month = oct,
	volume = {19},
	number = {10},
	eid = {149},
	pages = {149},
	doi = {10.1088/1674-4527/19/10/149},
	archivePrefix = {arXiv},
	eprint = {1905.05915},
	primaryClass = {astro-ph.IM},
	adsurl = {https://ui.adsabs.harvard.edu/abs/2019RAA....19..149W}
}

@ARTICLE{2022A&A...667A..62B,
       author = {{Brennan}, S.~J. and {Fraser}, M.},
        title = "{The Automated Photometry of Transients pipeline (AUTOPHOT)}",
      journal = {\aap},
         year = 2022,
        month = nov,
       volume = {667},
          eid = {A62},
        pages = {A62},
          doi = {10.1051/0004-6361/202243067},
archivePrefix = {arXiv},
       eprint = {2201.02635},
 primaryClass = {astro-ph.IM},
       adsurl = {https://ui.adsabs.harvard.edu/abs/2022A&A...667A..62B}
}

@ARTICLE{Yang2021A&A...655A..90Y,
       author = {{Yang}, S. and {Sollerman}, J. and {Strotjohann}, N.~L. and {Schulze}, S. and {Lunnan}, R. and {Kool}, E. and {Fremling}, C. and {Perley}, D. and {Ofek}, E. and {Schweyer}, T. and {Bellm}, E.~C. and {Kasliwal}, M.~M. and {Masci}, F.~J. and {Rigault}, M. and {Yang}, Y.},
        title = "{A low-energy explosion yields the underluminous Type IIP SN 2020cxd}",
      journal = {\aap},
         year = 2021,
        month = nov,
       volume = {655},
          eid = {A90},
        pages = {A90},
          doi = {10.1051/0004-6361/202141244},
archivePrefix = {arXiv},
       eprint = {2107.13439},
 primaryClass = {astro-ph.HE},
       adsurl = {https://ui.adsabs.harvard.edu/abs/2021A&A...655A..90Y}
}

@ARTICLE{WangZeyi2026ApJ..1001..156W,
       author = {{Wang}, Zeyi and {Zhang}, Jujia and {Zhai}, Qian and {Li}, Liping and {Valerin}, G. and {Reguitti}, A. and {Pastorello}, A. and {Wang}, Zhenyu and {Zhao}, Zeyi and {Song}, Tengfei and {Cai}, Yongzhi},
        title = "{SN 2023axu: A Type IIP Supernova Interacted with a Low-density Stellar Wind}",
      journal = {\apj},
         year = 2026,
        month = apr,
       volume = {1001},
       number = {2},
          eid = {156},
        pages = {156},
          doi = {10.3847/1538-4357/ae4fae},
archivePrefix = {arXiv},
       eprint = {2603.11646},
 primaryClass = {astro-ph.HE},
       adsurl = {https://ui.adsabs.harvard.edu/abs/2026ApJ..1001..156W}
}

@INCOLLECTION{Blinnikov2017hsn..book..843B,
       author = {{Blinnikov}, Sergei},
        title = "{Interacting Supernovae: Spectra and Light Curves}",
    booktitle = {Handbook of Supernovae},
         year = 2017,
       editor = {{Alsabti}, Athem W. and {Murdin}, Paul},
        pages = {843},
          doi = {10.1007/978-3-319-21846-5_31},
       adsurl = {https://ui.adsabs.harvard.edu/abs/2017hsn..book..843B}
}

@ARTICLE{Blinnikov2006A&A...453..229B,
       author = {{Blinnikov}, S.~I. and {R{\"o}pke}, F.~K. and {Sorokina}, E.~I. and {Gieseler}, M. and {Reinecke}, M. and {Travaglio}, C. and {Hillebrandt}, W. and {Stritzinger}, M.},
        title = "{Theoretical light curves for deflagration models of type Ia supernova}",
      journal = {\aap},
         year = 2006,
        month = jul,
       volume = {453},
       number = {1},
        pages = {229-240},
          doi = {10.1051/0004-6361:20054594},
archivePrefix = {arXiv},
       eprint = {astro-ph/0603036},
 primaryClass = {astro-ph},
       adsurl = {https://ui.adsabs.harvard.edu/abs/2006A&A...453..229B}
}

@ARTICLE{Blinnikov1993A&A...273..106B,
       author = {{Blinnikov}, S.~I. and {Bartunov}, O.~S.},
        title = "{Non-equilibrium radiative transfer in supernova theory : models of linear type II supernovae.}",
      journal = {\aap},
         year = 1993,
        month = jun,
       volume = {273},
        pages = {106-122},
       adsurl = {https://ui.adsabs.harvard.edu/abs/1993A&A...273..106B}
}

@ARTICLE{Paxton2019ApJS..243...10P,
       author = {{Paxton}, Bill and {Smolec}, R. and {Schwab}, Josiah and {Gautschy}, A. and {Bildsten}, Lars and {Cantiello}, Matteo and {Dotter}, Aaron and {Farmer}, R. and {Goldberg}, Jared A. and {Jermyn}, Adam S. and {Kanbur}, S.~M. and {Marchant}, Pablo and {Thoul}, Anne and {Townsend}, Richard H.~D. and {Wolf}, William M. and {Zhang}, Michael and {Timmes}, F.~X.},
        title = "{Modules for Experiments in Stellar Astrophysics (MESA): Pulsating Variable Stars, Rotation, Convective Boundaries, and Energy Conservation}",
      journal = {\apjs},
         year = 2019,
        month = jul,
       volume = {243},
       number = {1},
          eid = {10},
        pages = {10},
          doi = {10.3847/1538-4365/ab2241},
archivePrefix = {arXiv},
       eprint = {1903.01426},
 primaryClass = {astro-ph.SR},
       adsurl = {https://ui.adsabs.harvard.edu/abs/2019ApJS..243...10P}
}

@ARTICLE{Paxton2018ApJS..234...34P,
       author = {{Paxton}, Bill and {Schwab}, Josiah and {Bauer}, Evan B. and {Bildsten}, Lars and {Blinnikov}, Sergei and {Duffell}, Paul and {Farmer}, R. and {Goldberg}, Jared A. and {Marchant}, Pablo and {Sorokina}, Elena and {Thoul}, Anne and {Townsend}, Richard H.~D. and {Timmes}, F.~X.},
        title = "{Modules for Experiments in Stellar Astrophysics (MESA): Convective Boundaries, Element Diffusion, and Massive Star Explosions}",
      journal = {\apjs},
         year = 2018,
        month = feb,
       volume = {234},
       number = {2},
          eid = {34},
        pages = {34},
          doi = {10.3847/1538-4365/aaa5a8},
archivePrefix = {arXiv},
       eprint = {1710.08424},
 primaryClass = {astro-ph.SR},
       adsurl = {https://ui.adsabs.harvard.edu/abs/2018ApJS..234...34P}
}

@ARTICLE{Paxton2013ApJS..208....4P,
       author = {{Paxton}, Bill and {Cantiello}, Matteo and {Arras}, Phil and {Bildsten}, Lars and {Brown}, Edward F. and {Dotter}, Aaron and {Mankovich}, Christopher and {Montgomery}, M.~H. and {Stello}, Dennis and {Timmes}, F.~X. and {Townsend}, Richard},
        title = "{Modules for Experiments in Stellar Astrophysics (MESA): Planets, Oscillations, Rotation, and Massive Stars}",
      journal = {\apjs},
         year = 2013,
        month = sep,
       volume = {208},
       number = {1},
          eid = {4},
        pages = {4},
          doi = {10.1088/0067-0049/208/1/4},
archivePrefix = {arXiv},
       eprint = {1301.0319},
 primaryClass = {astro-ph.SR},
       adsurl = {https://ui.adsabs.harvard.edu/abs/2013ApJS..208....4P}
}

@ARTICLE{Paxton2011ApJS..192....3P,
       author = {{Paxton}, Bill and {Bildsten}, Lars and {Dotter}, Aaron and {Herwig}, Falk and {Lesaffre}, Pierre and {Timmes}, Frank},
        title = "{Modules for Experiments in Stellar Astrophysics (MESA)}",
      journal = {\apjs},
         year = 2011,
        month = jan,
       volume = {192},
       number = {1},
          eid = {3},
        pages = {3},
          doi = {10.1088/0067-0049/192/1/3},
archivePrefix = {arXiv},
       eprint = {1009.1622},
 primaryClass = {astro-ph.SR},
       adsurl = {https://ui.adsabs.harvard.edu/abs/2011ApJS..192....3P}
}

@ARTICLE{Barbon1979A&A....72..287B,
       author = {{Barbon}, R. and {Ciatti}, F. and {Rosino}, L.},
        title = "{Photometric properties of type II supernovae.}",
      journal = {\aap},
         year = 1979,
        month = feb,
       volume = {72},
        pages = {287-292},
       adsurl = {https://ui.adsabs.harvard.edu/abs/1979A&A....72..287B}
}

@BOOK{Vink2020pesr.book.....V,
       author = {{Vink}, Jacco},
        title = "{Physics and Evolution of Supernova Remnants}",
         year = 2020,
          doi = {10.1007/978-3-030-55231-2},
       adsurl = {https://ui.adsabs.harvard.edu/abs/2020pesr.book.....V}
}

\onecolumn
\appendix

\section{Supplementary material}

\begin{figure}[htbp]
    \centering
    \includegraphics[width=0.48\linewidth]{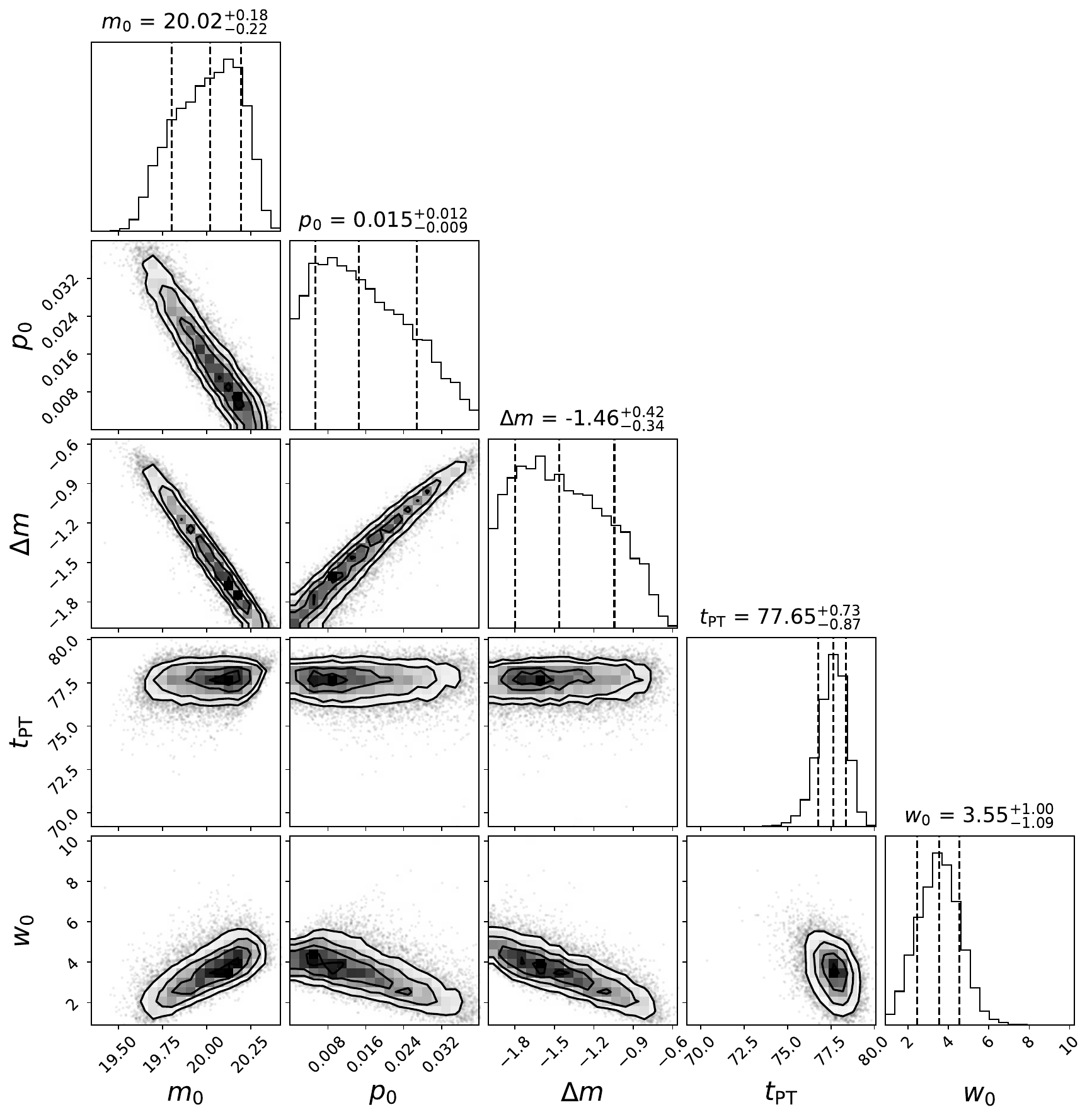}
    \hfill
    \includegraphics[width=0.48\linewidth]{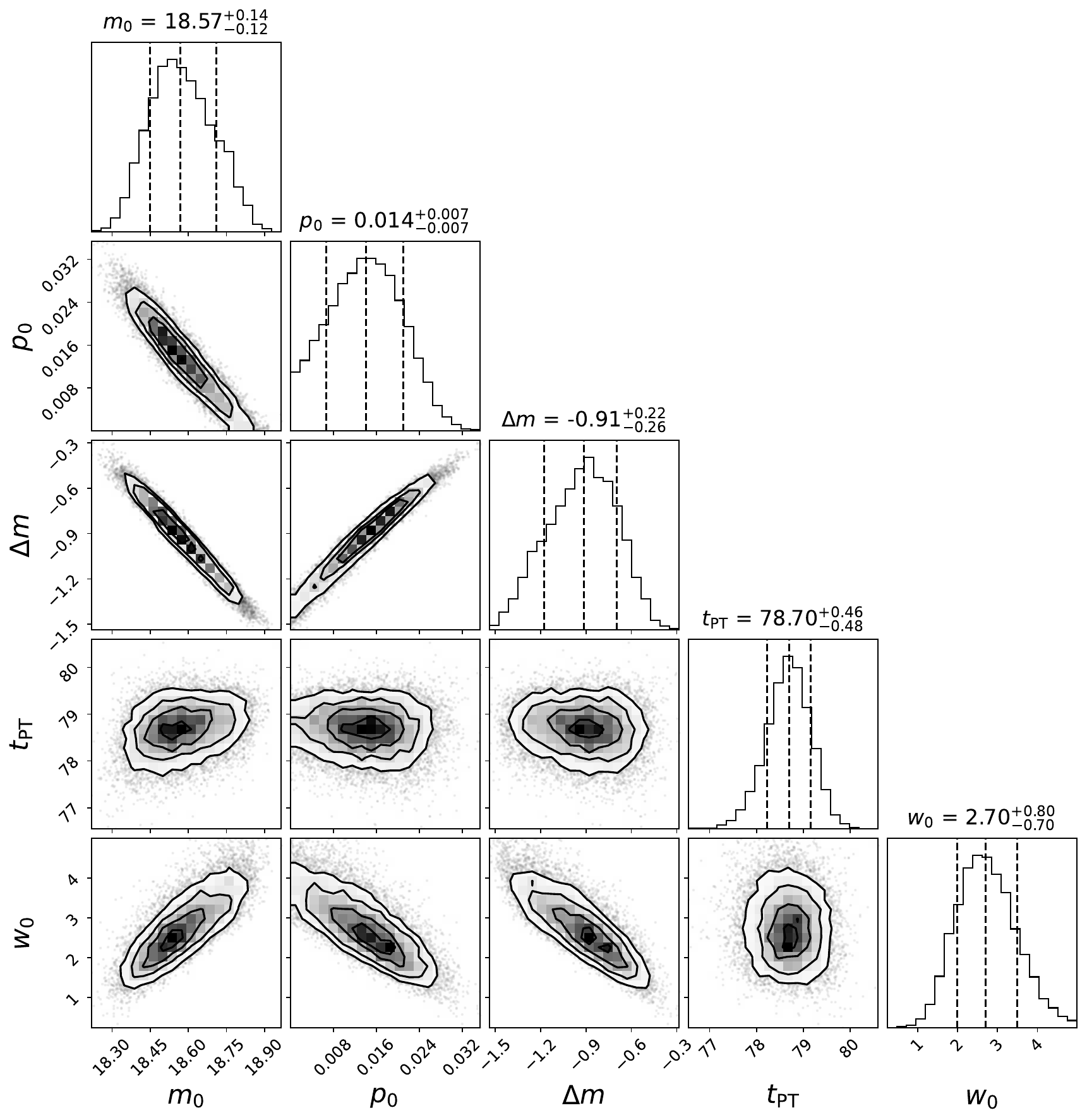}
    \caption{Corner plot of the posterior distributions for the Fermi-Dirac transition fit to the SN\,2025aedz $g$-band (left) and $r$-band (right) light curve.
   The diagonal panels show the marginalized 1D posteriors for $m_0$, $p_0$, $\Delta m$, $t_{\rm PT}$ and $w_0$, with titles reporting the median and 16th/84th percentile uncertainties. }
    \label{Fig:FDmodel_corner}
\end{figure}

\begin{table}[htbp]
\centering
\caption{SN\,2025aedz photometry in $ugriz$ bands based on the LJT and MATCH.}
\label{tab:phot_data}
\begin{tabular}{cccccc}
\hline\hline
MJD & $u$ & $g$ & $r$ & $i$ & $z$ \\
\hline
61008.7  & $17.55 \pm 0.26$ & $17.19 \pm 0.06$ & $16.91 \pm 0.05$ & $16.80 \pm 0.09$ & $16.89 \pm 0.05$ \\
61009.7  & $17.60 \pm 0.49$ & $17.26 \pm 0.05$ & $16.96 \pm 0.04$ & $16.87 \pm 0.09$ & $16.93 \pm 0.04$ \\
61011.6  & $17.78 \pm 0.15$ & $17.33 \pm 0.06$ & $17.02 \pm 0.04$ & $16.96 \pm 0.07$ & $16.99 \pm 0.05$ \\
61012.6  & $17.82 \pm 0.13$ & $17.37 \pm 0.07$ & $17.02 \pm 0.05$ & $16.98 \pm 0.04$ & $17.04 \pm 0.06$ \\
61012.9  & -- & $17.34 \pm 0.05$ & $17.04 \pm 0.05$ & $16.99 \pm 0.09$ & -- \\
61016.7  & -- & $17.48 \pm 0.07$ & $17.09 \pm 0.05$ & $17.07 \pm 0.06$ & -- \\
61018.8  & $18.64 \pm 0.13$ & $17.53 \pm 0.05$ & $17.09 \pm 0.07$ & $17.05 \pm 0.07$ & $17.07 \pm 0.05$ \\
61022.8  & -- & $17.67 \pm 0.03$ & $17.14 \pm 0.04$ & $17.09 \pm 0.06$ & $17.15 \pm 0.05$ \\
61026.7  & $19.46 \pm 0.12$ & $17.80 \pm 0.07$ & $17.22 \pm 0.05$ & $17.20 \pm 0.07$ & $17.23 \pm 0.05$ \\
61031.7  & -- & $17.91 \pm 0.05$ & $17.25 \pm 0.05$ & $17.22 \pm 0.08$ & $17.42 \pm 0.19$ \\
61038.7  & $20.28 \pm 0.26$ & $18.02 \pm 0.08$ & $17.29 \pm 0.05$ & $17.27 \pm 0.07$ & $17.30 \pm 0.04$ \\
61045.7  & -- & $18.12 \pm 0.06$ & $17.34 \pm 0.06$ & $17.32 \pm 0.07$ & $17.33 \pm 0.06$ \\
61053.8  & -- & $18.25 \pm 0.05$ & $17.35 \pm 0.06$ & $17.35 \pm 0.08$ & $17.36 \pm 0.05$ \\
61059.6  & -- & $18.43 \pm 0.05$ & $17.49 \pm 0.05$ & $17.46 \pm 0.08$ & $17.42 \pm 0.07$ \\
61065.6  & -- & $18.66 \pm 0.08$ & $17.62 \pm 0.06$ & $17.59 \pm 0.07$ & $17.50 \pm 0.06$ \\
61072.7  & -- & $19.19 \pm 0.13$ & $17.98 \pm 0.06$ & $17.93 \pm 0.12$ & $17.78 \pm 0.07$ \\
61073.6  & -- & $19.26 \pm 0.12$ & -- & -- & -- \\
61074.7  & -- & $19.30 \pm 0.48$ & $17.95 \pm 0.28$ & $17.78 \pm 0.32$ & -- \\
61076.6  & -- & $19.66 \pm 0.09$ & $18.35 \pm 0.05$ & $18.26 \pm 0.07$ & -- \\
61077.6  & -- & $19.84 \pm 0.07$ & $18.46 \pm 0.05$ & $18.35 \pm 0.11$ & -- \\
61078.6  & -- & $19.79 \pm 0.11$ & $18.52 \pm 0.08$ & $18.43 \pm 0.11$ & -- \\
61080.8  & -- & $19.99 \pm 0.09$ & $18.56 \pm 0.10$ & $18.56 \pm 0.09$ & -- \\
61088.6  & -- & $20.22 \pm 0.09$ & $18.77 \pm 0.06$ & $18.76 \pm 0.11$ & -- \\
61091.6  & -- & $20.26 \pm 0.10$ & $18.75 \pm 0.07$ & $18.80 \pm 0.08$ & -- \\
61096.6  & -- & -- & $18.86 \pm 0.13$ & $18.92 \pm 0.17$ & -- \\
61097.6  & -- & $20.05 \pm 0.33$ & $18.90 \pm 0.14$ & $18.86 \pm 0.24$ & -- \\
61101.6  & -- & $<20.34$ & $19.04 \pm 0.20$ & $19.07 \pm 0.30$ & -- \\
\hline
\end{tabular}
\end{table}

\begin{table}[t]
\caption{Log of spectroscopic observations of SN\,2025aedz based on LJT.}
\label{table:specinfo}
\centering
\small
\setlength{\tabcolsep}{6pt}
\begin{tabular}{c c c c c c c} 
\hline\hline
Date & MJD  & Phase$^*$ & Telescope/Instrument & Grism+Slit & Airmass & Exp.time(s) \\
\hline
2025-11-29 & 61008.8 & 13.2 & LJT/YFOSC & G3+2.51" & 1.19 & 600*2 \\
2025-11-30 & 61009.8 & 14.2 & LJT/YFOSC & G3+2.51" & 1.19 & 600*2 \\
2025-12-02 & 61011.7 & 16.1 & LJT/YFOSC & G3+2.51" & 1.36 & 700*2 \\
2025-12-03 & 61012.9 & 17.3 & LJT/YFOSC & G3+2.51" & 1.37 & 800*2 \\
2025-12-07 & 61016.8 & 21.2 & LJT/YFOSC & G3+2.51" & 1.18 & 2000 \\
2025-12-09 & 61018.8 & 23.2 & LJT/YFOSC & G3+2.51" & 1.21 & 2000 \\
2025-12-12 & 61021.7 & 26.1 & LJT/YFOSC & G3+2.51" & 1.20 & 2000 \\
2025-12-17 & 61026.7 & 31.1 & LJT/YFOSC & G3+2.51" & 1.18 & 2000 \\
2025-12-22 & 61031.7 & 36.1 & LJT/YFOSC & G3+2.51" & 1.20 & 2000 \\
2025-12-29 & 61038.7 & 43.1 & LJT/YFOSC & G3+2.51" & 1.20 & 2000 \\
2026-01-04 & 61044.8 & 49.2 & LJT/YFOSC & G3+2.51" & 1.43 & 2000 \\
2026-01-13 & 61053.8 & 58.2 & LJT/YFOSC & G3+2.51" & 1.41 & 2400 \\
2026-01-19 & 61059.7 & 64.1 & LJT/YFOSC & G3+2.51" & 1.19 & 1900 \\
2026-01-25 & 61065.6 & 70.0 & LJT/YFOSC & G3+2.51" & 1.18 & 2400 \\
2026-02-02 & 61073.6 & 78.0 & LJT/YFOSC & G3+2.51" & 1.21 & 3000 \\
\hline
\multicolumn{7}{l}{{$^*$Phases are relative to the estimated explosion epoch (MJD = $60995.6$) in observer frame.}}
\end{tabular}
\end{table}

\begin{table}[t]
    \centering
    \small
    \setlength{\tabcolsep}{7pt} 
    \caption{Properties of the comparison SNe IIP.}
    \begin{tabular}{cccccc}
        \hline\hline
        SN & Explosion epoch [MJD] & Redshift & Distance [Mpc] & $E(B-V)_\mathrm{tot}$ & References \\
        \hline
        \multicolumn{6}{c}{Normal SNe IIP} \\ \hline
        1999em & 51475.1$\pm$1.4 & 0.0024 & 8.2$\pm$2.6 & 0.10$\pm$0.05 & 1 \\
        2009N &  54847.6 & 0.0035 & 21.6$\pm$1.1 & 0.13$\pm$0.02 & 2 \\
        2017eaw & 57886.5$\pm$0.1 & 0.00013 & 6.85$\pm$0.63 & 0.41 & 3 \\
        \hline
        \multicolumn{6}{c}{Short-plateau SNe IIP} \\ \hline
        2016egz & 57588.2$\pm$ 2.0 & 0.0232 & 100 & 0.014 & 4 \\ 
        2017ahn & 57791.8$\pm$0.5& 0.009 & $33.0\pm6.5$ & 0.196$\pm$0.054 & 5 \\
        2018gj &  58127.3$\pm$1.4& 0.00454 & $19.61\pm1.37$ & 0.08$\pm$0.02 & 6 \\
        2020jfo & 58973.7 & 0.0052 & $14.5$ & 0.065 & 7 \\
        2023ufx & 60223.3$\pm$0.5 & 0.0146 & 63.2$\pm$3.1 & 0.04 & 8 \\
        \hline
    \end{tabular}%
    \begin{flushleft}
    \small
    References: 
    1. \citet{Leonard2002PASP..114...35L,Elmhamdi2003MNRAS.338..939E},
    2. \citet{Takats2014MNRAS.438..368T},
    3. \citet{Szalai2019ApJ...876...19S,VanDyk2019ApJ...875..136V},
    4. \citet{Hiramatsu2021ApJ...913...55H},
    5. \citet{Tartaglia2021ApJ...907...52T},
    6. \citet{Teja2023ApJ...954..155T},
    7. \citet{Sollerman2021A&A...655A.105S,Teja2022ApJ...930...34T,Kilpatrick2023MNRAS.524.2161K,
    Ailawadhi2023MNRAS.519..248A},
    8. \citet{2024ApJ...976..178T,Ravi2025ApJ...982...12R}.
    \end{flushleft}
    \label{tab:SNe_II_info}
\end{table}

\end{document}